\documentclass[preprint, aps, prd, superscriptaddress, nofootinbib, floatfix]{revtex4-2}

\usepackage[T1]{fontenc}
\usepackage[utf8]{inputenc}
\usepackage{amsmath, amssymb, physics, bm}
\usepackage{graphicx, xcolor}
\usepackage{siunitx}
\usepackage{ulem}
\usepackage{booktabs}
\usepackage{comment}
\usepackage{stackengine}
\usepackage{adjustbox}
\usepackage{multirow}
\usepackage{caption}
\usepackage{geometry}
\usepackage{accents}
\usepackage{trimclip}

\newcommand\ringring[1]{%
  {\mathop{\kern0pt #1}\limits^{
    \vbox to-1.85ex{
      \kern-2ex
      \hbox to 0pt{\hss\normalfont\kern.1em \r{}\kern-.45em \r{}\hss}
      \vss
    }
  }}
}
\usepackage[colorlinks=true, linkcolor=blue, citecolor=blue, urlcolor=blue]{hyperref}
\usepackage{cleveref}

\begin{document}

\flushbottom % important for better vertical spacing control

\title{\Large Large Scale White Noise\\ and Cosmology}

\author{Gabriela Barenboim}
\affiliation{Instituto de F\'{i}sica Corpuscular, CSIC-Universitat de Val\`{e}ncia, Paterna 46980, Spain}
\affiliation{Departament de F\'{i}sica Te\`{o}rica, Universitat de Val\`{e}ncia, Burjassot 46100, Spain}

\author{Aurora Ireland}
\affiliation{Department of Physics, University of Chicago, Chicago, IL 60637, USA}
\affiliation{Leinweber Institute for Theoretical Physics, Stanford University, Stanford, CA 94305, USA}

\author{Albert Stebbins}
\affiliation{Fermi National Accelerator Laboratory, Theoretical Astrophysics Group, Batavia, IL 60510, USA}

\email{gabriela.barenboim@uv.es}
\email{anireland@stanford.edu}
\email{stebbins@fnal.gov}

\begin{abstract}
The generation of white noise on large scales is a generic property of the dynamics of physical systems described by local non-linear partial differential equations. Non-linearities prevent the small scale dynamics from being erased by smoothing. Unresolved small scale dynamics act as an uncorrelated (white or Poissonian) noise (seemingly stochastic but actually deterministic) contribution to large scale dynamics. This white noise exists even when the dynamics is very nearly linear. In cases where the power spectrum is sub-Poissonian on large scales, this noise will dominate on the largest scale power no matter the amplitude of the inhomogeneities. Such is the case in the standard model of cosmology, where the primordial density power spectrum is expected to have an almost Harrison-Zel'dovich, $P[k]\sim k$, spectrum on a much broader range of scales than can be observed. Even though linear gravitational evolution dominates non-linear corrections by a factor $\sim10^5$, the non-observation of white noise on the Hubble scale precludes the extrapolation of this power law below the comoving $1\,$pc scale. More generally, observation or non-observation of large scale white noise provides a powerful probe of the universe on very small scales in the early  universe. Gravitational radiation, phase transitions, vorticity, and running of the spectral index are all phenomena that can be probed with large scale white noise. Large scale white noise is a non-optional feature of all cosmological models but one which has not heretofore been appreciated.
\end{abstract}

\maketitle

\pagebreak

%%%%%%%%%%%%%%%%%%%%%%%%%%%%%%%%%%%%%%%%%%%%%%%%%%%%%%%%%%%%
%%%%%%%%%%%%%%%%%%%%%%%%%%%%%%%%%%%%%%%%%%%%%%%%%%%%%%%%%%%%
\section{Introduction}\label{sec:introduction}

We observe a universe which is nearly homogeneous and isotropic.  On large scales or at early times, the inhomogeneities are apparently extremely small, a fractional perturbation of only $\sim10^{-5}$.  The evolution of the inhomogeneities are given by equations of hydrodynamics and general relativity.  These equations are non-linear but, as the inhomogeneities are so small, one might think that one could accurately linearize these equations about a homogeneous fluid to describe the evolutions of the inhomogeneities. This is not true!

The purpose of this paper is to justify the last statement.  This statement should more precisely say that generically one cannot accurately use linear theory to describe the evolution on {\it all} scales.  Linear theory may be accurate on a large range of scales but not on the largest scales.  The evolutions of inhomogeneities with the longest wavelengths depends strongly on small non-linearities in the evolution of inhomogeneities with much shorter wavelengths.  This is true no matter how small the amplitude of inhomogeneities are. The ratio of the long to the short wavelengths will be larger the smaller the amplitude of inhomogeneities is.  Thus, for the extremely small inhomogeneities observed, one would have to cover a huge dynamic range of scales in order to uncover the failure of linear theory.

This failure of linear theory is inevitable for {\it any} dynamical system with properties similar to those of the standard model of cosmology (by system, we mean both the equations of motion and the initial conditions).  A primary manifestation of the non-linear effects described here is the inevitability of large scale white noise (LSWN).

%%%%%%%%%%%%%%%%%%%%%%%%%%%%%%%%%%%%%%%%%%%%%%%%%%%%%%%%%%%%
%%%%%%%%%%%%%%%%%%%%%%%%%%%%%%%%%%%%%%%%%%%%%%%%%%%%%%%%%%%%
\subsection{Linear, Non-linear, and Pseudo-Linear}
\label{sec:PseudoLinearity}

An equation is said to be linear if multiplying any solution by a constant factor yields another solution. We often use linear or linearized equations to describe the evolution of physical systems.  Linearized here means that the equations of motion which are known not to be linear are approximated by the first term in a Taylor series about a known solution (usually zero).  This is the usual description of early universe cosmological inhomogeneities.  Again, this seems reasonable because the higher order terms in the Taylor series are at least $\sim10^{-5}$ times smaller than the linear terms.

A solution or class of solutions where the non-linear terms are much smaller than the linear terms we say is in the {\it linear regime}.  Where the non-linear terms are much larger than the linear terms we call the {\it non-linear regime} and the intermediate case we call the {\it quasi-linear regime}.  The boundary between these regimes is admittedly rather fuzzy, but the standard model of cosmological inhomogeneities in the early universe lies squarely in the linear regime.

No matter which regime the system is in, we refer to any consequence of the non-linear terms in the equation of motion as {\it non-linearities}. Thus, there are non-linearities in the linear regime as well as in the quasi-linear and non-linear regimes. Since the precise equations of motion of cosmological inhomogeneities are not strictly linear, the evolution of inhomogeneities in the early universe has non-linearities.  One might think that non-linearities in the linear regime are necessarily unimportant because they are small, but this is not necessarily the case. This is the main topic of this paper.

The type of non-linearities in the linear regime which are most familiar are slow drifts in the parameters of linear solution.  These drifts are slow because the non-linearities are small. One can usually use two-timing methods to model such systems; separating the fast dynamical timescale from the slow drift timescale. This is {\it not} the situation of interest in this paper.

A system exhibiting important rapid deviations from linear solutions while in the linear regime is what we call a {\it pseudo-linear} system. It is ``pseudo'' because it is not what one expects from any system in the linear regime. Here ``rapid'' means on the dynamical timescale of the system and ``important'' is a subjective criterion based on what is physically or observationally important.\footnote{One can always define importance such that any system is pseudo-linear e.g. by defining importance as any deviation from linear behavior.}  The topic of this paper is one type of pseudo-linearity.

A generic way that a system becomes pseudo-linear is through {\it spillover}.  This occurs when some part of the Hilbert space of linear solutions has very little ``power'' in it becomes coupled through non-linearities with other parts of Hilbert space which are not so empty.  Even very small non-linearities can cause a transfer of power which overwhelms the linear theory prediction of the power spectrum in these initially empty regions.  This is only possible because they are initially empty and not because the non-linearities are large.  Heuristically, pseudo-linearity occurs when the smallness of the non-linearities which couples empty and non-empty regions of the Hilbert space exceeds the power imbalance in these two regions.

As we will show, the standard model of cosmology is just such a pseudo-linear system when one considers the spatial power spectrum of inhomogeneities to be important.  The inhomogeneities are very small on all scales so the system is in the linear regime and non-linear terms in the equations of motion are much smaller than linear term.  However, non-linearities couple short wavelength modes to long-wavelength modes.  The standard model of cosmology has a power spectrum with more power at small wavelengths than at large wavelengths and this is enough to allow the power spilled from short to long wavelengths to dominate the initial long wavelength power (or lack thereof); thus creating large scale white noise (LSWN). \footnote{This is true for the relevant definition of inhomogeneity, as we will show.}  More specifically the generic white noise ($k^0$) spillover to large scales has more power on large scales than the standard model power spectrum which is close to a Harrison-Zel'dovich (HZ) spectrum ($k^1$).  Thus, we should expect cosmological large scale white noise (LSWN).

%%%%%%%%%%%%%%%%%%%%%%%%%%%%%%%%%%%%%%%%%%%%%%%%%%%%%%%%%%%%
%%%%%%%%%%%%%%%%%%%%%%%%%%%%%%%%%%%%%%%%%%%%%%%%%%%%%%%%%%%%
\subsection{Misconceptions}
\label{sec:Misconceptions}

Spillover from short to long wavelength inhomogeneities in cosmology has been considered previously.  It has erroneously been thought that small scale causal processes can at most produce a $k^4$ long wavelength tail to the power spectrum (see \cite{Peebles1971,Zeldovich1972,Peebles1980,Weinberg2008}).  A basis for this misconception relies on the idea that there is large scale (global) conservation for energy and momentum.  This goes beyond the local energy momentum conservation provided by GR.  In linear theory, local conservations laws extend to global conservations laws.  This leads to the circular argument: if linear theory is accurate then there are conservations laws and one can thereby show that linear theory is accurate.  Linear theory is a consistent model, but one that is inevitably inaccurate on the largest scales.  To see the inaccuracy, one must include non-linearities. One should not, as we often do, take it as self-evident that just because the inhomogeneities are small and the non-linearities even smaller that they are unimportant in parts of the spectrum where the power is also small.

Another component of the argument that one cannot spontaneously  produce a significant long wavelength tail is causality: a large long wavelength tail would require acausal correlations on super-horizon scales.  There is a causality argument that applies directly to spatial correlations, but its implications to the power spectrum is more subtle.  We concur that there should be zero acausal spatial correlations of locally measurable quantities. However, there is no inconsistency between spontaneously creating a white noise power spectrum on large scales and not spontaneously creating acausal correlations. Indeed, a white noise power spectrum is consistent with a \textit{lack} of spatial correlations.

Another misconception is the argument that since we measure temperature fluctuations in the CMBR correlated at large angles that this implies acausal correlations in the cosmological fluid.  The observed temperature correlations as viewed from Earth does not imply that at all.  In the standard model, the temperature correlations we measure at large angles are primarily due to gravitational potential differences between us and the surface of last scattering at two different points.  While it is true that gravitational potentials have large correlations outside the horizon this is not a sign of acausality because gravitational potentials are not locally measurable quantities. Linearized general relativity (GR) predicts instantaneous action-at-a-distance changes in gravitational potentials just as Newtonian gravity does.

In summary, one can spontaneously produce much more large wavelength inhomogeneity than is implied by the incorrect $k^4$ argument whilst respecting causality.

%%%%%%%%%%%%%%%%%%%%%%%%%%%%%%%%%%%%%%%%%%%%%%%%%%%%%%%%%%%%
%%%%%%%%%%%%%%%%%%%%%%%%%%%%%%%%%%%%%%%%%%%%%%%%%%%%%%%%%%%%
\subsection{Pseudo-Linear Cosmology}
\label{sec:RoughQuantitativeEstimates}

Here we give the heuristic numerology of cosmological inhomogeneities with large scale white noise (LSWN).  A more rigorous discussion is given in \S\ref{sec:LSWNinCosmology}.  One can assume an approximate Harrison-Zel'dovich (HZ) + white noise power spectrum: 
\begin{equation}\label{eq:Pdelta}
	P[t,k] \simeq A[t]^2 \left( k + k_{\rm LSWN} \right) \,,
\end{equation}
assuming the HZ component is primordial and the LSWN component is generated by non-linearities in the evolution of the primordial component.  Here $A[t]$ gives the linear rate of growth of inhomogeneities and $k_{\rm LSWN}$ is the wavenumber where the two components give equal power.  $k_{\rm LSWN}$ parametrizes the amplitude of white noise generated by non-linearities on some small scale $k\sim k_\mathrm{nl}$.  The amplitude of inhomogeneities at horizon crossing is $\sim10^{-5}$ corresponding to a ``power'' $\sim10^{-10}$.  The dominant non-linearities are quadratic, so at $k\sim k_\mathrm{nl}$ the leading order non-linearities are $\sim10^{-10}$ times smaller than the HZ spectrum. Thus $k_\mathrm{nl}\sim 10^{10}\,k_\mathrm{LSWN}$.  Since the inhomogeneities are so small, there is a huge difference between the scale of non-linearity that generates LSWN and the scale where LSWN becomes dominant.  In general, pseudo-linearity of the type described here is only important for sufficiently large systems.  Cosmology is such a spatially large system.

On observable scales, we see an HZ --- not a LSWN --- spectrum.  This implies that the $k_\mathrm{LSWN}$ must correspond to wavelengths longer than we have observed.  Roughly speaking, the largest observed wavelength is the current horizon size, or $\sim10\,\mathrm{Gpc}$, so $k_\mathrm{LSWN}\,\mathrm{Gpc} \lesssim 0.1$.  Given the relation between $k_\mathrm{LSWN}/k_\mathrm{nl} \sim 10^{-10}$, this requires that $k_\mathrm{nl}\,\mathrm{pc} \lesssim 1$.  In other words, the dominant non-linearities are on comoving scales greater then $\sim 1\,\mathrm{pc}$.

This numerology extrapolates the observed HZ spectrum down to the pc which is far smaller than our observational sensitivity.  Thus, this extrapolation may or may not be valid.  However, if one can extrapolate, then one can only extrapolate the HZ spectrum to the pc scale. If one extrapolated to scales smaller than a pc, the non-linearities on these smaller scales would produce too much LSWN, causing $k_\mathrm{LSWN}$ to be larger than $(10\,\mathrm{Gpc})^{-1}$. This is inconsistent with observations 
 so one \emph{cannot} extrapolate the observed HZ spectrum to scales smaller than $\sim1\,\mathrm{pc}$.  
This cutoff on the short wavelength power spectrum is a new constraint on viable cosmological models.

\subsection{Outline}
\label{sec:Outline}

The rest of the paper is divided into the following sections and appendices
\begin{itemize}
    \item[{\bf\S\ref{sec:LSWN}}] shows mathematically the inevitably of LSWN in a broad range of systems (PDEs and initial conditions). Assumptions and approximations are stated explicitly.  These systems are then specialized to classes  of systems which include models of cosmological inhomogeneities.  This section is long, technical, and only about mathematics rather than physics.  The mathematics itself is straightforward and not at all subtle. A list of results and concepts from \S\ref{sec:LSWN} are given in \S\ref{sec:LSWNreview}.
    \item[{\bf\S\ref{sec:LargeScaleNoise}}] uses the results of \S\ref{sec:LSWN} to give a general accurate approximate and succinct model for LSWN in self-similar systems where the LSWN is generated early.  While this is again mathematics, not physics, it can be directly applied to cosmological inhomogeneities in the radiation era.
    \item[{\bf\S\ref{sec:PlanarNewtonianCosmology}}] specializes to the physical but idealized case of Newtonian cosmology with planar perturbations during radiation domination and during matter domination.  We choose this idealization because it is tractable analytically or numerically and avoids the coordinate/gauge issues of GR.  We use numerical results to validate the accuracy of the approximations of \S\ref{sec:LSWN}\,\&\,\ref{sec:LargeScaleNoise}.
    \item[{\bf\S\ref{sec:LSWNinCosmology}}] addresses directly the production of LSWN in a cosmological setting.  The emphasis is on the inevitability of LSWN from principles developed in \S\ref{sec:LSWN}\,\&\,\ref{sec:LargeScaleNoise}. This section is more deductive than constructive.  We leave the construction of a quantitative model for LSWN to Paper 2.  Nevertheless, this section formulates the constraint discussed in \S\ref{sec:RoughQuantitativeEstimates} in a more rigorous and precise way.
    \item[{\bf\S\ref{sec:RelatedPhenomena}}] illustrates the universality of LSWN by listing similar phenomena in a number of other (non-cosmological) systems.
    \item[{\bf\S\ref{sec:Synopsis}}]  summarizes the results of this paper. 
    \item[\S{\bf\ref{app:conventions}}] specifies the mathematical and notational conventions used in this paper.
   \item[\S{\bf\ref{app:EquivalentSystems}}] defines equivalent systems through allowed transformations (T1-T5) that preserve the physical properties of LSWN while allowing changes of temporal variable, spatial coordinates, dependent variables, basis solutions, and growing mode normalizations. Deformations that fall outside these transformations create non-equivalent systems with potentially different LSWN characteristics.
    \item[\S{\bf\ref{sec:calManalytic}}] provides analytical expressions for the LSWN kernel functions $\mathcal{M}_\pm[\varphi]$ in  2nd order self-similar systems with quadratic non-linearities. These functions, expressed in terms of regularized hypergeometric functions, determine the amplitude of LSWN generated by mode coupling and exhibit characteristic power-law behavior at small and large scales with a transition (``knee'') at horizon crossing.
\end{itemize}
An overview of the main results of this paper has already been summarized in \S\ref{sec:introduction}, the Introduction. For the shortest read, one could skip to \S\ref{sec:Synopsis} from here.\\  

%%%%%%%%%%%%%%%%%%%%%%%%%%%%%%%%%%%%%%%%%%%%%%%%%%%%%%%%%%%%
%%%%%%%%%%%%%%%%%%%%%%%%%%%%%%%%%%%%%%%%%%%%%%%%%%%%%%%%%%%%
\section{Large Scale White Noise}
\label{sec:LSWN}

%In this section, we discuss the initial generation and growth of LSWN in several increasing levels of complexity. 
%
%We begin in \S\ref{sec:InstantaneousGeneration} by deriving expressions for the initial LSWN growth rate. We also identify the two exceptions to this phenomenon, as well as comment on ...
%
%In \S\ref{sec:LeadingOrderPerturbationTheory} we give a more complete picture of the development of LSWN in time specializing to the case where the non-linearities are small. 
%
%In \S\ref{sec:CosmologyLike} we further specialize to systems which have properties usually assumed for cosmological inhomogeneities: systems that are 2nd order in time, have statistical isotropy and positive parity, have 2nd order non-linearities and are initially Gaussian.
%
%The goal is to provide analytic expressions/intuition for the existence of LSWN phemenon. This will be supported by explicit numerical simulations in \S\ref{sec:PlanarNewtonianCosmology}. 

%%%%%%%%%%%%%%%%%%%%%%%%%%%%%%%%%%%%%%%%%%%%%%%%%%%%%%%%%%%%
%%%%%%%%%%%%%%%%%%%%%%%%%%%%%%%%%%%%%%%%%%%%%%%%%%%%%%%%%%%%
\subsection{Initial Generation of LSWN}
\label{sec:InstantaneousGeneration}

In this sub-section, we derive expressions for the initial growth rate of LSWN, starting from sub-Poissonian initial conditions at $t = t_i$. We defer expressions for the subsequent evolution to \S\ref{sec:LeadingOrderPerturbationTheory}. Here we restrict ourselves to PDEs which are both spatially homogeneous --- in the sense that if $q[t,\vec{x}]=f[t,\vec{x}]$ is a solution so is $q[t,\vec{x}]=f[t,\vec{x}+\Delta \vec{x}]$ --- and homogeneous --- admitting $q[t,\vec{x}]=0$ as a solution. A $p^{\rm th}$ order spatially homogeneous and homogeneous PDE (\textit{shhPDE}) can be written in the form
\begin{equation}
    \frac{\partial^p}{\partial t^p} q[t,\vec{x}] + \cdots = 0 \,,
\end{equation}
where $\cdots$ encompass spatial and temporal derivatives of order $p-1$ or smaller. The PDE will be accompanied by initial conditions specified on some constant time slice $t = t_i$. We will typically take these initial conditions to be stochastic. The shhPDE will then propagate the statistics of the initial conditions to statistics of the solutions.  

To demonstrate how LSWN comes about for such a PDE, consider perhaps the simplest class of non-linear shhPDEs, which take the form
\begin{equation}
    \dot{q}[t,\vec{x}] = q[t,\vec{x}]^n \,,
\end{equation}
for integer $n > 1$. Fourier transforming both sides of the equation, the Fourier amplitude evolves according to\footnote{We work in $d$-dimensional Euclidean spatial coordinates $\vec{x}=\{x_1,\cdots,x_d\}\in\mathbb{R}^d$ for complete generality.}
\begin{equation}\label{eq:tildeqdot}
    \dot{\tilde{q}}[t,\vec{k}] = \int\frac{d^d\vec{x}}{(2\pi)^{d/2}} \, e^{-i \vec{k} \cdot \vec{x}} \prod_{i=1}^n \left( \bar{q}[t] + \int\frac{d^d\vec{k}_i}{(2\pi)^{d/2}} \tilde{q}[t,\vec{k}_i] e^{i\,\vec{k}_i\cdot\vec{x}} \right) \,,
\end{equation}
where $\bar{q}[t]$ denotes the spatial average. This is an order $p = 1$ PDE, so we require just one initial condition, which we take to be sub-Poissonian, i.e. $\tilde{q}[t_i,\vec{0}^+] = 0$. More generally for a $p^{\rm th}$ order shhPDE, we will define sub-Poissonian initial conditions to mean 
\begin{equation}\label{eq:subPoissonianBC}
    \frac{d^m}{dt^m}\tilde{q}[t,\vec{0}^+] \bigg|_{t = t_i}=0 \,\,\, \forall \,\,\, m < p \,.
\end{equation}
Note that sub-Poissonian initial conditions also imply that $\bar{q}[t_i] = \tilde{q}[t_i, \vec{0}^+] = 0$. Then evaluating Eq.~(\ref{eq:tildeqdot}) at $t = t_i$ with these initial conditions yields
\begin{equation}\label{eq:tildeqdot2}
    \dot{\tilde{q}}[t_i,\vec{k}] = \left( \prod_{j = 1}^n \int \frac{d^d k_j}{(2\pi)^{d/2}} \tilde{q}[t_i, \vec{k}_j] \right) (2\pi)^{d/2} \delta^{(d)} \bigg[ \sum_{j=1}^n \vec{k}_j - \vec{k} \bigg] \,,
\end{equation}
where we have made use of the integral representation of the delta function. Note that the presence of this delta function indicates that the $\vec{k}_i$ ``mix'' into $\vec{k} = \sum_i \vec{k}_i$.

Let us now determine the impact of this mode mixing on the statistical properties of the field $q$. We define the power spectrum $P_q(t,\vec{k})$ in the usual way 
\begin{equation}\label{eq:Pdef}
    \langle\tilde{q}[t,\vec{k}]\,\tilde{q}[t,\vec{k}']\rangle 
    \equiv (2\pi)^d \delta^{(d)}[\vec{k}+\vec{k}']\, P_q[t,\vec{k}] \,,
\end{equation}
where $\langle ... \rangle$ denotes an average over realizations and the $\delta$-function is guaranteed by homogeneity. The power spectrum is related to the real-space correlation function $\xi_q[t,\vec{x}-\vec{x}{\,}']$ as
\begin{equation}
\begin{split}
    \xi_q[t,\vec{x}-\vec{x}{\,}'] & \equiv
    \big\langle \big( q[t,\vec{x}]- \bar{q}[t] \big) \, \big( q[t,\vec{x}{\,}'] - \bar{q}[t] \big) \big\rangle\\
    & = \int d^d k \, P_q[t,\vec{k}] e^{i \vec{k} \cdot (\vec{x} - \vec{x}')} \,,
\end{split}
\end{equation}
where in the first line we have used that $\langle q[t,\vec{x}]\rangle = \bar{q}[t]$, which follows from homogeneity. 

We classify the distribution of the scalar field based on its large scale $\vec{k} \rightarrow 0$ properties. Consider the long-wavelength limit of the power spectrum,
\begin{equation}
    P_q[t,\vec{0}^+] \equiv \lim_{\vec{k} \rightarrow 0} P_q[t,\vec{k}] \,.
\end{equation}
As we have mentioned already, a distribution with $P_q[t,\vec{0}^+] = 0$ is sub-Poissonian. Meanwhile, a distribution with $0 <P_q[t,\vec{0}^+] <\infty$ is said to be Poissonian, or ``white noise on large scales''.\footnote{A distribution can also be super-Poissonian, $P_q[t,\vec{0}^+] = \infty$, though we will not consider this possibility here.} This can be seen by recalling the definition of a white noise spectrum, 
\begin{equation}
    P_q^{\rm WN}[t,\vec{k}] = N[t] \,,
\end{equation}
which is $k$-independent and so does not vanish in the $\vec{k} \rightarrow 0$ limit.\footnote{The corresponding real-space correlation function is $\xi_q^{\rm WN}[t,\vec{x}-\vec{x}'|] =(2\pi)^d\,N[t]\,\delta^{(d)}[\vec{x}-\vec{x}']$, which implies that fluctuations at separated spatial points are uncorrelated.}

We are now ready to see how the mode mixing of Eq.~(\ref{eq:tildeqdot2}) can lead an initially sub-Poissonian distribution to instantaneously evolve into a Poissonian distribution (and hence produce LSWN). On some level this is already obvious: if $P_q[t_i,\vec{k}]$ has initial support only in the band $k_{\rm min}\leq|\vec{k}|\leq k_{\rm max}$, the non-linearity instantaneously transfers power to the broader band $0\leq|\vec{k}|\leq n\,k_{\rm max}$.\footnote{If $n$ is even then the region will always extend to both $|\vec{k}|=0$ and  $|\vec{k}|=n\,k_{\rm max}$.  If $n$ is odd it will only extend to $|\vec{k}|=0$ if the initial support sufficiently encircles the origin. If the initial power spectrum is isotropic then the region will always extend to $k = 0$, regardless of whether $n$ is even or odd.} Since power is now non-vanishing in the large scale limit, $P_q^{\rm WN}[t,\vec{0}^+] \neq 0$, the distribution has become Poissonian, with some (possibly time-dependent) constant value on large scales. To derive the initial growth rate, consider taking time derivatives of the power spectrum, defined in Eq.~(\ref{eq:Pdef}). The first derivative vanishes in the large scale limit, since $\dot{P}[t_i,\vec{0}^+] \propto \langle \dot{\tilde{q}}[t_i,\vec{0}^+] \tilde{q}[t_i,\vec{0}^+] \rangle = 0$ for sub-Poissonian initial conditions. The second time derivative is
\begin{equation}\label{eq:Pdoubledot}
    \ddot{P}_q[t_i,\vec{0}^+] = 2 \lim_{\vec{k} \rightarrow 0} \frac{\langle \dot{\tilde{q}}[t_i,\vec{k}] \dot{\tilde{q}}[t_i,\vec{k}'] \rangle}{(2\pi)^d \delta^{(d)}[\vec{k} + \vec{k}']} \,,
\end{equation}
which is positive unless $\dot{\tilde{q}}[t,\vec{0}^+] = 0$. Using our result of Eq.~(\ref{eq:tildeqdot2}), the two point function appearing on the right hand side is explicitly
\begin{equation}\label{eq:qdotqdoti}
\begin{split}
    \langle \dot{\tilde{q}}[t_i,\vec{k}] \dot{\tilde{q}}[t_i,\vec{k}'] \rangle = \left( \prod_{j = 1}^n \int \! \frac{d^d k_j}{(2\pi)^{d/2}} \right)  \delta^{(d)} & \bigg[ \sum_{j=1}^n \vec{k}_j - \vec{k} \bigg] \left( \prod_{\ell = 1}^n \int \! \frac{d^d k_\ell}{(2\pi)^{d/2}} \right) \delta^{(d)} \bigg[ \sum_{j=\ell}^n \vec{k}_\ell - \vec{k} \bigg] \\
    & \textcolor{white}{.} \hspace{11mm} \times  (2\pi)^d \bigg\langle \bigg( \prod_{j = 1}^n \tilde{q}[t_i, \vec{k}_j] \bigg) \bigg( \prod_{\ell = 1}^n \tilde{q}[t_i, \vec{k}_\ell] \bigg) \bigg\rangle \,.
\end{split}
\end{equation}
Again assuming stochastic initial conditions (spatially homogeneous Gaussian noise), we can now evaluate the initial LSWN growth rate $\ddot{P}_q[t_i,\vec{0}^+]$ for sample values of $n$. For example for $n=2$, we find
\begin{equation}
    n=2: \,\,\,\,\,\, \ddot{P}_q[t_i,\vec{0}^+] = 4 \int d^d k_1 \, P_q[t_i,\vec{k}_1]^2 \,,
\end{equation}
while for $n=3$ we find
\begin{equation}
    n=3: \,\,\,\,\,\, \ddot{P}_q[t_i, \vec{0}^+] = 12 \int d^d k_1 \int d^d k_2 \, P_q[t_i, \vec{k}_1] P_{q}[t_i, \vec{k}_2] P_q[t_i, \vec{k}_1 + \vec{k}_2] \,,
\end{equation}
and so forth, where in deriving these expressions we have made extensive use of the fact that our sub-Poissonian initial conditions $P[t_i, \vec{0}^+]$ eliminate many terms in the sum. Clearly, $\ddot{P}_q[t_i, \vec{0}^+] \geq 0$ since $P_q[t_i,\vec{k}_1] \geq 0$. In the $n=2$ case, so long as $P_q[t_i, \vec{k}_1] \neq 0$ for some $\vec{k}_1$ in the integral, it follows that the initial growth rate is positive, $\ddot{P}_q[t_i, \vec{0}^+] > 0$. In the $n=3$ case, so long as there exist values of $\vec{k}_1$ and $\vec{k}_2$ for which all three power spectra are non-zero, this rate too will be non-vanishing and positive, $\ddot{P}_q[t_i, \vec{0}^+] > 0$.

Now having seen how LSWN arises from mode-mixing induced by non-linearities in the evolution equations, we can immediately identify two exceptions to this phenomenon:\newline

\label{sec:LinearException}
\subsubsection{The Linear Exception}
%\noindent \textbf{The Linear Exception}
%\vskip2pt

For linear shhPDEs, sub-Poissonian initial conditions of Eq.~(\ref{eq:subPoissonianBC}) are sufficient to guarantee that $P_q[t,\vec{0}^+] = 0$ for all $t$ --- i.e. the solution remains sub-Poissonian. Intuitively this is clear since for linear shhPDEs, $k$ modes remain uncoupled and so there can be no mixing to redistribute power to zero wavenumber. At the level of equations, it is clear to see that $\ddot{P}_q[t_i, \vec{0}^+]$ of Eq.~(\ref{eq:Pdoubledot}) vanishes for $n=1$. In a similar way, one can show that $\ddddot{P}_q[t,\vec{0}^+] = 0$ and so forth provided $\frac{d^m}{dt^m}\tilde{q}[t,\vec{0}^+] = 0$ for all $m < p$. A solution to a linear shhPDE with sub-Poissonian initial conditions will remain sub-Poissonian.\newline

\label{sec:ConservativeException}
\subsubsection{The Conservative Exception}
%\noindent \textbf{The Conservative Exception}
%\vskip2pt

We define a conservative $p^{\rm th}$ order shhPDE as one having the form
\begin{equation}\label{eq:conservativePDE}
    \frac{d^p}{dt^p}q[t,\vec{x}] + \mathrm{linear\,terms} + \vec{\alpha}[t] \cdot \nabla f\big[t,q[t,\vec{x}], \cdots \big] = 0 \,,
\end{equation}
where $f$ contains all of the non-linear terms. Recall that the spatial average of the field $q[t,\vec{x}]$ is $\bar{q}[t] = \tilde{q}[t, \vec{0}^+]$. Because the spatial average of a gradient is zero, it follows that the spatial average of Eq.~(\ref{eq:conservativePDE}) obeys a homogeneous ordinary differential equation (ODE) of the form
\begin{equation}
    \frac{d^p}{dt^p} \bar{q}[t] + \sum_{m=0}^{p-1} \beta_m[t] \frac{d^m}{dt^m} \bar{q}[t] = 0 \,,
\end{equation}
for some arbitrary coefficients $\beta_m[t]$. This equation is linear and admits $p$ independent solutions, each of which is a constant of motion (hence why we call shhPDEs of the form of Eq.~(\ref{eq:conservativePDE}) conservative). For sub-Poissonian initial conditions, i.e. $\frac{d^m}{dt^m} \bar{q}[t] \big|_{t = t_i} = 0$ for $0 \leq m < p$, then $\bar{q}[t] = 0 \,\, \forall \,\, t$. Thus, conservative shhPDEs do not produce LSWN.

To be clear the quantity that must be conserved is $q$, it does not matter if there is some other quantity which is conserved.  Furthermore since we are concerned with the power spectrum of the conventional Fourier transform in Euclidean $\vec{x}$ the conservations law is that the Euclidean volume weighted $q$ is conserved.  It does not matter if $q$ is conserved with some other weighting. \\

Other than these exceptions, we expect all sshPDEs to exhibit LSWN. These are exceptions {\it only} if the initial conditions are sub-Poissonian which, as emphasized in \S\ref{sec:GuidingVariables}, will often not be the case.  Now, one can always reparameterize a PDE via a change of variables (with corresponding changes to the initial conditions) in order to simplify the system.\footnote{We refer to the PDE with the statistical distribution of initial conditions as \textit{the system}.} This includes reparameterizations that transform non-linear PDEs, which feature LSWN, into either linear or conservative PDEs, which exhibit no LSWN.\footnote{For example, one can transform any system exhibiting LSWN into another system exhibiting no LSWN using the transformation $q[t,\vec{x}] \rightarrow \underline{q}[t,\vec{x}] = \nabla^2 q[t,\vec{x}]$.} In this sense, one may worry that the amount or even existence of LSWN is tied to a particular parametrization of the PDE, and therefore might seem mathematically ill-defined.\\

\label{sec:Tyranny}
\subsubsection{Tyranny of Observables}
%\noindent \textbf{Tyranny of Observables}
%\vskip2pt

While it is certainly true that these reparameterized systems are mathematically equivalent, physically they are not equivalent because in practice we only have experimental access to a certain class of observables. When this \textit{tyranny of observables} requires us to consider those quantities whose evolution is governed by non-linear PDEs, we should expect white noise on sufficiently large scales. Of course, if we could measure a quantity $q(t,\vec{x})$ precisely on all scales, then we could perform any mathematical transformation on the data to remove LSWN entirely. However, real world measurements are never infinitely precise and never have unlimited dynamical range. \textit{One only can measure what one can measure}, and what one can measure might inevitably exhibit LSWN.\\

\subsubsection{Equivalent Systems and Deformations}
\label{sec:Deformations}
%\noindent \textbf{Equivalent Systems and Deformations}
%\vskip2pt

For these reasons, the full set of mathematical reparameterizations may not be useful because one would need a larger dynamic range than the data provides to make the equivalent reparameterizations of the data. Here, we will only consider reparameterizations to give \textit{physically equivalent systems} if they preserve power spectrum ratios in the sense that the power spectra are only transformed by a time-dependent shift on a log-log plot. Allowed reparameterizations which give equivalent systems are listed in Appendix~\ref{app:EquivalentSystems}. Unallowed reparameterizations which do not yield equivalent systems we call {\it deformations}. There exist deformations that maintain homogeneity and spatial homogeneity, shhPDE$\rightarrow$\underline{shhPDE}, which nevertheless do not yield equivalent systems. Fluid dynamics in Lagrangian or Eulerian coordinates is an example of mathematically equivalent systems which are not equivalent in our sense. Finally, we comment that there are many physical systems which may be conservative and/or linear but for which the quantity measured is not, perhaps because of non-linearity in a sensor, because spatial averaging is not uniform, or for any number of other reasons.  In these cases, the measurements will exhibit LSWN even though the quantity under study does not. 

%The experimentally induced LSWN may be good or bad. Bad if it allows small scale phenomena to contaminate the large scales one is interested in.  Good if it allows one to probe phenomena on smaller scales than the naive resolution of your experiment.

%%%%%%%%%%%%%%%%%%%%%%%%%%%%%%%%%%%%%%%%%%%%%%%%%%%%%%%%%%%%
%%%%%%%%%%%%%%%%%%%%%%%%%%%%%%%%%%%%%%%%%%%%%%%%%%%%%%%%%%%%
\subsection{LSWN in Perturbation Theory}
\label{sec:LeadingOrderPerturbationTheory}

So far, we have demonstrated how LSWN arises for a generic class of PDEs due to non-linearity induced mode-mixing in Fourier space. This phenomenon is fully generic, and will come to dominate sub-Poissonian initial conditions on sufficiently large scales, even if deviations from linearity are small on all scales. We have also supplied expressions for the initial growth rate of this LSWN from sub-Poissonian initial conditions. Of course, studying the evolution of the system beyond this initial time is generically quite hard for non-linear PDEs which may not admit analytic solutions or be amenable to numerical methods. So long as the amplitude of the field $q$ remains small, however, one can still use perturbation theory to accurately estimate the amplitude of LSWN in the pseudo-linear system. In this section, we consider such a system and derive generic expressions for the subsequent LSWN evolution in a systematic perturbative expansion.\\ 

%%%%%%%%%%%%%%%%%%%%%%%%%%%%%%%%%%%%%%%%%%%%%%%%%%%%%%%%%%%%
%\label{sec:LeadingOrderBornApproximation}
\noindent \textbf{Leading-Order Born Approximation}
\vskip2pt

Consider a $p^{\rm th}$ order non-linear shhPDE of the form
\begin{equation}
    \hat{\cal D}\,q \equiv \frac{\partial^p}{\partial t^p}q[t,\vec{x}] + \cdots = 0 \,,
\label{eq:qPDE}
\end{equation}
where we have defined the differential operator $\hat{\mathcal{D}}$ and the $\cdots$ encode spatial and temporal derivatives of order $p-1$ or smaller. Taylor expanding the PDE in the amplitude $q$ yields
\begin{equation}\label{eq:Taylor}
    \hat{\mathcal{D}} q = \sum_{m=1}^\infty \, {}_{(m)}\hat{\mathcal{D}} q \,,
\end{equation}
where ${}_{(m)}\hat{\cal D}$ are partial differential operators on $q$. To illustrate our notation, consider the following PDE and its corresponding perturbative expansion:
\begin{equation*}
\begin{split}
    \textit{Example:} \,\,\,\,\,\, & \hat{\mathcal{D}}q = \ddot{q} - \alpha[t] \nabla \cdot (e^{- \beta[t] q^2} \nabla q) = 0 \\
    & {}_{(1)}\hat{\mathcal{D}}q = \ddot{q} - \alpha[t] \nabla^2 q \\
    & {}_{(2)}\hat{\mathcal{D}}q = 0 \\
    & {}_{(3)}\hat{\mathcal{D}}q = \alpha[t] \beta[t] \left( q^2 \nabla^2 q + 2 q (\nabla q)^2 \right) \\
    & {}_{(4)}\hat{\mathcal{D}}q = 0 \\
    & ...
\end{split}
\end{equation*}
This expansion is useful if for $|q| \ll 1$ the series converges rapidly. If only the $m=1$ term is non-zero, then the system is linear. For a non-linear PDE, we refer to the smallest $m > 1$ for which ${}_{(m)}\hat{\mathcal{D}} q \neq 0$ as the \textit{leading order non-linearity} and denote it by $n$. In most cases, the leading order non-linearity is quadratic and so $n=2$. For $|q| \ll 1$, the system is well-approximated by considering only the leading order non-linearity,
\begin{equation}\label{LeadingOrderApproximation}
    \hat{\mathcal{D}}q \simeq \left( {}_{(1)}\hat{\mathcal{D}} + {}_{(n)}\hat{\mathcal{D}} \right) q = 0 \,.
\end{equation}

At linear order, the equation ${}_{(1)}\hat{\mathcal{D}}q = 0$ admits a general solution of the form
\begin{equation}\label{eq:linearsol}
    {}_{(1)}q[t,\vec{x}] = \int \frac{d^d k}{(2\pi)^{d/2}}\,e^{i \vec{k}\cdot\vec{x}} \, {}_{(1)}\tilde{q}[t,\vec{k}] \,, \,\,\,\,\,\, {}_{(1)}\tilde{q}[t,\vec{k}] = \sum_{\alpha = 1}^p q_\alpha[\vec{k}] \, Q_\alpha[t,\vec{k}] \,,
\end{equation}
where $Q_\alpha[t,\vec{k}]$ are a set of $p$ independent Fourier mode solutions and $q_\alpha[\vec{k}]$ are the mode amplitudes. The solution to the full PDE can formally be written in terms of this linear order solution as
\begin{equation}\label{eq:FormalGreenSolution}
    q[t,\vec{x}] = {}_{(1)}q[t,\vec{x}] - \int \frac{d^d k}{(2\pi)^{d/2}} \, e^{i\,\vec{k}\cdot\vec{x}} \int_{t_\mathrm{i}}^t dt' \, \tilde{\mathcal{G}}[t,t',\vec{k}] \int \frac{d^d x'}{(2\pi)^{d/2}} \, e^{-i \vec{k}\cdot\vec{x}'} \left(\hat{\mathcal{D}}-{}_{(1)}\hat{\mathcal{D}} \right) q[t',\vec{x}'] \,,
\end{equation}
where $\hat{\mathcal{D}}-{}_{(1)}\hat{\mathcal{D}}$ is the non-linear part of the operator and $\tilde{\mathcal{G}}$ is the Fourier transform of the Green's function. This can generically be constructed from the Wronskian $w_\alpha$ as
\begin{equation}\label{eq:GreenFunction}
    \tilde{\mathcal{G}}[t,t',\vec{k}] = \sum_{\alpha=1}^p w_\alpha[t',\vec{k}] Q_\alpha[t,\vec{k}] \,, \,\,\,\,\,\, w_{\alpha_p} = \frac{\sum_{\alpha_1 = 1}^p \cdots \sum_{\alpha_{p-1} = 1}^p \epsilon^{\alpha_1 \cdots \alpha_p} \prod_{j=1}^{p-1} Q_{\alpha_j}^{(j-1)}[t,\vec{k}]}{\sum_{\alpha_1 = 1}^p \cdots \sum_{\alpha_{p} = 1}^p \epsilon^{\alpha_1 \cdots \alpha_p} \prod_{j=1}^{p} Q_{\alpha_j}^{(j-1)}[t,\vec{k}]} \,,
\end{equation}
where $\epsilon^{\alpha_1 \cdots \alpha_p}$ is the Levi-Civita symbol and $Q^{\ell} = \frac{\partial^\ell}{\partial t^\ell} Q$. This expression looks quite complicated, but the first few are familiar
\begin{equation}
\begin{split}
    & p=1: \,\,\,\,\,\, \tilde{\mathcal{G}}[t,t',\vec{k}] = \frac{Q_{1}[t,\vec{k}]}{Q_1[t',\vec{k}]} \,, \\
    & p=2: \,\,\,\,\,\, \tilde{\mathcal{G}}[t,t',\vec{k}] = \frac{Q_1[t',\vec{k}] Q_2[t,\vec{k}] - Q_2[t',\vec{k}] Q_1[t,\vec{k}]}{Q_1[t',\vec{k}] Q_2^{(1)}[t',\vec{k}] - Q_2[t',\vec{k}] Q_1^{(1)}[t',\vec{k}]} \,, \\
    & \cdots \,\,\,\,\,\,\,\,\,\,\,\,\,\,\, \cdots
\end{split}
\end{equation}

Eq.~(\ref{eq:FormalGreenSolution}) can be considered a formal solution because $q$ appears on both the left- and right-hand sides. Equivalently, it can be considered as a reformulation of the non-linear PDE as an integral equation. The \textit{Born approximation} replaces $q$ in the integral of the right-hand side by the first-order solution ${}_{(1)}q$, 
\begin{equation}
    \left( \hat{\mathcal{D}}-{}_{(1)}\hat{\mathcal{D}} \right) q \, \rightarrow \, \left( \hat{\mathcal{D}}-{}_{(1)}\hat{\mathcal{D}} \right) {}_{(1)} q \,.
\end{equation}
The \textit{leading-order Born approximation} (LOBA) goes one step further and substitutes the leading order non-linearity for the full non-linearity in the Born approximation, 
\begin{equation}
    \left( \hat{\mathcal{D}}-{}_{(1)}\hat{\mathcal{D}} \right) q \, \rightarrow \, \left( \hat{\mathcal{D}}-{}_{(1)}\hat{\mathcal{D}} \right) {}_{(1)}q \, \rightarrow \, {}_{(n)}\hat{\mathcal{D}} \, {}_{(1)}q \,.
\label{eq:LOBA1}
\end{equation}
This approximation will be accurate so long as the magnitude of $q$ remains small. More concretely, for LOBA to be accurate, the system should be in the \textit{linear regime}, which is defined by the condition that all terms in the non-linear part of the PDE, $(\hat{\mathcal{D}}-{}_{(1)}\hat{\mathcal{D}})\,q$, 
be much smaller than those in the linear part, ${}_{(1)}\hat{\mathcal{D}}\,q$. Note that when we speak about \textit{non-linearities} in this paper, we are referring to the non-linear part of the PDE. The presence of non-linearities has no bearing on whether the system is in the linear or non-linear regime. Indeed as we will show, non-linearities can significantly affect the solution even when the system is in the linear regime.  

Going forward, we will largely be working in the LOBA, for which the formal solution of Eq.~(\ref{eq:FormalGreenSolution}) becomes
\begin{equation}\label{eq:LOBAGreenSolution}
    q[t,\vec{x}] = {}_{(1)}q[t,\vec{x}] - \int \frac{d^d k}{(2\pi)^{d/2}} \, e^{i\,\vec{k}\cdot\vec{x}} \int_{t_\mathrm{i}}^t dt' \, \tilde{\mathcal{G}}[t,t',\vec{k}] \int \frac{d^d x'}{(2\pi)^{d/2}} \, e^{-i \vec{k}\cdot\vec{x}'} {}_{(n)}\hat{\mathcal{D}} \, {}_{(1)}q[t',\vec{x}'] \,.
\end{equation}
The corresponding Fourier amplitude is
\begin{equation}\label{eq:LOBA}
    \begin{split}
    \tilde{q}[t,\vec{k}] & = {}_{(1)}\tilde{q}[t,\vec{k}] - \int_{t_i}^t dt' \, \tilde{\mathcal{G}}[t,t',\vec{k}] \, {}_{(n)} \hat{\tilde{\mathcal{D}}} \, {}_{(1)}\tilde{q}[t',\vec{k}] \\
    & = {}_{(1)}\tilde{q}[t,\vec{k}] - (2\pi)^{d/2} \int d^n \mathbf{K} \delta^{(d)}[\vec{K} - \vec{k}] \int_{t_i}^t dt' \, \tilde{\mathcal{G}}[t,t',\vec{k}] N_{(n)}[t', {}_{(1)}\mathbf{Q}, \mathbf{K}] \,,
\end{split}
\end{equation}
where ${}_{(n)}\hat{\tilde{\mathcal{D}}}$ in the first line is the Fourier transform of the leading order non-linearity differential operator. In order to write the second line compactly, we have introduced a number of notational conveniences; we define
\begin{equation}
    \mathbf{K} \equiv \vec{k}_1, \cdots \vec{k}_n \,, \,\,\,\,\,\, \vec{K} \equiv \sum_{i=1}^n \vec{k}_i \,, \,\,\,\,\,\, \int d^n \mathbf{K} f[\mathbf{K}] \equiv \int \! \frac{d^d k_1}{(2\pi)^{d/2}} \cdots \int \! \frac{d^d k_n}{(2\pi)^{d/2}} f[\vec{k}_1, \cdots \vec{k}_n] \,.
\end{equation}
Additionally, the order $n$ non-linearity will depend on Fourier amplitudes and their time derivatives $\tilde{q}^{(l)}[t,\vec{k}] \equiv
\frac{\partial^l}{\partial t^l}q[t,\vec{k}]$; this set of quantities we denote compactly as $\mathbf{Q}$.\footnote{Because we only make use of first-order solutions in the LOBA, we have replaced ${\bf Q} \rightarrow {}_{(1)}{\bf Q}$ on the right-hand side of Eq.~(\ref{eq:LOBA}).} Finally, for regular differential operators, each element of the operator Taylor series can be decomposed into a sum of terms, each of which can be factorized as  
\begin{equation}\label{eq:NonlinearityExpansion}
    {}_{(n)}\hat{\tilde{\cal D}}\,\tilde{q}[t,\vec{k}] = (2\pi)^{d/2} \int d^n{\bf K}\,\delta^{(d)}[\vec{k}-\vec{K}]\, N_{(n)}[t,{\bf Q},{\bf K}] \,,
\end{equation}
where
\begin{equation}
    N_{(n)}[t,{\bf Q},{\bf K}] = \sum_{j=1}^{m_n} c_{(n,j)}[t]\,d_{(n,j)}[{\bf K}]\, \prod_{i=1}^n \tilde{q}^{(l_{(n,j,i)})}[t,\vec{k_i}] \,,
\label{eq:Ndefinition}
\end{equation}
and $\hat{d}_{(n,j)}[\bf{K}]$ is a multinomial in $\vec{k}_i$. The number of terms in the sum, $m_n$, depends on the form of the non-linearity. To illustrate the usage of this notation, consider revisiting the example of the previous section, for which the leading-order non-linearity is cubic ($n=3$):
\begin{equation*}
\begin{split}
    \textit{Example:} \,\,\,\,\,\, & \hat{\mathcal{D}}q = \ddot{q} - \alpha[t] \nabla \cdot (e^{- \beta[t] q^2} \nabla q) = 0 \\
    & {}_{(3)}\hat{\tilde{\mathcal{D}}} \tilde{q}[t,\vec{k}] = (2\pi)^{d/2} \!\! \int \! \frac{d^d k_1}{(2\pi)^{d/2}} \int \! \frac{d^d k_2}{(2\pi)^{d/2}} \int \! \frac{d^d k_3}{(2\pi)^{d/2}} \delta^{(d)}[\vec{k} - \vec{k_1} - \vec{k}_2 - \vec{k}_3] N_{(3)}[t,\mathbf{Q},\mathbf{K}] \,,\\
    & N_{(3)}[t,\mathbf{Q}, \mathbf{K}] = - \alpha[t] \beta[t] (\vec{k}_1^{\, 2} + 2 \vec{k}_1 \cdot \vec{k}_2) \tilde{q}^{(0)}[t,\vec{k}_1] \tilde{q}^{(0)}[t,\vec{k}_2] \tilde{q}^{(0)}[t,\vec{k}_3] \\
    & \text{Coefficients:}\\
    & \textcolor{white}{.} \hspace{8mm} c_{(3,1)}[t] = - \alpha[t] \beta[t] \,, \,\,\, d_{(3,1)}[\mathbf{K}] = \vec{k}_1^{\, 2} \,, \,\,\, \ell_{(3,1,1)} = \ell_{(3,1,2)} = \ell_{(3,1,3)} = 0 \,, \\
    & \textcolor{white}{.} \hspace{8mm} c_{(3,2)}[t] = - 2 \alpha[t] \beta[t] \,, \,\,\, d_{(3,2)}[\mathbf{K}] = \vec{k}_1 \cdot \vec{k}_2 \,, \,\,\, \ell_{(3,2,1)} = \ell_{(3,2,2)} = \ell_{(3,2,3)} = 0 \,.
\end{split}
\end{equation*}

%%%%%%%%%%%%%%%%%%%%%%%%%%%%%%%%%%%%%%%%%%%%%%%%%%%%%%%%%%%%
\subsubsection{Linear Relics of Non-Linearities}
\label{sec:LinearRelics}
%\noindent \textbf{Linear Relics}
%\vskip2pt

The general formula of Eq.~(\ref{eq:LOBA}) is the main result of this section. We now specialize to conditions which will be relevant for cosmological systems. First, in some cases non-linear dynamics may cease to be important at late times, in which case it becomes an accurate approximation to extend the time integral in Eq.~(\ref{eq:LOBA}) to infinity. In such a case, we can define a new quantity $\tilde{q}_{\rm relic}[t,\vec{k}]$ and write
\begin{equation}
    \tilde{q}[t,\vec{k}] \simeq {}_{(1)}\tilde{q}[t,\vec{k}] + \tilde{q}_{\rm relic}[t,\vec{k}] \,,
\end{equation}
where
\begin{equation}
    \tilde{q}_{\rm relic}[t,\vec{k}] \equiv \sum_{\alpha = 1}^p \tilde{r}_\alpha[\vec{k}] Q_\alpha[t,\vec{k}] \,, 
\end{equation}
and
\begin{equation}
    \tilde{r}_\alpha[\vec{k},\mathbf{Q}] \equiv - (2\pi)^{d/2} \int d^n \mathbf{K} \delta^{(d)}[\vec{K} - \vec{k}] \int_{t_i}^\infty dt' \, w_\alpha[t',\vec{k}] N_{(n)}[t', {}_{(1)}\mathbf{Q}, \mathbf{K}] \,.
\end{equation}

The coefficients $\tilde{r}_\alpha[\vec{k},\mathbf{Q}]$ play an analogous role to the coefficients $q_\alpha[\vec{k}]$ introduced in the decomposition of Eq.~(\ref{eq:linearsol}) in that they act as amplitudes for the strictly linear dynamics described by the mode functions $Q_\alpha[t,\vec{k}]$. In the case of $\tilde{r}_\alpha[\vec{k},\mathbf{Q}]$, they come from early-time non-linearities, and in the case of $q_\alpha[\vec{k}]$, they are set by initial conditions. For a late-time observer who has no way of knowing the initial conditions, there is a degeneracy between these two contributions since there is no way to distinguish between them from the late-time behavior of $q$ alone. This is quite important in the case of sub-Poissonian initial conditions, since in this case the relic term will dominate on sufficiently large scales for almost any non-linearity. We describe a system for which the non-linear dynamics are most important at early times as \textit{leaving linear relics}. As we will demonstrate in \S\ref{sec:RadiationLike}, a broad class of systems of interest in cosmology leave linear relics. We will refer to other systems for which non-linear dynamics are also important at late times as {\it maintaining active non-linearities}. Our focus will mostly be on the former.\\

\subsubsection{Growing Mode Initial Conditions}
\label{sec:GrowingModes}
%\noindent \textbf{Growing Mode Initial Conditions}
%\vskip2pt

This brings us to the question of initial conditions. In cosmology, one typically assumes \textit{growing mode} initial conditions, since growing modes tends to quickly come to dominate over other decaying modes (as the names would suggest).\footnote{The non-linearities which produce LSWN may occur at very early times where the argument that the fastest growing mode dominates may fail. In this case one should include other, non-growing, modes to compute the LSWN.  For illustrative purposes, we will assume growing mode initial conditions.} We thus assume initial conditions for $\tilde{q}$ and all derivatives $l < p$
\begin{equation}\label{eq:GrowingModeIC}
    \tilde{q}[t_{\rm i},\vec{k}] = \tilde{g}[\vec{k}]\,G [t_{\rm i},\vec{k}] \,, \,\,\,\,\,\, \tilde{q}^{(l)}[t_{\rm i},\vec{k}] = \tilde{g}[\vec{k}]\,G^{(l)} [t_{\rm i},\vec{k}] \,,
\end{equation}
where $\tilde{g}[\vec{k}]$ is the growing mode amplitude and $G[t,\vec{k}]$ are the growing mode coefficients, formally given by
\begin{equation}\label{eq:GrowingMode}
G[t,\vec{k}]=
    \sum_{\alpha=1}^p g_\alpha[\vec{k}]\,Q_\alpha[t,\vec{k}] \,,
\end{equation}
for some choice of $g_\alpha[\vec{k}]$. Under the LOBA, linear evolution is given by
\begin{equation}\label{eq:linLOBA}
    {}_{(1)} \tilde{q}[t,\vec{k}] = \tilde{g}[\vec{k}] G[t,\vec{k}] \,.
\end{equation}

In a perturbative expansion initial conditions are also perturbative (see \S\ref{sec:PerturbedInitialConditions}) and this growing mode prescription is to all orders, which we call {\it pure growing mode initial conditions}. For $\mathcal{O}[n]$ initial conditions it only matters that this is true to $\mathcal{O}[n]$.  Any $q$ that has this type of initial condition we call a {\it guiding variable} for reasons described in \S\ref{sec:GuidingVariables}.  There it is shown that if $q$ as guiding variable then any for deformation $\underline{q}=F[q]$ non-linear at $\mathcal{O}[n]$ that if $q$ is a guiding variable then $\underline{q}$ cannot be.  Guiding variables are special and must be carefully chosen for your application.

The growing mode amplitudes $\tilde{g}[\vec{k}]$ are typically drawn from a statistical distribution with zero mean. In anticipation of simplifying ensemble averages, it is convenient to factor out these random variables, writing
\begin{equation}
    N_{(n)}[t,\mathbf{Q},\mathbf{K}] = N_{(n)}[t,\mathbf{G},\mathbf{K}] \prod_{j=1}^n \tilde{g}[\vec{k}_j] \,,
\end{equation}
where ${\bf G}$ is the shorthand for
$G^{(l_1)}[t,\vec{k}_1],\cdots,G^{(l_n)}[t,\vec{k}_n]$. Substituting this decomposition and Eq.~(\ref{eq:linLOBA}) into Eq.~(\ref{eq:LOBA}), we find
\begin{equation}\label{eq:LOBAgrow}
    \tilde{q}[t,\vec{k}] \simeq \tilde{g}[\vec{k}]\,G[t,\vec{k}]- (2\pi)^d\, \int d^n{\bf K}\,\delta^{(d)}[\vec{K}-\vec{k}]\, \left(\prod_{j=1}^n\tilde{g}[\vec{k}_j]\right) \,M[t,\vec{k},{\bf G},{\bf K}] \,,
\end{equation}
where
\begin{equation}\label{eq:Mdefinition}
    M[t,\vec{k},{\bf G},{\bf K}]\equiv \int_{t_\mathrm{i}}^t dt'\, \tilde{\cal G}[t,t',\vec{k}]\, N_{(n)}[t',{\bf{G}},{\bf K}] \,.
\end{equation}
In the case of systems leaving linear relics (\textit{i.e.} $t \rightarrow \infty$ in $M[t,\vec{k},\mathbf{G}, \mathbf{K}]$), Eq.~(\ref{eq:LOBAgrow}) simplifies to
\begin{equation}\label{eq:qLOBArelicGrowingMode}
    \tilde{q}[t,\vec{k}] \simeq \tilde{g}[\vec{k}]\,G[t,\vec{k}] +\sum_{\alpha=1}^p \tilde{r}_\alpha[\vec{k},{\bf G}]\,Q_\alpha[t,\vec{k}] \,,
\end{equation}
where 
\begin{equation}
    \tilde{r}_\alpha [\vec{k},{\bf G}] \equiv -(2\pi)^{d/2}\,\int d^n{\bf K}\,\delta^{(d)}[\vec{K}-\vec{k}]\, \left(\prod_{j=1}^n\tilde{g}[\vec{k}_j]\right) \left(\int_{t_\mathrm{i}}^\infty dt'\, w_\alpha[t',\vec{k}]\,N_{(n)}[t',{\bf{G}},{\bf K}]\right) \,.
\end{equation}
%We see that the relics may not consist only of the growing mode.  If one is certain of the initial growing mode this can allow one to distinguish the relic contribution, consisting of all $Q_i[t,\vec{k}]$, from the initial condition contribution, consisting only of all $G[t,\vec{k}]$.  This might be difficult if there is a single mode which grows much faster than other modes at late times in which case this {\it late time growing mode} will dominate both terms.

\subsubsection{Homogeneous Stochastic Distributions}
\label{sec:HomogeneousStochasticDistribution}
%\noindent \textbf{Homogeneous Stochastic Distributions}
%\vskip2pt

Here, we presume that the initial conditions for $q$ are drawn from some spatially homogeneous stochastic distribution. Since Eq.~(\ref{eq:qPDE}) is assumed to contain only spatially homogeneous differential operators the stochastic distribution of $q$ will remain statistically homogeneous. We wish to determine the 1- and 2-point functions of $\tilde{q}[t,\vec{k}]$, which are guaranteed by spatial homogeneity to take the form
\begin{subequations}\label{eq:qPointFunctions}
\begin{equation}
    \langle \tilde{q}[t,\vec{k}] \rangle = (2\pi)^d\,\delta^{(d)}[\vec{k}]\,P_{(1)}[t] \,,
\end{equation}
\begin{equation}
    \langle \tilde{q}[t,\vec{k}]\,\tilde{q}[t,\vec{k}']\rangle  = (2\pi)^d\,\delta^{(d)}[\vec{k}+\vec{k}']\,P_{(2)}[t,\vec{k}] \,,
\end{equation}
\end{subequations}
where $\langle \cdots \rangle$ denotes expectation values of realizations under this distribution. The function $P_{(2)}$ is the {\it power spectrum} and we call $P_{(1)}$ the {\it secular term}.  In real space the secular term is 
$\langle q[t,\vec{x}]\rangle=(2\pi)^{d/2}\,P_{(1)}[t]$.

Under the LOBA with growing mode initial conditions, the relevant $1$-, $2$-, $n$-, $n+1$- and $2n$-point functions of the growing mode amplitudes $\tilde{g}$ are given by $R_{(1)}$, $R_{(2)}$, $R_{(n)}$, $R_{(n+1)}$, and $R_{(2\times n)}$, respectively, defined by
\begin{subequations}\label{eq:gPointFunctions}
\begin{equation}
    \langle\tilde{g}[\vec{k}]\rangle = (2\pi)^d\,\delta^{(d)}[\vec{k}]\,R_{(1)} \,,
\end{equation}
\begin{equation}
    \langle\tilde{g}[\vec{k}]\,\tilde{g}[\vec{k}']\rangle = (2\pi)^d\,\delta^{(d)}[\vec{k}+\vec{k}']\,R_{(2)}[\vec{k}] \,,
\end{equation}
\begin{equation}
    \langle\prod_{j=1}^n \tilde{g}[\vec{k}_j]\rangle = (2\pi)^d\,\delta^{(d)}[\vec{K}]\,R_{(n)}[t,{\bf K}] \,,
\end{equation}
\begin{equation}
    \langle\tilde{g}[\vec{k}]\prod_{j=1}^n\tilde{g}[\vec{k}_j]\rangle = (2\pi)^d\,\delta^{(d)}[\vec{k}+\vec{K}]\,R_{(1,n)}[\vec{k},{\bf K}] \,,
\end{equation}
\begin{equation}
    \langle\prod_{i=j}^n\tilde{g}[\vec{k}_j ] \,
       \prod_{l=1}^n\tilde{g}[\vec{k}_l']\rangle = (2\pi)^d\,\delta^{(d)}[\vec{K}+\vec{K}']\,
R_{(n,n)}[{\bf K},{\bf K}'] \,.
\end{equation}
\end{subequations}
One can express $P_{(1)}$ and $P_{(2)}$ in terms of $R_{(1)}$, $R_{(2)}$, $R_{(n)}$, $R_{(n+1)}$ and $R_{(n+1)}$ as
\begin{subequations}\label{eq:qPointFunctionsLOBA}
\begin{equation}
    P_{(1)}[t] \simeq R_{(1)}\,G[t,\vec{0}]- (2\pi)^{d/2}\,\int d^n{\bf K}\, \delta^{(d)}[\vec{K}]\,R_{(n)}[{\bf K}]\,M[t,\vec{0},{\bf G},{\bf K}] \,,
\end{equation}
and
\begin{equation}
\begin{split}
    P_{(2)}& [t,\vec{k}] \simeq R_{(2)}[\vec{k}]\,G[t,\vec{k}]\,G[t,-\vec{k}] \\
    & -2\,(2\pi)^{d/2}\! \int \!d^n{\bf K}\,\delta^{(d)}[\vec{k}+\vec{K}]\, R_{(1,n)}[\vec{k},{\bf K}]\, \,G[t,\vec{k}]\,M[t,-\vec{k},{\bf G},{\bf K}]]) \\
    & +(2\pi)^d \!\! \int \!\! d^n{\bf K } \!\!\int \!\! d^n{\bf K'}\, \delta^{(d)}[\vec{k}-\vec{K} ]\, \delta^{(d)}[\vec{k}+\vec{K}']\, R_{(n,n)}[{\bf K},{\bf K}']\, M[t,\vec{k},{\bf G},{\bf K}]\,M[t,-\vec{k},{\bf G},{\bf K}'] \,.
\end{split}
\end{equation}
\end{subequations}
The first term in these expressions is the linear approximation; the last terms are the purely non-linear contribution to the LOBA approximation; and the second term in the expression for $P_{(2)}$ is the interference term between linear and non-linear evolution under the LOBA.\\ 

%which in magnitude is never greater than either the linear or non-linear terms. In some cases this interference term will be zero but in others both non-linear term and the interference term may dominate the linear term (but not the non-linear term) on the largest scales.

\subsubsection{LSWN with Sub-Poissonian Initial Conditions}
\label{sec:LSWNLOBA}
%\noindent \textbf{LSWN with Sub-Poissonian Initial Conditions}
%\vskip2pt

In this paper, we are interested in systems where linear theory predicts sub-Poissonian power spectra so that non-linearity-induced white noise can dominate on large scales, i.e. systems with LSWN. For growing mode initial conditions, the definition of sub-Poissonian initial conditions is that $R_{(2)}[\vec{0}^+]=0$, which also implies that $R_{(n+1)}[\vec{0}^+,{\bf K}]=0$. Thus in the LOBA, the LSWN is given by $P_\mathrm{LSWN}[t] \equiv P_{(2)}[t,\vec{0}^+]$, or explicitly
\begin{equation}\label{eq:LSWNgeneral}
    P_\mathrm{LSWN}[t] \simeq (2\pi)^d \int d^n{\bf K }\,\delta^{(d)}[\vec{K} ]\,M[t,\vec{0}^+,{\bf G},{\bf K} ] \int d^n{\bf K'}\,\delta^{(d)}[\vec{K}']\,M[t,\vec{0}^+,{\bf G},{\bf K}']\, R_{(n,n)}[{\bf K},{\bf K}'] \,.
\end{equation}
This is the main result of this section. It is a very general expression, and many details must be filled in to obtain numerical values for $P_\mathrm{LSWN}$. In particular, linear theory combined with the choice of the growing mode and the form of the leading order non-linearity give $M[t,\vec{0},{\bf G},{\bf K}]$, and initial conditions for the growing mode give $R_{(2n)}[{\bf K},{\bf K}]$. Our goal in the remainder of this section will be to specify these quantities for cosmology-like systems.\\ 

%\label{sec:RelicLSWN}
%\noindent \textbf{Relic LSWN I}
%\vskip2pt

Before doing so, we comment that for a system with sub-Poissonian initial conditions that only leaves linear relics, Eq.~(\ref{eq:LSWNgeneral}) may be approximated by
\begin{equation}\label{eq:LSWNrelic}
    P_\mathrm{LSWN}[t] \simeq \sum_{i=1}^p Q_i[t,\vec{0}^+]\,\sum_{j=1}^p Q_j[t,\vec{0}^+]\,W_{i,j}[{\bf G}] \,,
\end{equation}
where
\begin{subequations}
\begin{equation}
    W_{i,j}[{\bf G}] \equiv (2\pi)^d\, \int d^n{\bf K}\,\delta^{(d)}[\vec{K}]\,W_i[{\bf G},{\bf K}] \int d^n{\bf K'}\,\delta^{(d)}[\vec{K'}]\,W_j[{\bf G},{\bf K}']\, R_{(n,n)}[{\bf K},{\bf K}'] \,,
\end{equation}
\begin{equation}
    W_i[{\bf G},{\bf K}] \equiv \int_{t_\mathrm{i}}^\infty d\bar{t}\, w_i[\bar{t},\vec{0}^+]\,N_{(n)}[\bar{t},{\bf{G}},{\bf K}] \,.
\end{equation}
\end{subequations}
%In this expression, the initial conditions which are encoded in $R_{(n,n)}$ combined with the choice of growing mode, the linear part of the PDE is encoded in $w_i[\bar{t},\vec{0}^+]$ and the non-linear part of the PDE including the choice of growing modes are encoded in $N_{(n)}[t,{\bf{G}},{\bf K}]$.
A sufficient condition for relic LSWN at some late time is generally that the $W_i$ integrals converge for all finite ${\bf K}$ with $\vec{K}=0$.  The late time where relic LSWN is accurate are values of the upper limit where these integrals have accurately converged.\\ 

\subsubsection{Summary}
%\noindent \textbf{Summary}
%\vskip2pt

\textit{In this subsection we developed a perturbative framework for LSWN in non-linear systems. Starting from a general $p^{\rm th}$ order spatially homogeneous PDE, we expanded the dynamics in powers of the field amplitude and identified the leading order non-linearity. Using the Green function formalism we derived the leading order Born approximation (LOBA), which replaces the full non-linear dynamics with insertions of the linear solution. This approximation is valid when the system remains perturbative, i.e.~when the non-linear contributions remain finite and subdominant compared to the linear terms.} 

\textit{We showed that even under these conditions, non-linearities generically induce LSWN on large scales. Depending on the system, the effect may survive only as a linear relic of early-time dynamics, or remain active at late times. For growing mode initial conditions, we derived a compact expression for the LSWN amplitude in terms of the mode functions, the Green function, and the statistical correlators of the initial fluctuations. This general formula (Eq.~\eqref{eq:LSWNgeneral}) encapsulates the main result: any non-linear, non-conservative system with sub-Poissonian initial conditions will generate a universal white-noise contribution to the large-scale power spectrum.}

%%%%%%%%%%%%%%%%%%%%%%%%%%%%%%%%%%%%%%%%%%%%%%%%%%%%%%%%%%%%
%%%%%%%%%%%%%%%%%%%%%%%%%%%%%%%%%%%%%%%%%%%%%%%%%%%%%%%%%%%%
\subsection{Cosmology-Like Systems}\label{sec:CosmologyLike}

In this section, we specialize the results to systems like those encountered in simple models of cosmological inhomogeneity, which typically assume isotropy, positive parity, Gaussianity, and growing mode initial conditions. The simplest models of cosmological inhomogeneities are governed by 2$^{\rm nd}$ order PDE ($p=2$) and have quadratic leading-order non-linearities ($n=2$). Combined with the assumptions of isotropy and homogeneity, this leads to a very small class of differential operators and simple expressions for relic LSWN, which one can normalize with the fastest growing mode at late times. 

%%%%%%%%%%%%%%%%%%%%%%%%%%%%%%%%%%%%%
%\subsubsection{Isotropy}\label{sec:Isotropy}

The Cosmological Principle requires both homogeneity and isotropy on large scales. So far, we have only assumed spatial homogeneity. The assumption of \textit{spatial isotropy} puts additional constraints on the linear and non-linear part of the PDE, as well as on the distribution functions of initial conditions. In particular, this constraint manifests in that the $\vec{k}$ dependence of $Q_\alpha$, $G$, $\tilde{\cal G}$, $P_{(2)}$, and $R_{(2)}$ reduces to a dependence on $k \equiv |\vec{k}|$. On a practical level, this allows to make the substitutions $Q_\alpha[t,\vec{k}] \rightarrow Q_\alpha[t,k]$, $\tilde{\cal G}[t,\bar{t},\vec{k}]\rightarrow \tilde{\cal G}[t,\bar{t},k]$, $g_\alpha[\vec{k}]\rightarrow g_\alpha[k]$ $G[t,\vec{k}]\rightarrow G[t,k]$, $M[t,\vec{k},{\bf G},{\bf K}] \rightarrow M[t,k,{\bf G},{\bf K}]$, $P_{(2)}[t,\vec{k}] \rightarrow P_{(2)}[t,k]$, $R_{(2)}[\vec{k}]\rightarrow R_{(2)}[k]$, etc. Isotropy also guarantees that the ${\bf K}$ and ${\bf K}'$ dependence of the functions $N_{(m)}$, $d_{(m,j)}$, and $M$ will depend only on rotationally-invariant combinations of the vector wavenumbers $\vec{k}_i$. These functions will be further simplified by making the assumption of \textit{positive parity}, which enforces $d_{(m,j)}[{\bf K}]=d_{(m,j)}[-{\bf K}]$, such that
$N_{(m)}[t,{\bf G},{\bf K}]=N_{(m)}[t,{\bf G},-{\bf K}]$
and $M[t,k,{\bf G},{\bf K}]=M[t,k,{\bf G},-{\bf K}]$. 

%%%%%%%%%%%%%%%%%%%%%%%%%%%%%%%%%%%%%
%\subsubsection{Positive Parity}\label{sec:Parity}

%%%%%%%%%%%%%%%%%%%%%%%%%%%%%%%%%%%%%
%\subsubsection{Gaussian Initial Conditions}\label{sec:GaussianInitialConditions}

In addition to these assumptions of isotropy and positive parity, which simplify the PDEs, we will also simplify the initial conditions by taking them to be \textit{Gaussian}. We assume the distribution of $\tilde{g}[\vec{k}]$ to be zero-mean homogeneous, isotropic, Gaussian random noise. Consequently, all $m$-point functions of $\tilde{g}$ will vanish for odd $m$, and will be able to be decomposed as the sum of products of 2-point functions according to Wick's theorem for even $m$. In the LOBA, this can be used to express $R_{(2n)}[{\bf K},{\bf K}']$ in terms of $R_{(2)}[\vec{k},\vec{k}']$.\\ 

%%%%%%%%%%%%%%%%%%%%%%%%%%%%%%%%%%%%%
\label{sec:p2n2}
\subsubsection{$\mathbf{2^{\rm nd}}$ Order with Quadratic Non-Linearities}
%\noindent \textbf{$\mathbf{2^{\rm nd}}$ Order PDE with Quadratic Non-Linearities}
%\vskip2pt

Let us now see how these assumptions impact LSWN, focusing for concreteness on a 2$^{\rm nd}$ order PDE with quadratic non-linearities. A general spatially homogeneous and isotropic 2$^{\rm nd}$ order PDE has the linear operator
\begin{equation}
    {}_{(1)}\hat{\cal D}q =\frac{d^2q}{dt^2}+\alpha[t]\,\frac{dq}{dt}+\beta[t]\,q+\gamma[t]\,\nabla^2q \,,
\label{eq:order2linear}
\end{equation}
corresponding to the linear coefficients
\begin{equation}
\begin{matrix}
    c_{(1,1)}[t]=1& \qquad &
    d_{(1,1)}[{\bf K}]=1 & \qquad &
    l_{(1,1,1)}=2 \\
    c_{(1,2)}[t]=\alpha[t] & \qquad &
    d_{(1,2)}[{\bf K}]=1 & \qquad &
    l_{(1,2,1)}=1 \\
    c_{(1,3)}[t]=\beta[t] & \qquad &
    d_{(1,3)}[{\bf K}]= 1 & \qquad &
    l_{(1,3,1)}=0 \\
    c_{(1,4)}[t]=\gamma[t] & \qquad &
    d_{(1,4)}[{\bf K}]=-\vec{k}_1\cdot\vec{k}_1& \qquad &
    l_{(1,4,1)}=0 \,.
\end{matrix}
\end{equation}
A general spatially homogeneous and isotropic quadratic operator with terms containing no more than 2 derivatives is
\begin{equation}\label{eq:quadraticNL}
    {}_{(2)}\hat{\cal D}q =\xi_1[t]\, q^2 +\xi_2[t]\,q\,\frac{dq}{dt} +\xi_3[t]\,\left(\frac{dq}{dt}\right)^2 +\xi_4[t]\,q\,\nabla^2q +\xi_5[t]\,\vec{\nabla}q\cdot\vec{\nabla}q \,.
\end{equation}
Unless $\xi_j=0$ $\forall \, j$, the leading order non-linearity is quadratic ($n=2)$. The quadratic operator is represented by the coefficients
\begin{equation}
\begin{matrix}
    c_{(2,1)}[t]=\xi_1[t] & \qquad &
    d_{(2,1)}[{\bf K}]=1 & \qquad &
    l_{(2,1,1)}=0 &\quad& l_{(2,1,2)}=0 \\
    c_{(2,2)}[t]=\xi_2[t] & \qquad &
    d_{(2,2)}[{\bf K}]=1 & \qquad &
    l_{(2,1,1)}=0 &\quad& l_{(2,1,2)}=1 \\
    c_{(2,3)}[t]=\xi_3[t] & \qquad &
    d_{(2,3)}[{\bf K}]=1 & \qquad &
    l_{(2,3,1)}=1 &\quad& l_{(2,3,2)}=1 \\
    c_{(2,4)}[t]=\xi_4[t] & \qquad &
    d_{(2,4)}[{\bf K}]=-\vec{k}_2\cdot\vec{k}_2 & \qquad &
    l_{(2,4,1)}=0 &\quad& l_{(2,4,2)}=0 \\
    c_{(2,5)}[t]=\xi_5[t] & \qquad &
    d_{(2,5)}[{\bf K}]=-\vec{k}_1\cdot\vec{k}_2 & \qquad &
    l_{(2,5,1)}=0 &\quad& l_{(2,5,2)}=0 \,. \\
\end{matrix} 
\end{equation}
Since $n=2$, the Gaussian correlator relevant for LSWN is
\begin{equation}\label{eq:p2Correlations}
\begin{split}
\langle\tilde{g}[\vec{k}_1]\tilde{g}[\vec{k}_2]\,
       \tilde{g}[\vec{k}_1']\tilde{g}[\vec{k}_2']\rangle = \langle\tilde{g}[\vec{k}_1 ]\tilde{g}[\vec{k}_2 ]\rangle \langle\tilde{g}[\vec{k}_1']\tilde{g}[\vec{k}_2']\rangle & + \langle\tilde{g}[\vec{k}_1 ]\tilde{g}[\vec{k}_1']\rangle \langle\tilde{g}[\vec{k}_2 ]\tilde{g}[\vec{k}_1']\rangle \\
       & + \langle\tilde{g}[\vec{k}_1]\tilde{g}[\vec{k}_2']\rangle \langle\tilde{g}[\vec{k}_2]\tilde{g}[\vec{k}_1']\rangle \,.
\end{split}
\end{equation}
Since $\langle\tilde{g}[\vec{k}]\,\tilde{g}[\vec{k}']\rangle
=(2\pi)^d\,\delta^{(d)}[\vec{k}+\vec{k}']\,R_{(2)}[k]$,
we find from Eq.~(\ref{eq:gPointFunctions}) that
\begin{equation}
    R_{(2,2)}[{\bf K},{\bf K}']\equiv R_{(2,2)}[\vec{k}_1,\vec{k}_2,\vec{k}_1',\vec{k}_2']= (2\pi)^d\, (\delta^{(d)}[\vec{k}_1+\vec{k}_1'] +\delta^{(d)}[\vec{k}_1+\vec{k}_2']) R_{(2)}[k_1]\,R_{(2)}[k_2] \,.
\label{eq:R22Gaussian}
\end{equation}
Finally invoking also positive parity, the LOBA expression for LSWN in Eq.~(\ref{eq:LSWNgeneral}) becomes\footnote{Without positive parity $2\,M[t,\vec{0},{\bf G},{\bf K}]^2\rightarrow M[t,\vec{0},{\bf G},{\bf K}]^2+
M[t,\vec{0},{\bf G},{\bf K}]\,M[t,\vec{0},{\bf G},-{\bf K}]$. For completely negative parity non-linearities $M[t,\vec{0},{\bf G},-{\bf K}]=-M[t,\vec{0},{\bf G},{\bf K}]$, so under the LOBA the LSWN is zero.}
\begin{equation}\label{eq:LSWNp2}
    P_\mathrm{LSWN}[t] \simeq 2\int d^d\vec{k}_1\, M[t, 0,{\bf G},\vec{k}_1 ,-\vec{k}_1 ]^2\, R_{(2)}[k_1]^2 \,,
\end{equation}
where
\begin{equation}\label{eq:M0GK}
    M[t,0^+,{\bf G},\vec{k}_1,-\vec{k}_1] = \int_{t_\mathrm{i}}^t d\bar{t}\, \tilde{\cal G}[t,\bar{t},0^+] N_{(2)}[\bar{t},{\bf G},\vec{k}_1,-\vec{k}_1] \,,
\end{equation}
and
\begin{equation}
    N_{(2)}[\bar{t},{\bf G},\vec{k}_1,-\vec{k}_1] \!=\! \big(\xi_1[\bar{t}]-(\xi_4[\bar{t}]-\xi_5[\bar{t}]) k_1^2 \big) G[\bar{t},\vec{k_1}]^2 \! +\!  \xi_2[\bar{t}] G[\bar{t},\vec{k_1}] G^{(1)}[\bar{t},\vec{k_1}] + \xi_3[\bar{t}] G^{(1)}[\bar{t},\vec{k_1}]^2 \,.
\end{equation}
Note that if $\xi_4[t]=\xi_5[t]$, then the non-linearity from these two terms sums to $\xi_4[t]\,\vec{\nabla}\cdot(q\,\vec{\nabla}q)$
which is a conservative non-linearity and therefore does not contribute to the LSWN.\\

%%%%%%%%%%%%%%%%%%%%%%%%%%%%%%%%%%%%%
\label{sec:p2n2relicGaussian}
\subsubsection{Relic LSWN}
%\noindent \textbf{Relic LSWN II}
%\vskip2pt

We can also specialize the results of relic LSWN to 2$^{\rm nd}$ order isotropic Gaussian systems with positive parity and quadratic non-linearities. Isotropy requires that the function $W_i[{\bf G},\vec{k}_1,-\vec{k}_1]$ depends only on $k_1$ through $|\vec{k}_1|$.  Defining
\begin{equation}
    W_i[{\bf G},\vec{k}]\equiv W_i[{\bf G},\vec{k},-\vec{k}] =\int_{t_\mathrm{i}}^\infty d\bar{t}\,w_i[\bar{t},\vec{0}^+]\, N_{(2)}[\bar{t},{\bf{G}},\vec{k},-\vec{k}] \,,
\end{equation}
the relic LSWN is given by 
\begin{equation}\label{eq:RelicLSWN2}
    P_\mathrm{rLSWN}[t] \simeq 2\int_0^\infty dk\,k^{d-1}\,U[t,{\bf G},k]\,R_{(2)}[k]^2 \,,
\end{equation}
where
\begin{equation}
    U[t,{\bf G},k] \equiv \int d^{d-1}\hat{n}\, \left(\sum_i Q_i[t,0^+]\, W_i[{\bf G},k\,\hat{n}]\right)^2 \,.
\end{equation}
The function $U$ encodes the PDE, linear and non-linear, while the function $R_{(2)}$ encodes the initial conditions. More generally, for any $n$ there is an analogous expression
\begin{equation}\label{eq:RelicLSWNn}
    P_\mathrm{rLSWN}[t] \simeq n!\int_0^\infty dk\,k^{d-1}\,U[t,{\bf G},k]\,R_{(2)}[k]^n \,,
\end{equation}
but with $U$ given by a higher dimensional $k$-space integral. The $n!$ combinatorial factor is the number of different ways to pair the $k_i$ with the $k_i'$.\\

\label{sec:Fastest}
\subsubsection{Late Time Fastest Growing Mode}
%\noindent \textbf{Late-Time Fastest Growing Mode}
%\vskip2pt

These results can be further simplified by another well-motivated cosmological assumption --- growing-mode initial conditions. We assume that at small $k$ and late times, there is a non-oscillatory {\it fastest growing mode}. More concretely, we assume there is a function $F[t]$ where $f_i \equiv \mathrm{lim}_{t\rightarrow\infty}\frac{Q_i[t,0^+]}{F[t]}$ is bounded for all modes and finite for some. i.e. $\forall i\ |f_i|<\infty$  and $\exists i\ f_i\ne0$.  Similarly define
$f_G\equiv \mathrm{lim}_{t\rightarrow\infty} \frac{G[t,0^+]}{F[t]} =\sum_i g_i[0^+]\,f_i$.  Hence $0\le f_G<\infty$.  Here we assume that $f_G\ne0$, which is often (but not always) the case. Often, the fastest growing mode at late times is, apart from a constant factor, the same as the mode which grows fastest at early times.  If  one chooses $G$ to be this early time fastest growing mode, then $F[t] \propto G[t,0^+]$, so $f_i\propto g_i[0^+]$.

If there is relic LSWN and $f_G\ne0$, then
\begin{equation}
    \mathrm{lim}_{t\rightarrow\infty}\, U[t,{\bf G},k] = f_G^2 F[t]^2 V[{\bf G},k] \,,
\end{equation}
where 
\begin{equation}
    V[{\bf G},k] \equiv \int \frac{d^{d-1}\hat{n}}{V_{S^{d-1}}}\, \left(\sum_i \frac{f_i}{f_G}\,W_i[{\bf G},k\,\hat{n}] \right)^2 \,,
\end{equation}
and so 
\begin{equation}\label{eq:relicLSWNasymptotic}
    \mathcal{P}_\mathrm{aLSWN} \equiv \mathrm{lim}_{t\rightarrow\infty} \frac{P_\mathrm{rLSWN}[t]}{F[t]^2} \simeq 2\,V_{S^{d-1}} \int_0^\infty dk\,k^{d-1}\,V[{\bf G},k]\,R_{(2)}[k]^2 \,.
\end{equation}
Thus, at late times, $P_\mathrm{rLSWN}[t] = \mathcal{P}_\mathrm{aLSWN} \,F[t]^2$. We refer to this form of LSWN as asymptotic relic LSWN. $\mathcal{P}_\mathrm{aLSWN}$ is constant in time but only because it is normalized to a reference growth rate, $F[t]$, whose normalization can be chosen arbitrarily.  One can expect such an asymptote to exist for most systems with relic LSWN since this asymptote only requires there to be a late time fastest growing mode and $f_G\ne0$.\\

%%%%%%%%%%%%%%%%%%%%%%%%%%%%%%%%%%%%%
\label{sec:NoNoScale}
\subsubsection{No-No-Scale Theorem}
%\noindent \textbf{No-No-Scale Theorem}
%\vskip2pt

It is convenient to rewrite Eq.~(\ref{eq:relicLSWNasymptotic}) as
\begin{equation}\label{eq:relicLSWNasymptoticAlt}
    \mathcal{P}_\mathrm{aLSWN} = \int_0^\infty \frac{dk}{k}\,Y[k]\,<\,\infty \,, \qquad Y[k]\equiv V_{S^{d-1}}\,k^d\,V[{\bf G},k]\,R_{(2)}[k]^2 \,,
\end{equation}
since for equivalent systems $Y[k]$ is invariant under reparameterizations.  For the LOBA to be accurate, one requires finite asymptotic LSWN, i.e.
$\mathcal{P}_\mathrm{aLSWN}<\infty$. This convergence requirement requires the dimensionless non-negative function $Y[k]$ to go to zero at both large and small $k$.  If there any non-linearities at all, $Y[k]$ is positive for some $k$ and therefore has one (or more) maxima.  These maxima define a particular length scale so the system (PDE+initial conditions) cannot be ``scale free''.  We call this result the {\it No-No-Scale Theorem}.

For definiteness, we assume a unique maxima at wavenumber $k=k_\mathrm{max}$.  In cosmological systems (and probably most physical systems) the integral of Eq.~(\ref{eq:relicLSWNasymptoticAlt}) is dominated by a narrow range in $\mathrm{ln}[k]$.  One should therefore identify the characteristic cosmological LSWN wavenumber $k_*$ with $k_\mathrm{max}$, i.e. $k_*\sim k_\mathrm{max}$.\\

\noindent \textbf{Summary}
\vskip2pt

\textit{In this subsection, we specialized the general results of \S\ref{sec:LeadingOrderPerturbationTheory} to systems that share the main properties of cosmological inhomogeneities. Beyond homogeneity, we imposed isotropy, positive parity, Gaussian initial conditions, and the growing mode assumption. These simplifications reduce the dependence of all mode functions and correlators to the wavenumber magnitude $k$, and constrain the form of the differential operators. We emphasized that for $p=2$ equations with quadratic leading order non-linearities ($n=2$), the perturbative treatment yields compact expressions for relic LSWN.} 

\textit{The main conclusion is that even in such highly constrained systems, sub-Poissonian initial conditions inevitably evolve into large scale white noise. The amplitude of this LSWN depends on the chosen growing mode, the statistical properties of the initial fluctuations, and the detailed form of the quadratic non-linearities. In this way, cosmology-like systems provide a concrete realization of the general phenomenon that non-linear dynamics populates the largest scales with a stochastic white-noise floor, regardless of linear-theory expectations.}

%%%%%%%%%%%%%%%%%%%%%%%%%%%%%%%%%%%%%
\subsection{Cosmology-Like with Sound}
\label{sec:RadiationLike}

%%%%%%%%%%%%%%%%%%%%%%%%%%%%%%%%%%%%%
%\subsubsection{Sound Horizon}\label{sec:SoundHorizon}

In this subsection, we further specialize to systems similar to cosmic inhomogeneities during radiation domination --- i.e., ones where $\gamma[t]\ne0$ in Eq.~(\ref{eq:order2linear}). If $\gamma[t]>0$, then the PDE is exponentially unstable for large wavenumbers. If on the other hand $\gamma[t]<0$, then the propagation speed of large $k$ modes is given by the sound speed $c_\mathrm{s}[t] \equiv \sqrt{-\gamma[t]}$. From the sound speed, we can define the sound horizon as
\begin{equation}\label{eq:SoundHorizon}
    h[t]\equiv \int_{t_\mathrm{i}}^t d\bar{t}\,c_\mathrm{s}[\bar{t}] \,.
\end{equation}
One can use $h$ as the temporal variable so long as $\gamma[t] < 0 \, \forall t$, which we will do for the remainder of this section.  Where $\gamma[t] < 0$, we sometimes refer to $q[t,\vec{x}]$ as {\it sound waves}.  At a ``time'' $h$ we refer to wavenumbers with $h k\ll1$ as {\it super-horizon}, wavenumbers with $hk\gg1$ as {\it sub-horizon}, and the condition $h k\sim 1$ as {\it horizon crossing}.\\

%%%%%%%%%%%%%%%%%%%%%%%%%%%%%%%%%%%%%
\label{sec:SelfSimilar}
\subsubsection{Self-Similar Evolution}
%\noindent \textbf{Self-Similar Evolution}
%\vskip2pt

One will often encounter systems which, at least approximately, behave in a self-similar way. More concretely, PDE will be called self-similar if from any solution to the PDE one can obtain another solution by linearly rescaling the spatial and temporal coordinates by time-dependent factors, i.e. if $q[t,\vec{x}]=f[t,\vec{x}]$ is solution then so is
$q[t,\vec{x}]=f[a[t] t,b[t]\,\vec{x}]$. If the temporal coordinate is taken to be the sound horizon $h$, then temporal and spatial factors must be equal and constants, i.e. if $q[h,\vec{x}]=f[n,\vec{x}]$ is solution then so is $q[t,\vec{x}]=f[a\,h,a\,\vec{x}]$. Self-similarity imposes restrictions on the PDE; in particular, the time dependent factors in Eqs.~(\ref{eq:order2linear}) and (\ref{eq:quadraticNL}) must be of the form
\begin{eqnarray}\label{eq:xiSelfSimilar}
    &&\alpha[h]=\frac{\zeta_1}{h} \,, \qquad
    \beta[h]=\frac{\zeta_2}{h^2} \,, \qquad
    \gamma[h]=-1 \,, \\
    &&\xi_1[h]=\frac{\zeta_3}{h^2} \,, \qquad
    \xi_2[h]=\frac{\zeta_4}{h}  \,, \qquad
    \xi_3[h]=\zeta_5 \,, \qquad
    \xi_4[h]=\zeta_6 \,, \qquad
    \xi_5[h]=\zeta_7 \,, \nonumber
\end{eqnarray}
where all the $\zeta_i$ are time-independent. Thus, one has
\begin{eqnarray}\label{eq:DSelfSimilar}
    {}_{(1)}\hat{\cal D}q &=&\ \frac{\partial^2q}{\partial h^2} +\frac{\zeta_1}{h}\,\frac{\partial q}{\partial h} +\frac{\zeta_2}{h^2}\,q -\nabla^2q \nonumber\\
    {}_{(2)}\hat{\cal D}q &=& \frac{\zeta_3}{h^2}\,q^2 +\frac{\zeta_4}{h}\,q\,\frac{\partial q}{\partial h} +\zeta_5\,\left(\frac{\partial q}{\partial h}\right)^2 +\zeta_6\,q\,\nabla^2q +\zeta_7\,\vec{\nabla}q\cdot\vec{\nabla}q \,.
\end{eqnarray}
If $q$ is dimensionless, then so are the $\zeta_i$.  We refer to this sort of non-linearities as {\it canonical} self-similar 2$^{\rm nd}$ order non-linearities.  Non-canonical non-linearities can arise when there is deviation from locality, homogeneity, isotropy, self-similarity, or if one allows for higher than 2$^{\rm nd}$ order spatial derivatives.

The linear equation ${}_{(1)}\hat{\cal D}q=0$ has two solutions in Fourier space
\begin{equation}\label{eq:SelfSimilarModes}
    Q_1[h,k]={\cal Q}_+[\varphi]=\varphi^\mu\,J_\nu[\varphi] \,,
    \qquad
    Q_2[h,k]={\cal Q}_-[\varphi]=\varphi^\mu\,Y_\nu[\varphi] \,,
\end{equation}
where $\varphi=k h$, $\mu=\frac{1}{2}(1-\zeta_1)$ and 
$\nu=\frac{1}{2}\sqrt{(1-\zeta_1)^2-4\,\zeta_2}$.  The Green's function is
\begin{equation}
\begin{split}
\tilde{\cal G}[h,\bar{h},k]&=\frac{1}{k}\,
{\cal G}[k\,h,k\,\bar{h}] \\
{\cal G}[\varphi,\bar{\varphi}]&\equiv
\frac{\pi}{2}\,
\frac{\varphi^\mu}{{\bar{\varphi}}^{\mu-1}}
(J_\nu[\bar{\varphi}]\,Y_\nu[\varphi]-Y_\nu[\bar{\varphi}]\,J_\nu[\varphi])
=\sum_\pm\omega_\pm[\bar{\varphi}]\,{\cal Q}_\pm[\varphi] \\
\omega_\pm[\bar{\varphi}]&\equiv
\mp\frac{\pi}{2}\,\bar{\varphi}^{1-2\mu}\,{\cal Q}_\mp[\bar{\varphi}]
\end{split}\ ,
\label{eq:GreenFunction}
\end{equation}
i.e. $w_1[h,k]=\omega_+[h\,k]/k$ and $w_2[h,k]=\omega_-[h\,k]/k$.  Here
$J_\nu$ and $Q_\nu$ are Bessel functions of the first and second kind so long as $\nu \ge 0$ ($\zeta_2<\frac{1}{4}\,(1-\zeta_1)^2$).  For convenience we assume 
\begin{equation}
  \nu>0
\label{eq:nu0}
\end{equation}
Results for $\nu=0$ can be recovered as the $\nu\rightarrow0$ limit.

The long wavelength limit needed in Eq.~(\ref{eq:M0GK}) is given by
\begin{equation}\label{eq:GreenFunction0}
    \tilde{\cal G}[h,\bar{h},0^+] =\frac{\bar{h}}{2\,\nu}\, \left( \left(\frac{h}{\bar{h}}\right)^{\mu+\nu}- \left(\frac{h}{\bar{h}}\right)^{\mu-\nu} \right) \,.
\end{equation}
In order to respect self-similarity the growing mode coefficients must be $k$-independent, i.e. $g_1[k]\rightarrow g_1$ and $g_2[k]\rightarrow g_2$ so the growing mode depends only on $\varphi$, i.e. $G[h,k]\rightarrow G[\varphi]$.  Thus the LSWN of Eq.~(\ref{eq:M0GK}) becomes
\begin{equation}\label{eq:M0GKG}
    {\cal M}[\bar{\varphi}_{\mathrm{i}},\varphi] \equiv M[h,\vec{0}^+,{\bf G},\vec{k},-\vec{k}]= \frac{1}{2\,\nu} \int_{\bar{\varphi}_{\mathrm{i}}}^{\varphi} d\bar{\varphi}\,\bar{\varphi}\, \left( \left(\frac{\varphi}{\bar{\varphi}}\right)^{\mu+\nu}- \left(\frac{\varphi}{\bar{\varphi}}\right)^{\mu-\nu} \right)\,\Psi[\bar{\varphi}] \,,
\end{equation}
where $\varphi = k h$, $\bar{\varphi} = k \bar{h}$,
$\bar{\varphi}_{\mathrm{i}} =
k \bar{h}_\mathrm{i}$,
and
\begin{equation}\label{eq:Psidef}
    \Psi[\bar{\varphi}]\equiv \frac{N_{(2)}[\bar{h},{\bf G},\vec{k},-\vec{k}]} {|\vec{k}|^2} = (\frac{\zeta_3}{{\bar{\varphi}}^2} -(\zeta_6-\zeta_7))\, {\cal Q}_+[\bar{\varphi}]^2 + \frac{\zeta_4}{\bar{\varphi}}\, {\cal Q}_+[\bar{\varphi}_1]\,{{\cal Q}_+}'[\bar{\varphi}] + \zeta_5\, {{\cal Q}_+}'[\bar{\varphi}]^2 \,.
\end{equation}
LSWN in this case is approximately
\begin{equation}\label{eq:LSWNrad}
\begin{split}
    P_\mathrm{LSWN}[h] &\approx 2\,V_{S^{d-1}}\int_0^\infty dk\,k^{d-1}\, {\cal M}[k\,h_\mathrm{i},k\,h]^2\,R_{(2)}[k]^2 \\
    &= 2\,\frac{V_{S^{d-1}}}{h^d}\int_0^\infty d\varphi\,\varphi^{d-1}\, {\cal M}[\varphi_\mathrm{i},\varphi]^2\,R_{(2)}[\varphi/h]^2 \,,
\end{split}
\end{equation}
which can be compared to the superhorizon linear theory power spectrum
\begin{equation}\label{eq:PlinearSuperHorizon}
    P_{(2)}^\mathrm{lin}[h,k] \xrightarrow{k\rightarrow0} \frac{(k\,h)^{2(\mu+\nu)}}{2^{2\nu}\,\Gamma[1+\nu]^2}\,R_{(2)}[k] \,.
\end{equation}

%%%%%%%%%%%%%%%%%%%%%%%%%%%%%%%%%%%%%
%\label{sec:AsymptotesAmplitudesPhases}
\subsubsection{Asymptotes, Amplitudes, and Phases}
%\noindent \textbf{Asymptotes, Amplitudes, and Phases}
%\vskip2pt

The super-horizon and sub-horizon limit of linear basis solutions are
\begin{eqnarray}
{\cal Q}_\pm[\varphi]
&\xrightarrow{\varphi\rightarrow0}&
\pm\frac{\pi^{\pm\frac{1}{2}}\,\varphi^{\mu\pm\nu}}
{\sqrt{\pi}\,(2^\nu\,\Gamma[\nu])^{\pm1}}\,
\nonumber \\
{\cal Q}_\pm[\varphi]
&\xrightarrow{\varphi\rightarrow\infty}&
\sqrt{\frac{2}{\pi}}\,\varphi^{\frac{1}{2}-2\mu}\,
\cos[\varphi-(\nu+1)\,\frac{\pi}{2}\pm\frac{\pi}{4}]\ .
\label{eq:asymptotes}
\end{eqnarray}
Since $\nu\ge0$, the ${\cal Q}_+[\varphi]$ grows faster than ${\cal Q}_-[\varphi]$ for $\varphi\ll1$, so we call 
${\cal Q}_+[\varphi]$ the {\it growing mode} and 
${\cal Q}_-[\varphi]$ the {\it decaying mode}. For $\varphi\gg1$, however, the two solutions oscillate with the same relative amplitude but with a phase difference of $\frac{\pi}{2}$.  These oscillations we call {\it sound waves}.  There is no similar distinction on sub-horizon scales.

The Green function of Eq.~(\ref{eq:GreenFunction}) gives a useful decomposition of the time evolution of the Fourier amplitudes of $q$,
\begin{eqnarray}
    \tilde{q}[h,\vec{k}]&=&
    \sum_\pm a_\pm[h,\vec{k}]\,{\cal Q}_\pm[h,|\vec{k}|] \,,
    \nonumber\\
    \dot{\tilde{q}}[h,\vec{k}]&=&
    \sum_\pm a_\pm[h,\vec{k}]\,|k|\,
    {{\cal Q}_\pm}'[h\,|\vec{k}|] \ .
\label{eq:nonLinearModeDecomposition}
\end{eqnarray}
When $a_-[h,\vec{k}]=0$, we say $\tilde{q}[h,\vec{k}]$ is a {\it pure growing mode}, and when $a_-[h,\vec{k}]=0$, we say $\tilde{q}[h,\vec{k}]$ is a {\it pure decaying mode}.  The $a_\pm$ evolve as
\begin{equation}
\dot{a}_\pm[h,\vec{k}]=-\omega_\pm[|\vec{k}|\,h]\,
(\hat{\tilde{\cal D}}-{}_{(1)}\tilde{\hat{\cal D}})\,q \,.
\label{eq:GrowDecayEvolution}
\end{equation}
The rate of change of growing and decaying modes are given by $\omega_\pm$ times a non-linear forcing term. One can dissect the $a_\pm[h,\vec{k}]$ as follows
\begin{equation}
\left(\begin{matrix}
a_+[h,\vec{k}]\\
a_-[h,\vec{k}]
\end{matrix}\right)
=a[h,\vec{k}]\,
\left(\begin{matrix}
e^{i\,\psi_+[h,\vec{k}]}\,\cos[\Delta\varphi[h,\vec{k}]]\\
e^{i\,\psi_-[h,\vec{k}]}\,\sin[\Delta\varphi[h,\vec{k}]]
\end{matrix}\right) \ .
\label{eq:dissection}
\end{equation}
where
\begin{equation}
a[h,\vec{k}]\equiv\sqrt{\sum_\pm |a_\pm[h,\vec{k}]|^2} \,, \qquad
\Delta\varphi[h,\vec{k}]\equiv
\mathrm{tan}^{-1}[\frac{|a_-[h,\vec{k}]|}{|a_+[h,\vec{k}]|}]\,, \qquad
\psi_\pm[h,\vec{k}]\equiv\,\mathrm{arg}[a_\pm[h,\vec{k}]] \,.
\end{equation}
We call $a[h,\vec{k}]>0$ the {\it amplitude}, $\Delta\varphi[h,\vec{k}]\in[0,{\frac{\pi}{2}}]$ the {\it temporal phase}\footnote{The temporal phase of sound waves play a special role in cosmology because it determines the peaks and troughs in the angular power spectrum of the CMBR.}, and the $\psi_\pm[\vec{k}]$ {\it spatial phases}. $\tilde{q}[h,\vec{k}]$ is a pure growing mode or a pure decaying mode when 
$\Delta\varphi[h,\vec{k}]=0$ or when $\Delta\varphi[h,\vec{k}]=\frac{\pi}{2}$, respectively.  Since $q$ and $\dot{q}$ are real, $a[h,-\vec{k}]=a[h,\vec{k}]$, $\Delta\varphi[h,-\vec{k}]=\Delta\varphi[h,\vec{k}]$ and 
$e^{i\,\psi_\pm[h,-\vec{k}]}=e^{-i\,\psi_\pm[h,\vec{k}]}$. To the extent the non-linearities are negligible, $a[h,\vec{k}]$, $\Delta\varphi[h,\vec{k}]$ and $\psi_\pm[h,\vec{k}]$ are constant in time.

In the sub-horizon limit,
\begin{eqnarray}
\tilde{q}[h,|\vec{k}|]
&\xrightarrow{\varphi\rightarrow\infty}&+
\frac{a[h,\vec{k}]}{\sqrt{2\pi}}\,\varphi^{\frac{1}{2}-2\mu}\,
\sum_\pm
(e^{i\,\psi_-[h,\vec{k}]}\mp e^{i\,\psi_+[h,\vec{k}]})\,
\cos
[\varphi\pm\Delta\varphi[h,\vec{k}]-(2\,\nu+1)\,\frac{\pi}{4}] \,,
\nonumber\\
\frac{\dot{\tilde{q}}[h,|\vec{k}|]}{|\vec{k}|}
&\xrightarrow{\varphi\rightarrow\infty}&-
\frac{a[h,\vec{k}]}{\sqrt{2\pi}}\,\varphi^{\frac{1}{2}-2\mu}\,
\sum_\pm
(e^{i\,\psi_-[h,\vec{k}]}\mp e^{i\,\psi_+[h,\vec{k}]})\,
\sin
[\varphi\pm\Delta\varphi[h,\vec{k}]-(2\,\nu+1)\,\frac{\pi}{4}]
\,.
\end{eqnarray}
This is the sum of two sound waves --- one with phase advanced by $+\Delta\varphi$ with respect to the growing mode and the other with phase retarded by $-\Delta\varphi$ with respect to the growing mode.  The amplitude of the former is 
$\propto\sqrt{1-\cos[\psi_+-\psi_-]}$
and the latter 
$\propto\sqrt{1+\cos[\psi_+-\psi_-]}$.
So if the spatial phase of the growing mode is positively correlated with that of the decaying mode, corresponding to $\cos[\psi_+-\psi_-]>0$,
the retarded phase wave is larger; if it is negatively correlated, i.e. $\cos[\psi_+-\psi_-]<0$, it is the advanced wave that is larger. For pure growing modes ($\Delta\varphi=0$) or pure decaying modes ($\Delta\varphi=\frac{\pi}{2}$), the two sound waves merge into one.\footnote{If $\Delta\varphi=\frac{\pi}{2}$, the two waves have opposite sign, which can be absorbed into the spatial phase}

Using the mode decomposition of Eq.~(\ref{eq:nonLinearModeDecomposition}), the $q$ power spectrum can be decomposed into
\begin{equation}
    P_q[h,k]=
    P_q^{++}[h,k]+P_q^{+-}[h,k]+P_q^{+-}[h,k]+P_q^{--}[h,k] \,,
\label{eq:PowerSpectrumDecomposition}
\end{equation}
where
\begin{equation}
    {\cal Q}_{\pm_1}[|\vec{k}|\,h]\,{\cal Q}_{\pm_2}[|\vec{k}'|\,h]\,
    \langle a_{\pm_1}[h,\vec{k}]a_{\pm_2}[h,\vec{k}']\rangle=
    (2\pi)^d\,\delta^{(d)}[\vec{k}+\vec{k}']\,P_q^{\pm_1\pm_2}[h,|\vec{k}|] \,,
\end{equation}
and $\pm_1$ and $\pm_2$ take on the values $+$ and $-$.  Note that 
$P_q^{+-}[h,k]=P_q^{-+}[h,k]$. The mode functions ${\cal Q}_\pm[\varphi]$ have zeros on sub-horizon scales and so will the $P_q^{\pm_1\pm_2}[h,k]$ at specific values of $kh$.  Identifying these zeros is one way of distinguishing the growing and decaying mode. Since non-linearities generate sound waves with no definite phase, i.e. a mixture of growing and decaying modes, the true $P_q[h,k]$ will be no zeros. If either the growing or decaying mode dominate, however, there will remain approximate zeros of $P_q[h,k]$.  In any cases, there are no zeros on super-horizon scales where sound wave oscillation has not yet begun.

Ratios of the different power spectra give weighted averages of the temporal phase
\begin{equation}
\overline{\tan[\Delta\varphi]^2}[h,k]
\equiv\frac{P_q^{--}[h,k]}{P_q^{++}[h,k]}
=\frac
{\langle a^2\,\sin[\Delta\varphi]^2\rangle}
{\langle a^2\,\cos[\Delta\varphi]^2\rangle} \,,
\end{equation}
and the relative spatial phase of the growing and decaying modes
\begin{equation}
\overline{\cos[\psi_+-\psi_-]}[h,k]
\equiv\frac{P_q^{+-}[h,k]}
{\sqrt{P_q^{++}[h,k]\,P_q^{--}[h,k]}}
=\frac
{\langle  a^2\,\cos[\Delta\varphi]\,\sin[\Delta\varphi]\,\cos[\psi_+-\psi_-]
 \rangle}
{\sqrt{\langle a^2\,\cos[\Delta\varphi]^2\rangle\,
       \langle a^2\,\sin[\Delta\varphi]^2\rangle}} \,.
\end{equation}
If the temporal phase $\Delta\varphi$ is the same for all modes with the same $h$ and $|\vec{k}|$, then 
$\overline{\tan[\Delta\varphi]^2}[h,k]=\tan[\Delta\varphi[h,k]]^2$ and
$\overline{\cos[\psi_+-\psi_-]}[h,|\vec{k}|]=
\langle a[h,\vec{k}]^2\cos[\psi_+[h,\vec{k}]-\psi_-[h,\vec{k}]]\rangle/
\langle a[h,\vec{k}]^2\rangle \ .$

The large scale $\vec{k}\rightarrow0$ limit of Eq.~(\ref{eq:PowerSpectrumDecomposition}) corresponds to the $\varphi\ll0$ limit of ${\cal Q}_\pm[\varphi]$ given in Eq.~(\ref{eq:asymptotes}). Thus using the decomposition of Eq.~(\ref{eq:nonLinearModeDecomposition}) one finds that
\begin{equation}
\begin{split}
\lim_{\vec{k}\rightarrow\vec{0}}\,\langle a_+[h,\vec{k}]a_+[h,\vec{k}']\rangle
&=
(2\pi)^d\,\delta^{(d)}[\vec{k}+\vec{k}']\,
\frac{P_\mathrm{LSWN}^{++}[h]}
{\left(\frac{1}{2^\nu\,\Gamma[\nu]}\,(|\vec{k}|\,h)^{\mu+\nu}\right)^2} \,,
\\
\lim_{\vec{k}\rightarrow\vec{0}}\,\langle a_+[h,\vec{k}]a_-[h,\vec{k}']\rangle
=
\lim_{\vec{k}\rightarrow\vec{0}}\,\langle a_-[h,\vec{k}]a_+[h,\vec{k}']\rangle
&=
(2\pi)^d\,\delta^{(d)}[\vec{k}+\vec{k}']\,
\frac{P_\mathrm{LSWN}^{+-}[h]}{\frac{1}{\pi}(|\vec{k}|\,h)^{2\mu}} \,,
\\
\lim_{\vec{k}\rightarrow\vec{0}}\,\langle a_-[h,\vec{k}]\,a_-[h,\vec{k}']\rangle
&=
(2\pi)^d\,\delta^{(d)}[\vec{k}+\vec{k}']\,
\frac{P_\mathrm{LSWN}^{--}[h]}
{\left(\frac{2^\nu\,\Gamma[\nu]}{\pi}\,(|\vec{k}|\,h)^{\mu-\nu}\right)^2} \,,
\end{split}
\end{equation}
where we have defined 
$P_\mathrm{LSWN}^{\pm_1\pm_2}[h]\equiv P_q^{\pm_1\pm_2}[h,0^+]$.
For Gaussian initial conditions
\begin{equation}
\begin{split}
P_\mathrm{LSWN}^{\pm_1\pm_2}[h]
&\approx
2\,\frac{V_{S^{d-1}}}{h^d}\int_0^\infty d\varphi\,\varphi^{d-1}\,
{\cal M}_{\pm_1}[\varphi_\mathrm{i},\varphi]\,
{\cal M}_{\pm_2}[\varphi_\mathrm{i},\varphi]\,R_{(2)}[\varphi/h]^2 \,,
\\
{\cal M}_\pm[\bar{\varphi}_{\mathrm{i}},\varphi]&\equiv
\frac{\pm1}{2\,\nu}
\int_{\bar{\varphi}_{\mathrm{i}}}^{\varphi}
d\bar{\varphi}\,\bar{\varphi}\,
\left(\frac{\varphi}{\bar{\varphi}}\right)^{\mu\pm\nu}
\,\Psi[\bar{\varphi}] \,,
\end{split}
\label{eq:M0GKGpm}
\end{equation}
which is a decomposition of Eq.~(\ref{eq:LSWNrad}).  For $\varphi>\varphi_\mathrm{i}$, note that
${\cal M}_+[\varphi_\mathrm{i},\varphi]>
-{\cal M}_-[\varphi_\mathrm{i},\varphi]>0$.  It then follows that
$P_\mathrm{LSWN}^{++}[h]\ge-P_\mathrm{LSWN}^{+-}[h]>P_\mathrm{LSWN}^{--}[h]>0$ and that
$\overline{\tan[\Delta\varphi]^2}_\mathrm{LSWN}[h]\equiv
\overline{\tan[\Delta\varphi]^2}[h,0^+]<1$
and
$\overline{\cos[\psi_+-\psi_-]}_\mathrm{LSWN}[h]\equiv
\overline{\cos[\psi_+-\psi_-]}[h,0^+]<0$. 

Previous arguments suggests that since $\cos[\psi_+-\psi_-]$ is preferentially negative, the wave advanced in phase with respect to the growing mode ($\varphi\rightarrow\varphi+\Delta\varphi$) tends to have larger amplitude than the wave retarded in phase with respect to the growing mode
($\varphi\rightarrow\varphi-\Delta\varphi$) for all $\vec{k}$.  If we are in a regime where the growing modes dominates $\Delta\varphi\ll1$, this suggests that eventually when the LSWN enters the horizon and starts to oscillate the near-zeros of $P_q$ will occur at slightly smaller $\varphi\equiv k\,h$ than for a pure growing mode.   This will remain true even at $k$ where the LSWN does not dominate the initial growing mode since it is only the LSWN which contributes decaying modes.\\

%%%%%%%%%%%%%%%%%%%%%%%%%%%%%%%%%%%%%
%\label{sec:InitialSingularity}
\subsubsection{Growing Mode with Initial Singularity}
%\noindent \textbf{Growing Mode with Initial Singularity}
%\vskip2pt

If one or more of  $\zeta_1$, $\zeta_2$, $\zeta_3$, and $\zeta_4$ is non-zero then the PDE becomes singular at $h=0$ which we call the {\it initial singularity}.  In cosmology this corresponds to the {\it Big Bang}.  If we want the initial conditions to be scale invariant, one should not introduce a spatial scale by choosing $h_\mathrm{i}\ne0$. We thus take $h_\mathrm{i}=0$. Since $Y_\nu[0]=-\infty$, in order to avoid singularities at the initial time one must choose $g_2[\vec{k}]=0$ --- i.e. pure growing mode initial conditions. If we choose the natural normalization $g_1[\vec{k}]=1$, we have $G[\varphi]=\varphi^\mu\,J_\nu[\varphi]$ which will be used henceforth.

If $\zeta_1>0$, then $\mu<\frac{1}{2}$, so the growing mode grows monotonically from zero for $\varphi\ll1$ and for $\varphi\gg1$ oscillates with decreasing amplitude .  This change from growth to decay we call {\it sound horizon crossing}.  The decay after sound horizon crossing may allow the rms $q$ to be dominated by modes which are just crossing the horizon. This depends on the parameters $\nu$ and $\mu$, however.

Since we are assuming $\varphi_\mathrm{1i}=0$ and $g_i[\vec{k}]=\delta_{i,1}$
\begin{equation}
\Psi[\varphi]\equiv 
{\varphi}^{2\,\mu}
\left(
\left(
\frac{\zeta_3+\mu\,\zeta_4+\mu^2\,\zeta_5}
{{\varphi}^2}
-(\zeta_6-\zeta_7)\right)\,
J_\nu[\varphi]^2
+
\frac{\zeta_4+2\,\mu\,\zeta_5}{\varphi}\,
J_\nu[\varphi]\,J_\nu'[\varphi]
+
\zeta_5\,
J_\nu'[\varphi]^2
\right) \,,
\label{eq:PsiJ}
\end{equation}
where 
$J_\nu'[\varphi]\equiv\frac{d}{d\varphi}J_\nu[\varphi]
=\frac{1}{2}(J_{\nu-1}[\varphi]-J_{\nu+1}[\varphi])$. Thus Eq.~(\ref{eq:LSWNrad}) becomes
\begin{equation}
P_\mathrm{LSWN}[h]
\approx
2\,\frac{V_{S^{d-1}}}{h^d}
\int_0^\infty d\varphi\,\varphi^{d-1}\,
{\cal M}[\varphi]^2\,
R_{(2)}[\varphi/h]^2 \,,
\label{eq:LSWNinitialSingularity}
\end{equation}
where
\begin{equation}
{\cal M}[\varphi]\equiv{\cal M}[0,\varphi]
=
\frac{1}{2\,\nu}\int_0^{\varphi}
d\bar{\varphi}\,\bar{\varphi}\,
\left(
\left(\frac{\varphi}{\bar{\varphi}}\right)^{\mu+\nu}-
\left(\frac{\varphi}{\bar{\varphi}}\right)^{\mu-\nu}
\right)\,\Psi[\bar{\varphi}] \,.
\label{eq:M0GKG0}
\end{equation}
This integral can be performed analytically, producing rather lengthy expressions involving generalized hypergeometric functions. At the lower limit,
\begin{equation}
\Psi[\varphi]=
\frac{\zeta_3+(\mu+\nu)\,(\zeta_4+(\mu+\nu)\,\zeta_5)}{2^{2\nu}\,\Gamma[1+\nu]^2}
\,\varphi^{2\,(\mu+\nu-1)} 
+{\cal O}[\varphi^{2\,(\mu+\nu)}] \ .
\end{equation}
For this system to be perturbative, one requires 
$|{\cal M}[\varphi]|<\infty$ for $0\le\varphi<\infty$
Using Eq.~(\ref{eq:nu0}) the condition for this is that\footnote{Unless $\zeta_3+(\mu+\nu)\,(\zeta_4+(\mu+\nu)\,\zeta_5)=0$ which we henceforth assume is not the case.} 
\begin{equation}
    \mu+\nu>0
\label{eq:mupnu0}
\end{equation}
The approximation of Eq.~(\ref{eq:M0GKG0}) is the main result of \S\ref{sec:RadiationLike}.\\

%%%%%%%%%%%%%%%%%%%%%%%%%%%%%%%%%%%%%
\label{sec:PowerLaw}
\subsubsection{Power Law Initial Conditions}
%\noindent \textbf{Power Law Initial Conditions}
%\vskip2pt

Setting a starting time at $h_\mathrm{i}=0$ does not impose any length scale on the system.  Similarly a power law initial spectrum 
\begin{equation}
 R_{(2)}[k]= A_\mathrm{fid}\,
 \left(\frac{k}{k_\mathrm{fid}}\right)^{s-2\,(\mu+\nu)}
\label{eq:R2powerLaw}
\end{equation}
for $\tilde{g}[\vec{k}]$ also does not impose any length scale on the system, since it gives the linear theory power spectrum
\begin{equation}
P_{(2)}^{\mathrm{lin}}[h,k]
=\frac{A_\mathrm{fid}}{(k_\mathrm{fid}\,h)^{s-2\,(\mu+\nu)}}\,
(k\,h)^{s-2\nu}\,J_\nu[k\,h]^2 \,.
\label{eq:PowerLawSpectrum}
\end{equation}

In the super-horizon limit,
\begin{equation}
\mathrm{lim}_{kh\rightarrow 0}P^\mathrm{lin}_{(2)}[h,k]=
\frac{A_\mathrm{fid}}{2^{2\nu}\,\Gamma[1+\nu]^2}\,
\frac{(kh)^s}{(k_\mathrm{fid} h)^{s-2\,(\mu+\nu)}} \,, 
\label{eq:Plinear2}
\end{equation}
such that $s=0$ corresponds to initial white noise.  Since we are interested in sub-Poissonian initial conditions, we have the additional requirement that $s>0$. In this case
\begin{equation}
    \Delta^2_\mathrm{lin}[h,k] =V_{S^{d-1}}\,A_\mathrm{fid}\, \frac{(k h)^{s+d-2\,\nu}} {{(k_\mathrm{fid} h})^{s-2\,(\mu+\nu)}\,h^d}\,J_\nu[k h]^2 \,.
\label{eq:Delta2LinearPL}
\end{equation}
%requirements for perturbativity are that one satisfy Eq.~\eqref{eq:Finite0} at lo-$k$ and since $\gamma[h]=1\ne0$ one must also satisfy Eq.~\eqref{eq:Finite4} at hi-$k$ or
In order to satisfy perturbativity, we demand
\begin{equation}
    0<s+d<2\,\nu-3 \,.
\label{eq:linearConvergencePL}
\end{equation}
The lower limit on $s$ is always satisfied for sub-Poissonian initial conditions while the upper limit requires $\nu>\frac{3}{2}$.

For the power law initial condition of Eq.~(\ref{eq:R2powerLaw}) the LOBA LSWN of Eq.~(\ref{eq:LSWNrad}) is
\begin{equation}
    P_\mathrm{LSWN}[h]
    \approx
    2\,\frac{{A_\mathrm{fid}}^2\,V_{S^{d-1}}}
    {h^d\,(k_\mathrm{fid}\,h)^{2(s-2(\mu+\nu))}}
    \int_0^\infty d\varphi\,\varphi^{2(s-2(\mu+\nu))+d-1}\,
    {\cal M}[\varphi]^2
    \propto h^{4(\mu+\nu)-s-d}\ .
\label{eq:LSWNradPowerLaw}
\end{equation}
Explicit expressions and limiting forms for ${\cal M}[\varphi]$ are given in Appendix \ref{sec:calManalytic} where it was shown that ${\cal M}[\varphi]$ approaches a power law at both large and small $\varphi$, with
the large $\varphi$ power law exponent greater than the small $\varphi$ power law exponent. The transition (the knee) occurs at $\varphi\sim1$ and allows power law initial conditions to give finite LSWN for a range of $s$:
\begin{equation}
\begin{cases}
0<s+\frac{1}{2}\,d<2\,\nu-1 & \mu+1>\nu \\
0<s+\frac{1}{2}\,d<\mu+\nu  & \mu+1<\nu \ .
\end{cases}
\label{eq:LSWNconvergencePL}
\end{equation}
This provides the remaining requirement for perturbativity. One can show that there exist parameter values which satisfy the many constraints of Eqs.~(\ref{eq:nu0}), (\ref{eq:mupnu0}), (\ref{eq:linearConvergencePL}), \& (\ref{eq:LSWNconvergencePL}), e.g. $d=3$, $\mu=\nu=\frac{9}{2}$ and $s=1$.\\

%%%%%%%%%%%%%%%%%%%%%%%%%%%%%%%%%%%%%
\label{sec:ActivePowerLawCorollary}
%\noindent \textbf{Active Power Law Corollary}
%\vskip2pt

Here we show by different arguments that self-similar systems with power law initial conditions cannot have relic LSWN, but rather there will necessarily have active non-linearities at all times. 

Using the formalism of ``Asymptotes, Amplitudes, and Phases'', we decompose the LOBA LSWN for initial power laws into growing and decaying modes. The decomposition of $P_\mathrm{LSWN}[h]$ is characterized by two additional quantities: $\overline{\tan[\Delta\varphi]^2}_\mathrm{LSWN}$
and
$\overline{\cos[\psi_+-\psi_-]}_\mathrm{LSWN}$. For any combination of signs, $P_\mathrm{LSWN}^{\pm_1\pm_2}\propto h^{4(\mu+\nu)-s-d}$ and accordingly these two quantities by ratios of $P_\mathrm{LSWN}^{\pm_1\pm_2}$ are time independent.  In particular
\begin{equation}
\overline{\tan[\Delta\varphi]^2}[h,0^+]
\equiv\frac{
\int_0^\infty d\varphi\,\varphi^{2(s-2(\mu+\nu))+d-1}\,
{\cal M}_-[\varphi]^2
}{
\int_0^\infty d\varphi\,\varphi^{2(s-2(\mu+\nu))+d-1}\,
{\cal M}_+[\varphi]^2
}\ne0 \ .
\end{equation}
This ratio is nonzero because ${\cal M}_-[\varphi]\ne0$. Since $\overline{\tan[\Delta\varphi]^2}\ne0$ it follows that $\Delta\varphi$ systematically differs from zero, meaning the temporal phase does not approach that of the growing mode, $\Delta\varphi=0$.  This indicates that the LSWN is active and not a relic since for relic LSWN the growing mode will dominate at late times.  The reason that the growing mode does not dominate for power law initial conditions of self-similar systems is that there is always the same amount of active non-linearities which continuously generates decaying modes. 

For self-similar systems with sound, the late-time fastest super-horizon (a.k.a. large scale or $\varphi\ll1$) growing mode is the same power law in $h$ as the initial growing mode, so $F[h]\propto h^{\mu+\nu}$.  One may use {\it growing mode} to refer to either. We have assumed $p=2$ so there is one other mode, which we call the {\it decaying mode}, where $\tilde{q}\propto D[h]\propto h^{\mu-\nu}$.  This decaying mode can increase or decrease, but always decreases relative to the growing mode since $\nu>0$.\footnote{In the limiting case $\nu\rightarrow0$ the ratio of decaying to growing mode decreases logarithmically in $h$.}

The ratio of the LSWN to the linear theory growing mode power spectrum is
\begin{equation}
\frac{P_\mathrm{LSWN}[h]}{F[h]^2}\propto h^{2(\mu+\nu)-s-d}
\ .
\end{equation}
For relic LSWN, this ratio would be constant and would be insensitive to the initial conditions.  In this case, for most parameters, the ratio is not constant.  Even if $s=2\,(\mu+\nu)-d$ and the ratio is constant, this constancy holds only for a particular initial power spectrum so it is not insensitive to initial conditions and therefore is not relic LSWN.  Thus, in agreement with the $\Delta\varphi\ne0$ argument, power law initial conditions lead to active LSWN.

The fact that power law initial conditions of self-similar PDEs cannot produce relic LSWN is actually a corollary of the No-No-Scale Theorem of \S\ref{sec:CosmologyLike}.  This is because neither the self-similar PDE nor the power law initial conditions introduce any particular spatial scale into the system.  The No-No-Scale theorem states that if there is relic LSWN, it must be characterized by a particular spatial scale. Since the systems which are self-similar with power law initial conditions have no scale, they cannot have relic LSWN and therefore they must have active LSWN.  We call this corollary the {\it Active Power Law Corollary}.  This corollary is just as general as the No-No-Scale Theorem and not restricted to the $p=n=2$ systems we are currently considering.\\

%%%%%%%%%%%%%%%%%%%%%%%%%%%%%%%%%%%%%
\label{sec:LateTimeLSWN}
\subsubsection{Late-Time LSWN}
%\noindent \textbf{Late-Time LSWN}
%\vskip2pt

A self-similar system with power law initial conditions is completely scale free and self-similar, and thus there is no distinction between late times and early times.  This is not the case if the initial conditions are not power laws because features in the initial power spectrum enter the sound horizon at different times.

As shown in appendix \S\ref{sec:calManalytic}, the function ${\cal M}[\varphi]$ is a power law at large and small $\varphi$,\footnote{The quantities ${\mathfrak A}$, ${\mathfrak B}$, and $\alpha$ are numbers of order unity that depend on $\mu$, $\nu$, and the $\zeta_i$.}
\begin{equation}
{\cal M}[\varphi]
\xrightarrow{\varphi\rightarrow0}
{\mathfrak A}\,\varphi^{\mu+\nu} \qquad
{\cal M}[\varphi]
\xrightarrow{\varphi\rightarrow\infty}
{\mathfrak B}\,\varphi^{\alpha} \ ,
\label{sec:calMasymptoticGeneral}
\end{equation}
with a knee at $\varphi\sim1$.  The knee generically indicates that most LSWN is generated by non-linearities at horizon crossing.  If in the initial power spectrum $R_{(2)}[k]$ is peaked near $k\sim k_*$, then much of the LSWN is generated by non-linearities at time $h_*\sim1/k_*$.  If $h\gg h_*$ then the integral in Eq.~(\ref{eq:LSWNinitialSingularity}) is dominated by 
$\varphi\sim k_*\,h\gg1$, where
${\cal M}[\varphi]\approx\mathfrak{B}\,\varphi^\alpha$, so that
\begin{equation}
P_\mathrm{LSWN}[h]
\xrightarrow{h\rightarrow\infty}
2\,V_{S^{d-1}}\,{\mathfrak B}^2\,h^{2\alpha}
\int_0^\infty dk\,k^{2\alpha+d-1}\,R_{(2)}[k]^2 \,.
\label{eq:LSWNasymptotic}
\end{equation}
The super-horizon fastest growing mode is $F[h]\propto h^{\mu+\nu}$, so
\begin{equation}
\lim_{h\rightarrow\infty}\frac{P_\mathrm{LSWN}[h]}{F[h]^2}\propto
\begin{cases}
\mathrm{constant} & \nu>1+\mu \qquad \mathrm{relic} 
\qquad \alpha=\mu+\nu \,, \\
h^{2(\mu-\nu+1)}  & \nu<1+\mu \qquad \mathrm{active} 
\qquad \hskip-7pt \alpha=2\mu+1 \,.
\end{cases}
\end{equation}
In the first case, this ratio is constant, giving relic LSWN generated by non-linearities at earlier times (nominally $h\sim h_*$).  Relic LSWN has negligible decaying mode content and evolves according to the linear theory of growing modes.  In the second case of active LSWN, small scale non-linearities continue to generate large scale power in the form of both growing and decaying modes, and the LSWN grows faster than the linear theory of both growing and decaying modes.  

Here, the discriminant between relic and active LSWN comes only from the linear PDE through $\mu-\nu$.  It does not depend on the non-linear part of the PDE or on the initial conditions except for the requirement that the integral in Eq.~(\ref{eq:LSWNasymptotic}) converges.\\

%%%%%%%%%%%%%%%%%%%%%%%%%%%%%%%%%%%%%
\label{sec:RelicLSWNIII}
\subsubsection{Relic LSWN with Truncated Power Law Initial Conditions}
%\noindent \textbf{Relic LSWN with Truncated Power Law Initial Conditions}
%\vskip2pt

Further restricting to the case of relic LSWN, i.e. $\nu>1+\mu$ so $\alpha=\mu+\nu$, we have that $\mathfrak{B}=\varsigma_+^2$ defined in Eq.~(\ref{eq:varsigmas}). Here we consider a truncated power law for the initial power spectra. In particular,
\begin{equation}
R_{(2)}[k]= A_\mathrm{fid}\,
\left(\frac{k}{k_\mathrm{fid}}\right)^{s-2\,(\mu+\nu)}\,C[k] \,,
\label{eq:R2truncatedPowerLaw}
\end{equation}
where $C[k]$ is some truncation function interpolating monotonically between $C[0]=1$ and $C[\infty]=0$.  LSWN occurs on large scales/small $k$ where $C[k]=1$ is a good approximation.  At small scales/large $k$, we assume the truncation is sufficiently abrupt that Eq.~(\ref{eq:LSWNasymptotic}) is finite. This system is perturbative if $\nu>0$ and $\mu+\nu>0$ and initially sub-Poissonian if $s>0$.  Perturbativity places no upper limit on $s$.  

Eq.~(\ref{eq:LSWNasymptotic}) may be written
\begin{equation}
P_\mathrm{LSWN}[h]
\xrightarrow{h\rightarrow\infty}
\frac{2\,V_{S^{d-1}}\,{\mathfrak B}^2}{d+2(s-(\mu+\nu))}\,
{A_\mathrm{fid}}^2{k_\mathrm{c}}^d\,
(k_\mathrm{c}\,h)^{2(\mu+\nu)}\,
\left(\frac{k_\mathrm{c}}{k_\mathrm{fid}}\right)^{2(s-2(\mu+\nu))}\,,
\label{eq:LSWNasymptoticCutoff}
\end{equation}
where
\begin{equation}
k_\mathrm{c}\equiv
\left((d+2(s-(\mu+\nu)))
\int_0^\infty\frac{dk}{k}\,k^{d+2(s-(\mu+\nu))}\,C[k]^2
\right)^{\frac{1}{d+2(s-(\mu+\nu))}} \,,
\end{equation}
is a characteristic truncation wavenumber. (For example, for a $\Theta$-function truncation $C[k]=\Theta[k_\mathrm{c}-k]$, $C[k]$ has to fall of fast enough that $k_\mathrm{c}<\infty$).

On super-horizon scales long after the LSWN was generated and (${k_\mathrm{c}}^{-1}\ll h\,\ll k^{-1}$), the linear theory power spectrum of Eq.~(\ref{eq:PlinearSuperHorizon}) is
\begin{equation}
P_{(2)}^\mathrm{lin}[h,k]
\xrightarrow{k\rightarrow0}
\frac{(k\,h)^{2(\mu+\nu)}}{2^{2\nu}\,\Gamma[1+\nu]^2}\,
A_\mathrm{fid}\,
\left(\frac{k}{k_\mathrm{fid}}\right)^{s-2\,(\mu+\nu)} \,,
\label{eq:PlinearSuperHorizonLate}
\end{equation}
which uses $C[k]\xrightarrow{k\ll k_\mathrm{c}}1$. Both the LSWN and linear theory power spectra grow at the same rate and the (constant in time) ratio of the two is
\begin{equation}
\frac{P_\mathrm{LSWN}[h]}{P_{(2)}^{\mathrm{lin}}[h,k]}
\xrightarrow{k\,h\ll1\ll k_\mathrm{c}\,h}
\left(\frac{k_\mathrm{LSWN}}{k}\right)^s \,,
\end{equation}
where
\begin{equation}
k_\mathrm{LSWN}\approx
\left(
\frac{2^{2\nu+1}\,\Gamma[1+\nu]^2\,V_{S^{d-1}}\,{\mathfrak B}^2}
{d+2(s-(\mu+\nu))}\,A_\mathrm{fid}\,
\left(\frac{k_{\mathrm{c}}}{k_\mathrm{fid}}\right)^{s-2(\mu+\nu)}\,
{k_\mathrm{c}}^{d+s}
\right)^{\frac{1}{s}} \,,
\label{eq:kLSWNofkc}
\end{equation}
is the cross-over wavenumber where the relic LSWN power equals that of super-horizon linear theory.  

With knowledge of the number of spatial dimensions ($d$), the linear PDE ($\mu$ and $\nu$), and the power law part of the initial conditions ($s$ and $A_\mathrm{fid}/{k_\mathrm{fid}}^{s-2(\mu+\nu)}$), one can determine the power law cutoff scale via
\begin{equation}
k_\mathrm{c}\approx
\left(
\frac{d+2(s-(\mu+\nu))}
{2^{2\nu+1}\,\Gamma[1+\nu]^2\,V_{S^{d-1}}\,{\mathfrak B}^2}
\frac{{k_\mathrm{fid}}^{s-2(\mu+\nu)}}{A_\mathrm{fid}}\,
{k_\mathrm{LSWN}}^s
\right)^{\frac{1}{d+2(s-(\mu+\nu))}} \,.
\label{eq:kcofkLSWN}
\end{equation}
%apart from an  order unity uncertainty coming from the nature of the non-linearities encoded in $\mathfrak{B}$.

%\label{sec:SoundSummary}
\noindent \textbf{Summary}
\vskip2pt
%\noindent \textbf{Summary}
%\vskip2pt

\textit{In systems with sound-like propagation (wave equations with pressure support), non-linearities inevitably transfer power from short to long wavelengths, generating LSWN. The resulting spectrum is insensitive to the detailed shape of the initial conditions, provided they are sub-Poissonian, and is determined instead by the interplay of the sound speed, the leading non-linearities, and the statistical distribution of initial fluctuations. At late times, the large-scale signal therefore acquires a stochastic white-noise floor even if linear theory would predict vanishing power there. This demonstrates that ``cosmology-like'' systems with acoustic propagation cannot remain free of contamination from small-scale physics on the largest observable scales.}

%%%%%%%%%%%%%%%%%%%%%%%%%%%%%%%%%%%%%
\subsection{Cosmology-Like without Sound}
\label{sec:MatterLike}

PDEs which contain no spatial gradients are referred to as {\it hyper-local} because $q$ at each $\vec{x}$ evolves independently of $q$ at any other point ($d_{(m,j)}=1$ in the expansion of Eq.~(\ref{eq:NonlinearityExpansion})).  The most general $p=2$ spatially hyper-local homogeneous isotropic linear operator is
\begin{equation}
{}_{(1)}\hat{\cal D}q
=\frac{\partial^2q}{\partial t^2}
+\alpha[t]\,\frac{\partial q}{\partial t}+\beta[t]\,q \,,
\label{eq:HyperLocal1}
\end{equation}
while the most general spatially quadratic, $m=2$, hyper-local homogeneous isotropic operator is 
\begin{equation}
{}_{(2)}\hat{\mathcal D}\,q
=\xi_1[t]\,q[t,\vec{x}]^2
+\xi_2[t]\,q[t,\vec{x}]\,\dot{q}[t,\vec{x}]
+\xi_3[t]\,\dot{q}[t,\vec{x}]^2
\,.
\label{eq:HyperLocal2}
\end{equation}
This is a general 2nd order quadratic ODE at each point. Since there are no gradients, there are no further restraints imposed by isotropy. Here we only consider systems with leading order non-linearity $n=2$.\\

%%%%%%%%%%%%%%%%%%%
%\label{sec:HyperLocal2equivalent}
\noindent \textbf{Canonical Equivalent Systems}
\vskip2pt

Solutions to Eq.~(\ref{eq:HyperLocal1}) can be oscillatory, but here we restrict to cases where there exist two linearly independent solutions $Q_\pm[t]$ to the local linear equation 
\begin{equation}
    \ddot{Q}_\pm[t]+\alpha[t]\,\dot{Q}_\pm[t]+\beta[t]\,Q_\pm[t]=0 \,,
\end{equation}
where $Q_+[t]\,\dot{Q}_-[t]\ne\,\dot{Q}_+[t]\,Q_-[t]$, $Q_\pm[t]>0$, $\frac{Q_+[t]}{Q_-[t]}>0$, and
$\frac{d}{dt}\frac{Q_+[t]}{Q_-[t]}>0$ for $t>t_\mathrm{i}$.  This may seem rather restrictive, but it is satisfied by most cosmological systems interest.

Using the transformations T3 and then T1 of \S\ref{app:EquivalentSystems}, one obtains the {\it canonical equivalent system} (henceforth referred to as the ``canonical system'')
\begin{equation}
q[t]\rightarrow\mathfrak{q}[\mathfrak{t}[t],\vec{x}]
=\frac{q[t,\vec{x}]}{Q_-[t]} \,,
\qquad
t\rightarrow\mathfrak{t}[t]=\frac{Q_+[t]}{Q_-[t]} \,,
\qquad
{}_{(1)}\hat{\mathfrak D}\,\mathfrak{q}
=\frac{\partial^2\mathfrak{q}[\mathfrak{t},\vec{x}]}{\partial \mathfrak{t}^2} \,,
\label{eq:HyperLocal2Transformation}
\end{equation}
with general linear solution and Green function
\begin{equation}
{}_{(1)}\mathfrak{q}[t,\vec{x}]=
\mathfrak{q}_\mathrm{i}[\vec{x}]
+\dot{\mathfrak{q}}_\mathrm{i}[\vec{x}]\,\mathfrak{t} \,,
\qquad
\tilde{\mathfrak G}[\mathfrak{t},\bar{\mathfrak{t}}]
=\mathfrak{t}-\bar{\mathfrak{t}} \,.
\end{equation}
The Green function does not depend on $\vec{k}$ so we have dropped this argument. The two linear solutions $\propto1$ and $\propto\mathfrak{t}$ we refer to as the {\it decaying} mode and {\it growing} mode, respectively. These correspond to $Q_-$ and $Q_+$, in the original system, respectively.

To leading order $n=2$ non-linearity, the equivalent $2^{\rm nd}$ order PDE is
\begin{equation}
({}_{(1)}\hat{\mathfrak D}+{}_{(2)}\hat{\mathfrak D})\,\mathfrak{q}
=\frac{\partial^2\mathfrak{q}[\mathfrak{t},\vec{x}]}{\partial \mathfrak{t}^2}
+\Xi_1[\mathfrak{t}]\,\mathfrak{q}[\mathfrak{t},\vec{x}]^2
+\Xi_2[\mathfrak{t}]\,\mathfrak{q}[\mathfrak{t},\vec{x}]\,
\frac{\partial\mathfrak{q}[\mathfrak{t},\vec{x}]}{\partial\mathfrak{t}}
+\Xi_3[\mathfrak{t}]\,
\left(\frac{\partial\mathfrak{q}[\mathfrak{t},\vec{x}]}{\partial\mathfrak{t}}
\right)^2 \,,
\label{eq:HyperLocal2CanonicalPDE}
\end{equation}
where
\begin{eqnarray}
\Xi_1[\mathfrak{t}[t]]
&=&Q_-[t]^2\,
\frac{Q_-[t]^2\,\xi_1[t]+Q_-[t]\,\dot{Q}_-[t]\,\xi_2[t]
      +\dot{Q}_-[t]^2\,\xi_3[t]}
     {(Q_-[t]\,\dot{Q}_+[t]-Q_+[t]\,\dot{Q}_-[t])^2} \,,
\nonumber \\
\Xi_2[\mathfrak{t}[t]]
&=&Q_-[t]\,
\frac{Q_-[t]\,\xi_2[t]+2\,\dot{Q}_-[t]\,\xi_3[t]}
   {Q_-[t]\,\dot{Q}_+[t]-Q_+[t]\,\dot{Q}_-[t]} \,,
\nonumber \\
\Xi_3[\mathfrak{t}[t]]
&=&\xi_3[t] \,,
\label{eq:Xis}
\end{eqnarray}
and so the LOBA is
\begin{align}
\mathfrak{q}[\mathfrak{t},\vec{x}]\approx
\mathfrak{q}_\mathrm{i}[\vec{x}]+
\dot{\mathfrak{q}}_\mathrm{i}[\vec{x}]\,
(\mathfrak{t}-\mathfrak{t}_\mathrm{i})
-\int_{\mathfrak{t}_\mathrm{i}}^\mathfrak{t}d\bar{\mathfrak{t}}
\,(\mathfrak{t}-\bar{\mathfrak{t}})
\,\bigl(
&\Xi_1[\bar{\mathfrak{t}}]\,
(\mathfrak{q}_\mathrm{i}[\vec{x}]+
\dot{\mathfrak{q}}_\mathrm{i}[\vec{x}]\,
(\bar{\mathfrak{t}}-\mathfrak{t}_\mathrm{i}))^2
\nonumber \\
+&\Xi_2[\bar{\mathfrak{t}}]\,
(\mathfrak{q}_\mathrm{i}[\vec{x}]+
\dot{\mathfrak{q}}_\mathrm{i}[\vec{x}]\,
(\bar{\mathfrak{t}}-\mathfrak{t}_\mathrm{i}))\,
\dot{\mathfrak{q}}_\mathrm{i}[\vec{x}])
\nonumber \\
+&\Xi_3[\bar{\mathfrak{t}}]\,\dot{\mathfrak{q}}_\mathrm{i}[\vec{x}]^2
\bigr) \,,
\label{eq:HyperlocalCanonicalLOBA}
\end{align}
where $\mathfrak{q}_\mathrm{i}[\vec{x}]$ and 
$\dot{\mathfrak{q}}_\mathrm{i}[\vec{x}]$ give the amplitudes of the decaying and growing modes at $\mathfrak{t}=\mathfrak{t}_\mathrm{i}$.\\

%%%%%%%%%%%%%%%%%%%
\label{sec:HyperLocal2equivalentSelfSimilar}
\noindent \textbf{Hyperlocal Self-Similarity}
\vskip2pt

Hyper-local PDEs, being local in $\vec{x}$, have no characteristic length scales whatsoever.  These PDEs are spatially self similar in the sense that if $q[t,\vec{x}]$ is a solution then so is $q[t,a \vec{x}]$.  Full self-similarity requires that $q[a t,\vec{x}]$ also be a solution. Just as with Eq.~(\ref{eq:DSelfSimilar}) this is obtained by
\begin{align}
\alpha[t]&=\frac{\zeta_1}{t} \,, &
\beta[t]&=\frac{\zeta_2}{t^2} \,, &
\xi_1[t]&=\frac{\zeta_3}{t^2} \,, &
\xi_2[t]&=\frac{\zeta_4}{t} \,, &
\xi_3[t]&=\zeta_5 \,, &
\label{eq:HyperlocalSelfSimilar}
\end{align}
except here there is no $\zeta_6$ or $\zeta_7$.

The self-similar hyper-local $p=n=2$ PDE without sound is the super-horizon limit of the self-similar $p=n=2$ PDE with sound in \S\ref{sec:SelfSimilar}.  Therefore, the solutions of this no sound PDE are the same as the super-horizon limit of the solutions of the corresponding PDE with sound.  Growing and decaying mode solutions to the linear equation are
\begin{equation}
Q_\pm[t]=\left(\frac{t}{t_\mathrm{fid}}\right)^{\mu\pm\nu} \,,
\label{eq:GrowDecayNoSound}
\end{equation}
where $t_\mathrm{fid}$ is some fiducial time, $\mu\equiv\frac{1}{2}(1-\zeta_1)$, and 
$\nu\equiv\frac{1}{2}\sqrt{(1-\zeta_1)^2-4\,\zeta_2}$, exactly as defined in 
Eq.~(\ref{eq:SelfSimilarModes}). For the solution to not be oscillatory, one requires that  $(1-\zeta_1)^2>4\,\zeta_2$ in order that $\nu$ be real, just as in \S\ref{sec:SelfSimilar}. 

The canonical PDE for this system is Eq.~(\ref{eq:HyperLocal2CanonicalPDE}) with
\begin{equation}
q[t,\vec{x}]
=\left(\frac{t}{t_\mathrm{fid}}\right)^{\mu-\nu}\,
\mathfrak{q}[\mathfrak{t}[t],\vec{x}] \,,
\qquad
\mathfrak{t}[t]=\left(\frac{t}{t_\mathrm{fid}}\right)^{2\nu} \,,
\qquad
\Xi_i[\mathfrak{t}]=\varkappa_i\,
\mathfrak{t}^{i+\frac{1}{2}(\frac{\mu}{\nu}-7)}
\,,
\label{eq:XiSelfSimilar}
\end{equation}
where
\begin{equation}
\varkappa_1=\frac{\zeta_3+(\zeta_4+\zeta_5\,(\mu-\nu))(\mu-\nu)}{4\,\nu^2} \,,
\quad
\varkappa_2=\frac{\zeta_4+2\,\zeta_5\,(\mu-\nu)}{2\,\nu} \,,
\quad
\varkappa_3=\zeta_5 \ .
\label{eq:XisAlt}
\end{equation}
Note that the equivalent canonical hyper-local system to a self-similar system is not itself self-similar unless $\mu+6\,\nu=7$. In this case, Eq.~(\ref{eq:HyperlocalCanonicalLOBA}) becomes
\begin{align}
\mathfrak{q}[\mathfrak{t},\vec{x}]\approx
\mathfrak{q}_\mathrm{i}[\vec{x}]+
\dot{\mathfrak{q}}_\mathrm{i}[\vec{x}]\,
(\mathfrak{t}-\mathfrak{t}_\mathrm{i})
-\int_{\mathfrak{t}_\mathrm{i}}^\mathfrak{t}d\bar{\mathfrak{t}}
\,(\mathfrak{t}-\bar{\mathfrak{t}})
\,\bigl(
&\varkappa_1\,
\bar{\mathfrak{t}}^{\frac{\mu-5\,\nu}{2\nu}}\,
(\mathfrak{q}_\mathrm{i}[\vec{x}]+
\dot{\mathfrak{q}}_\mathrm{i}[\vec{x}]\,
(\bar{\mathfrak{t}}-\mathfrak{t}_\mathrm{i}))^2
\nonumber \\
+&\varkappa_2\,
\bar{\mathfrak{t}}^{\frac{\mu-3\,\nu}{2\nu}}\,
(\mathfrak{q}_\mathrm{i}[\vec{x}]+
\dot{\mathfrak{q}}_\mathrm{i}[\vec{x}]\,
(\bar{\mathfrak{t}}-\mathfrak{t}_\mathrm{i}))\,
\dot{\mathfrak{q}}_\mathrm{i}[\vec{x}])
\nonumber \\
+&\varkappa_3\,
\bar{\mathfrak{t}}^{\frac{\mu-\nu}{2\nu}}\,\dot{\mathfrak{q}}_\mathrm{i}[\vec{x}]^2
\bigr)  .
\label{eq:HyperlocalSelfSimilarLOBA}
\end{align}
One can perform this integral, obtaining a finite results provided finite $\varkappa_i$ and $\mathfrak{t}_\mathrm{i}\le\mathfrak{t}<\infty$.  However, the exact solution to 
Eq.~(\ref{eq:HyperLocal2CanonicalPDE}) with $\Xi_i$ given by Eq.~(\ref{eq:XiSelfSimilar}) can diverge for finite $\mathfrak{t}$, as illustrated below. The LOBA is only accurate for small enough $\mathfrak{t}$, where the smallness criteria depends on $\mu$, $\nu$, and the $\varkappa_i$.\\

%%%%%%%%%%%%%%%%%%%
%\label{sec:HyperLocal2Singularity}
\noindent \textbf{Hyperlocal Initial Singularity}
\vskip2pt

The self-similar PDE is singular at $\mathfrak{t}=0$, and if $\mathfrak{t}_\mathrm{i}\rightarrow0$, both the exact solution and LOBA will diverge for certain values of $\mu$, $\nu$, $\varkappa_i$, $\mathfrak{q}_\mathrm{i}$, and $\dot{\mathfrak{q}}_\mathrm{i}$.  The initial values problem at the initial singularity is not well posed for parameters giving divergences.  We henceforth consider only pure growing mode initial conditions, which are least susceptible to divergences, i.e.~$\mathfrak{q}_\mathrm{i}[\vec{x}]=0$.  For pure growing modes Eq.~(\ref{eq:HyperlocalSelfSimilarLOBA}) converges for $\mathfrak{t}_\mathrm{i}=0$ only if $\mu+\nu>0$ (having already required $\nu>0$).  The LOBA for the canonical system is 
\begin{equation}
\mathfrak{q}[\mathfrak{t},\vec{x}]\approx
\dot{\mathfrak{q}}_\mathrm{i}[\vec{x}]\,\mathfrak{t}
-\frac{4\,\nu^2\,(\varkappa_1+\varkappa_2+\varkappa_3)}{(\mu+\nu)\,(\mu+3\,\nu)}\,
\dot{\mathfrak{q}}_\mathrm{i}[\vec{x}]^2\,
\mathfrak{t}^{\frac{\mu+3\,\nu}{2\,\nu}} \,,
\label{eq:HyperlocalSelfSimilarSingularCanonicalLOBA}
\end{equation}
which for the original system may be written
\begin{equation}
q[t,\vec{x}]\approx
g[\vec{x}]\,Q_+[t]\,
-C_\mathrm{LOBA}^\mathrm{hl}\,
g[\vec{x}]^2\,Q_+[t]^2 \,,
\label{eq:HyperlocalSelfSimilarSingularLOBA}
\end{equation}
where $g[\vec{x}]\equiv\dot{\mathfrak{q}}_\mathrm{i}[\vec{x}]$ is the amplitude for the growing mode $Q_+[t]\equiv(t/t_\mathrm{fid})^{\mu+\nu}$ and 
\begin{equation}
C_\mathrm{LOBA}^\mathrm{hl}\equiv
\frac{4\,\nu^2\,(\varkappa_1+\varkappa_2+\varkappa_3)}
{(\mu+\nu)\,(\mu+3\,\nu)}
=\frac{\zeta_3+(\mu+\nu)\,(\zeta_4+(\mu+\nu)\,\zeta_5)}
{(\mu+\nu)\,(\mu+3\,\nu)} \,.
\label{eq:ChlLOBA}
\end{equation}
There are parameter combinations which give $C_\mathrm{LOBA}^\mathrm{hl}=0$, even though only for $\zeta_3=\zeta_4=\zeta_5=0$ will the LSWN go to zero.  As $C_\mathrm{LOBA}^\mathrm{hl}\rightarrow0$, one will have to go to smaller wavenumbers for LOBA to be accurate.\\

%%%%%%%%%%%%%%%%%%%
%\label{sec:HyperLocal2PowerSpectra}
\noindent \textbf{Hyperlocal Power Spectra and LSWN}
\vskip2pt

Assuming the growing mode Gaussian initial conditions introduced in \S\ref{sec:LeadingOrderPerturbationTheory},
\begin{equation}
\tilde{g}[\vec{k}]\equiv
\int\frac{d^d\vec{x}}{(2\pi)^{d/2}}\,
e^{-i\,\vec{k}\cdot\vec{x}}g[\vec{x}] \,,
\qquad
\langle\tilde{g}[\vec{k}]\rangle=0 \,,
\qquad 
\langle\tilde{g}[\vec{k}]\,\tilde{g}[\vec{k}']\rangle
=(2\pi)^d\,\delta^{(d)}[\vec{k}+\vec{k}']\,R_{(2)}[\vec{k}] \,,
\end{equation}
the 1- and 2-point functions of $\tilde{q}[t,\vec{k}]$ are
\begin{align}
%\tilde{q}[t,\vec{k}]&\equiv \int\frac{d^d\vec{x}}{(2\pi)^{d/2}}\, e^{-i\,\vec{k}\cdot\vec{x}}q[t,\vec{x}] \nonumber \\
%\langle\tilde{q}[t,\vec{k}]\rangle&=(2\pi)^d\,\delta^{(d)}[\vec{k}]\,P_{(1)}[t] \qquad\qquad \langle\tilde{q}[t,\vec{k}]\,\tilde{q}[t,\vec{k}']\rangle =(2\pi)^d\,\delta^{(d)}[\vec{k}+\vec{k}']P_{(2)}[t,|\vec{k}|] \nonumber \\
P_{(1)}[t]&\approx
-C_\mathrm{LOBA}^\mathrm{hl}\,Q_+[t]^2\,
\int d^d\vec{k}_1\,R_{(2)}[|\vec{k}_1|] \,,
\nonumber \\
P_{(2)}[t,|\vec{k}|]&\approx R_{(2)}[|\vec{k}|]\,Q_+[t]^2
+2\,{C_\mathrm{LOBA}^\mathrm{hl}}^2\,Q_+[t]^4
\int d^d\vec{k}_1\,R_{(2)}[|\vec{k}_1|]\,
                   R_{(2)}[|\vec{k}-\vec{k}_1|]
\,.
\label{eq:HyperLocal2PowerSpectra}
\end{align}
For sub-Poissonian initial conditions ($R_{(2)}[0^+]=0$), the LSWN is then
\begin{equation}
P_\mathrm{LSWN}[t]
\approx2\,{C_\mathrm{LOBA}^\mathrm{hl}}^2\,Q_+[t]^4\,
\int d^d\vec{k}_1\,R_{(2)}[\vec{k}_1]^2 \,.
\label{eq:HyperLocal2LSWN}
\end{equation}
The LSWN power remains active, growing faster than the power in the fastest growing mode: $Q_+[t]^4$ vs $Q_+[t]^2$. Thus, there is never relic LSWN with these no-sound self-similar systems, just as there is not for self-similar systems with sound. Self-similar pure power law initial conditions are not viable, with or without sound, since they give divergent LSWN.\\

%%%%%%%%%%%%%%%%%%%
%\label{sec:HyperLocalSummary}
\noindent \textbf{Summary}
\vskip2pt

\textit{In the absence of sound-like restoring forces, the dynamics are qualitatively different. 
Growth is not wave-like, but instead dominated by gravitational-type or diffusive interactions, yet the same mode-mixing mechanism persists: non-linear couplings populate long wavelengths and produce LSWN. Compared to the sound case, the amplitude and scaling of the noise depend more directly on the growth rates of the underlying modes, but the key feature remains --- a universal large-scale white-noise spectrum that overrides linear predictions on the largest scales. Thus, even in non-acoustic cosmology-like systems, large-scale observables inevitably reflect the imprint of small-scale non-linear dynamics.}

\begin{comment}
%%%%%%%%%%%%%%%%%%%%%%%%%%%%%%%%%%%%%
\subsection{Linear Order with Quadratic Non-Linearities}
\label{sec:Order1PDEs}

A general homogeneous isotropic $p=1$, $n=2$ PDE (with spatial gradients no higher $1^{\rm st}$ order takes the form
\begin{equation}
{}_{(1)}\hat{\cal D}q=\frac{dq}{dt}+\beta[t]\,q \qquad
{}_{(2)}\hat{\cal D}q=\xi_1[t]\,q^2 \,,
\label{eq:Hybrid}
\end{equation}
which is hyper-local and has a single linear solution
\begin{equation}
Q[t]=e^{-\int_{t_\mathrm{i}}^t d\bar{t}\,\beta[\bar{t}]} \,.
\end{equation}
The equivalent canonical PDE is
\begin{equation}
q[t]\rightarrow\mathfrak{q}[\mathfrak{t}[t],\vec{x}]
=\frac{q[t,\vec{x}]}{Q[t]} \,,
\qquad
t\rightarrow\mathfrak{t}[t]=e^{\int_{t_\mathrm{i}}^t d\bar{t}\,\beta[\bar{t}]} \,,
\qquad
{}_{(1)}\hat{\mathfrak D}\,\mathfrak{q}
=\frac{\partial\mathfrak{q}[\mathfrak{t},\vec{x}]}{\partial \mathfrak{t}} \,,
\label{eq:HyperLocal2Transformation}
\end{equation}

\begin{equation}
    q[t,\vec{x}]
    =q[t_\mathrm{i} ,\vec{x}]\,e^{-\int_{t_\mathrm{i}}^t d\bar{t}\,\alpha[\bar{t}]}
    +\int_{t_\mathrm{i}}^t d\bar{t}\,
    e^{-\int_{\bar{t}}^t d\bar{\bar{t}}\,\alpha[\bar{\bar{t}}]}\,S[\bar{t},\vec{x}]
\end{equation}
\end{comment}

%%%%%%%%%%%%%%%%%%%%%%%%%%%%%%%%%%%%%
\subsection{Review}
\label{sec:LSWNreview}

This section has given a rather general exposition of the production of large scale white noise (LSWN).  This exposition makes use of a number of concepts and assumptions introduced in this section which are useful in understanding LSWN in cosmology: 
\begin{itemize}
\item locality (partial differential equations)
\item sub-Poissonian initial conditions
\item non-linearity
\item linear and non-linear regimes
\item conservative versus non-conservative systems
\item homogeneity and isotropy
\item growing and decaying modes
\item leading order non-linearities
\item leading order Born approximation (LOBA)
\item growing mode initial conditions and guiding variables
\item Gaussian initial conditions
\item self-similarity with an initial singularity
\item a No-No-Scale ``Theorem''
\item hyper-local systems and systems with sound
\item equivalent systems versus deformations
\item active versus relic non-linearities
\end{itemize}
One can refer back to this section for definitions of these concepts.

LSWN is a very general phenomena which typically only requires the first four items on this list.  The others are specializations related to cosmological applications. The LOBA, which is often analytically tractable, is quantitatively accurate in modeling LSWN in the linear regime. This is demonstrated in the coming sections for some simple cases.

%%%%%%%%%%%%%%%%%%%%%%%%%%%%%%%%%%%%%%%%%%%%%%%%%%%%%%%%%%%%
%%%%%%%%%%%%%%%%%%%%%%%%%%%%%%%%%%%%%%%%%%%%%%%%%%%%%%%%%%%%
\section{Large Scale Noise in Self-Similar Systems}
\label{sec:LargeScaleNoise}

Here we consider the power spectrum of late-time large scale noise in self-similar systems in the linear regime to obtain simple analytic approximations for the power spectrum in a broad region of $k$-space near horizon crossing. This differs from the results of the ``Asymptotes, Amplitudes, and Phases'' subsection of \S\ref{sec:RadiationLike}, which give the spectrum in the limit of infinite wavelength (i.e. on scales much larger than the sound horizon).  The long wavelength limit of the results derived here is usually, but not always, in agreement with the infinite wavelength limit. In systems of interest, LOBA predicts that while in the linear regime, the late-time large-scale dynamics is accurately described by linear theory, but linear theory with very different initial condition than actually existed.  This modification of the spectrum often causes the system to go non-linear in cases where one would only expect the inhomogeneities to decay.

Taylor expanding a general homogeneous $2^{\rm nd}$ order self-similar spatially homogeneous and isotropic PDE for $q[h,\vec{x}]$ in $d$ dimensions yields (see \S\ref{sec:SelfSimilar}) the PDE
\begin{eqnarray}
0&=&
\frac{\partial^2q}{\partial h^2}
+\frac{\zeta_1}{h}\,\frac{\partial q}{\partial h}
+\frac{\zeta_2}{h^2}\,q
-\nabla^2q
\nonumber\\
&&+
 \frac{\zeta_3}{h^2}\,q^2
+\frac{\zeta_4}{h}\,q\,\frac{\partial q}{\partial h}
+\zeta_5\,\left(\frac{\partial q}{\partial h}\right)^2
+\zeta_6\,q\,\nabla^2q
+\zeta_7\,\vec{\nabla}q\cdot\vec{\nabla}q
\nonumber\\
&&+\ldots
\label{eq:SelfSimilarp2n2}
\end{eqnarray}
where we are again using the sound horizon $h$ as the temporal variable. The first line gives the $m=1$ (linear) order differential operator, and the second line gives the $m=2$ order operator, etc. Assume that $|\zeta_3|,|\zeta_4|,|\zeta_5|,|\zeta_6|\,|\zeta_7|\lesssim1$ and at least one of these is non-zero.  In this case, the leading order non-linearity is quadratic, i.e.~$n=2$.

As before, we take the initial time to be at the singularity, $h=0$. We assume that initially $q$ is in the {\it linear regime}, i.e.~that $q$ and its derivatives are small enough that the $m=1$ terms dominate the differential equation. Finally, we assume that $q$ is initially described by spatially homogeneous and isotropic Gaussian random noise. Under these assumptions, we expect that the leading order Born approximation (LOBA) is an accurate representation of the system while the solution remains in the linear regimes.

Under these assumptions and using Eqs.~(\ref{eq:R22Gaussian}),
(\ref{eq:Mdefinition}),
(\ref{eq:Ndefinition}),
(\ref{eq:GreenFunction}),
(\ref{eq:qPointFunctionsLOBA}), and
(\ref{eq:SelfSimilarModes}), LOBA yields the following expression for the power spectrum
\begin{align}
P_{(2)}^\mathrm{LOBA}[h,\vec{k}]
&=
R_{(2)}[|\vec{k}|]\,{\cal Q}_+[|\vec{k}|\,h]^2
\nonumber \\
&+(2\pi)^{2\,d}\,
       \int\frac{d^d\vec{k}_1 }{(2\pi)^{d/2}}\int\frac{d^d\vec{k}_2 }{(2\pi)^{d/2}}\,
       \delta^{(d)}[\vec{k}-\vec{k}_1 -\vec{k}_2 ]\,
       R_{(2)}[|\vec{k}_1|]\,R_{(2)}[|\vec{k}_2|]
\nonumber \\
&\hskip35pt
\times\int\frac{d^d\vec{k}_1'}{(2\pi)^{d/2}}\int\frac{d^d\vec{k}_2'}{(2\pi)^{d/2}}
\,\delta^{(d)}[\vec{k}-\vec{k}_1'-\vec{k}_2']\,
(\delta^{(d)}[\vec{k}_1+\vec{k}_1']
+\delta^{(d)}[\vec{k}_1+\vec{k}_2'])
\nonumber\\
&\hskip35pt\times
{\cal M}[|\vec{k}|\,h,\frac{\vec{k}_1 }{|k|} ,\frac{\vec{k}_2 }{|k|} ]\,
{\cal M}[|\vec{k}|\,h,\frac{\vec{k}_1'}{|k|} ,\frac{\vec{k}_2'}{|k|} ] \,,
\label{eq:2PointFunctionsLOBAselfSimilar}
\end{align}
where
\begin{eqnarray}
{\cal M}[\varphi,\vec{r}_1,\vec{r}_2]&\equiv&
\sum_\pm\,{\cal Q}_\pm[\varphi]
{\cal W}_\pm[\varphi,\vec{r}_1,\vec{r}_2] \,,
\nonumber\\
{\cal W}_\pm[\varphi,\vec{r}_1,\vec{r}_2]&\equiv&
-\int_0^\varphi d\bar{\varphi}\,
\omega_\pm[\bar{\varphi}]\,
{\cal N}[\bar{\varphi},\vec{r}_1,\vec{r}_2] \,,
\nonumber\\
\omega_\pm[\varphi]&\equiv&
\mp\frac{\pi}{2}\,\varphi^{1-2\mu}\,{\cal Q}_\mp[\varphi] \,,
\nonumber\\
{\cal Q}_+[\varphi]&\equiv&\varphi^{\mu}\,J_\nu[\varphi] \,, \qquad
{\cal Q}_-[\varphi] \equiv \varphi^{\mu}\,Y_\nu[\varphi] \,,
\end{eqnarray}
$\mu=\frac{1}{2}(1-\zeta_1)$, $\nu=\frac{1}{2}\sqrt{(1-\zeta_1)^2-4\,\zeta_2}$,
and ${\cal N}[\bar{\varphi},\vec{r}_1,\vec{r}_2]$ gives the non-linearities, which are quadratic in ${\cal Q}_+[\bar{\varphi}]$ and ${\cal Q}_+'[\bar{\varphi}]$. For the canonical non-linearities of Eq.~(\ref{eq:DSelfSimilar}), 
\begin{align}
{\cal N}[\varphi,\vec{r}_1,\vec{r}_2]=&
\left(\frac{\zeta_3}{\varphi^2}-\zeta_6\,|\vec{r}_2|^2-\zeta_7\,\vec{r}_1\cdot\vec{r}_2
\right)\,
{\cal Q}_+[|\vec{r}_1|\,\varphi]\,
{\cal Q}_+[|\vec{r}_2|\,\varphi]
\nonumber\\
&+\frac{\zeta_4}{\varphi}\,|\vec{r}_2|\,{\cal Q}_+ [|\vec{r}_1|\,\varphi]\,{\cal Q}_+'[|\vec{r}_2|\,\varphi]
+\zeta_5\,|\vec{r}_1|\,|\vec{r}_2|\,{\cal Q}_+'[|\vec{r}_1|\,\varphi]\,{\cal Q}_+'[|\vec{r}_2|\,\varphi] \,.
\end{align}
Integrating over the $\delta$-functions, Eq.~(\ref{eq:2PointFunctionsLOBAselfSimilar}) reduces to
\begin{equation}
P_{(2)}[h,\vec{k}]\approx
R_{(2)}[|\vec{k}|]\,{\cal Q}_+[|\vec{k}|\,h]^2
\nonumber \\
+\int d^d\vec{k}_1\,R_{(2)}[|\vec{k}_1|]\,R_{(2)}[|\vec{k}-\vec{k}_1|]\,
{\cal U}[|\vec{k}|\,h,\frac{\vec{k}}{|k|},\frac{\vec{k}_1}{|k|}] \,,
\label{eq:2PointFunctionsLOBAselfSimilarAlt}
\end{equation}
where
\begin{equation}
{\cal U}[\varphi,\hat{\vec{n}},\vec{r}_1]\equiv
{\cal M}[\varphi,\vec{r}_1,\hat{\vec{n}}-\vec{r}_1]\,
\left(
{\cal M}[\varphi,-\vec{r}_1,\hat{\vec{n}}+\vec{r}_1]
+
{\cal M}[\varphi,-\hat{\vec{n}}-\vec{r}_1,2\,\hat{\vec{n}}+\vec{r}_1]
\right) \,.
\label{eq:Udef}
\end{equation}
In most cases, the two terms in parentheses will integrate to the same number, simply doubling the contribution of the first term.\\

%%%%%%%%%%%%%%%%%%%%%%%%%%%
%\vskip-25pt
%\noindent \textbf{Late-Time Relic Systems}
%\vskip2pt
\subsection{Late-Time Relic Systems}
\label{sec:LateTimeRelics}

Consider now relic self-similar systems in the linear regime, where non-linearities at early times dominate over those at later times --- specifically where the ${\cal W}_\pm$ integrals (or just the dominant of the two) converge rapidly enough so that at late times one can accurately extend the integral to infinity
\begin{equation}
{\cal M}[\varphi,\vec{r}_1,\hat{\vec{n}}-\vec{r}_1]\approx
\sum_\pm\,{\cal Q}_\pm[\varphi]\,
{\cal W}_\pm[\infty,\vec{r}_1,\hat{\vec{n}}-\vec{r}_1] \,.
\label{eq:Mrelic}
\end{equation}
Since the time dependence for a given $|\vec{k}|$ tracks the growing and decaying modes, this means that at late times, relic non-linearities result in inhomogeneities which evolve nearly as in linear theory but with a very different spectrum of initial inhomogeneities.

We are interested in the late-time power spectrum on scales much larger than the scale of the non-linearities, i.e. $|\vec{r}_1|\gg1$.  Then the  integrand ${\cal W}_\pm[\varphi,\vec{r}_1,\hat{\vec{n}}-\vec{r}_1]$ has two characteristic scales for $\bar{\varphi}$: $\bar{\varphi}\sim1$ and $\bar{\varphi}\sim1/|\vec{r}_1|$. If 
${\cal W}_\pm[\infty,\vec{r}_1,\hat{\vec{n}}-\vec{r}_1]<\infty$, the integral will converge above one of these two $\bar{\varphi}$ scales.  If it is the former, we call this {\it slow convergence}, and if it is the latter, {\it rapid convergence}.  For slow convergence, Eq.~(\ref{eq:Mrelic}) is only valid on scales much larger than the horizon, which is not the case we are interested in here. For relic large scale noise to occur on scales comparable to the horizon, we require rapid convergence. This puts an upper limit on $\mu$.

Since for rapid convergence $\bar{\varphi}\ll1$ on the convergence scale $\bar{\varphi}\sim1/|\vec{r}_1|$, we may use the approximation
\begin{equation}
\lim_{\varphi\rightarrow0}\omega_\pm[\varphi]
=2^{-1\pm\nu}\,\Gamma[\nu]^{\pm1}\,
\nu^{-\frac{1\mp1}{2}}\,\varphi^{1-\mu\mp\nu} \,,
\end{equation}
and
\begin{align}
{\cal N}[\varphi,\vec{r}_1,\hat{\vec{n}}-\vec{r}_1]
&\approx
{\cal N}[\varphi,\vec{r}_1,-\vec{r}_1]
\approx|\vec{r}_1|^2\,
\mathfrak{n}[|\vec{r}_1|\,\varphi] \,,
\nonumber\\
\mathfrak{n}[\psi]
&\equiv
\left(\frac{\zeta_3}{\psi^2}-\zeta_6+\zeta_7\right)\,
{\cal Q}_+[\psi]^2\,
+\frac{\zeta_4}{\psi}\,{\cal Q}_+ [\psi]\,{\cal Q}_+'[\psi]
+\zeta_5\,{\cal Q}_+'[\psi]^2 \,,
\nonumber\\
{\cal W}_\pm[\infty,\vec{r}_1,\hat{\vec{n}}-\vec{r}_1]
&\approx{\cal W}_\pm[\infty,\vec{r}_1,-\vec{r}_1]
\approx\,|\vec{r}_1|^{\mu\pm\nu}\,\mathfrak{W}_\pm \,,
\nonumber\\
\mathfrak{W}_\pm
&\equiv
-2^{-1\pm\nu}\,\Gamma[\nu]^{\pm1}\,\nu^{-\frac{1\mp1}{2}}
\int_0^\infty d\psi\,\,\psi^{1-\mu\mp\nu}\,\mathfrak{n}[\psi] \,,
\nonumber\\
{\cal M}[\varphi,\vec{r}_1,\hat{\vec{n}}-\vec{r}_1]
&\approx
{\cal M}[\varphi,\vec{r}_1,-\vec{r}_1]
\approx
\sum_\pm\,\mathfrak{W}_\pm\,
|\vec{r}_1|^{\mu\pm\nu}\,{\cal Q}_\pm[\varphi]
\approx
\mathfrak{W}_+\,|\vec{r}_1|^{\mu+\nu}\,{\cal Q}_+[\varphi] \,,
\nonumber\\
{\cal U}[\varphi,\hat{\vec{n}},\vec{r}_1]
&\approx
2\,{\cal M}[\varphi,\vec{r}_1,-\vec{r}_1]^2
\approx2\,
{\mathfrak{W}_+}^2\,|\vec{r}_1|^{2(\mu+\nu)}\,
{\cal Q}_+[\varphi]^2 \,,
\label{eq:aLOBA}
\end{align}
where we have used 
${\cal M}[\varphi,-\vec{r}_1,\vec{r}_1]
={\cal M}[\varphi,\vec{r}_1,-\vec{r}_1]$ by isotropy.  Since $\nu>0$, the growing mode coefficient is parametrically much larger than the decaying mode coefficient. These approximations are not appropriate for conservative non-linearities, which are discussed below.

With rapid convergence, the asymptotic form of the LOBA power spectrum Eq.~(\ref{eq:2PointFunctionsLOBAselfSimilarAlt}) is thus
\begin{eqnarray}
P_{(2)}[h,k]&\approx&
\left(
R_{(2)}[k]+R_{(2)}^\mathrm{nl}[k]
\right)\,{\cal Q}_+[k\,h]^2 \,,
\nonumber\\
R_{(2)}^\mathrm{nl}[k]
&\approx&
\frac{2\,{\mathfrak{W}_+}^2}{k^{2(\mu+\nu)}}
\int d^d\vec{k}_1\,k_1^{2(\mu+\nu)}\,
R_{(2)}[k_1]^2 \,.
\label{eq:2PointFunctionsLOBAselfSimilarAsymptotic}
\end{eqnarray}
In other words, the asymptotic (large scales, late times) LOBA form of the power spectrum is the same as linear theory, but with a modified initial power spectrum: 
$R_{(2)}[k]\rightarrow
R_{(2)}[k]+R_{(2)}^\mathrm{nl}[k]$. Thus, the power spectrum is approximately the sum
\begin{equation}
P_{(2)}[h,k]\approx
P_{(2)}^\mathrm{lin}[h,k]+P_{(2)}^\mathrm{nl}[h,k] \,,
\end{equation}
where
\begin{equation}
P_{(2)}^\mathrm{lin}[h,k]=R_{(2)}[k]\,{\cal Q}_+[k\,h]^2 \,,
\qquad
P_{(2)}^\mathrm{nl}[h,k]=R_{(2)}^\mathrm{nl}[k]\,
{\cal Q}_+[k\,h]^2 \,.
\end{equation}
Here $P_{(2)}^\mathrm{lin}$ is the linear theory prediction given the true initial conditions and $P_{(2)}^\mathrm{nl}$ is the corrections due to non-linearities. At the longest wavelengths, one has
\begin{equation}
P_{(2)}^\mathrm{nl}[0^+]\equiv
\lim_{k\rightarrow0}P_{(2)}^\mathrm{nl}[k]=
2\,
\left(
\frac{\mathfrak{W}_+}{2^\nu\,\nu\,\Gamma[\nu]}
\right)^2\,
V_{S^{d-1}} \int_0^\infty d k_1\,{k_1}^{d-1}\,
(k_1\,h)^{2\,(\mu+\nu)}\,R_{(2)}[k_1]^2 \,,
\end{equation}
where $V_{S^d}$ is the volume of the unit $d$-sphere.  If $\mathfrak{W}_+\ne0$, this is positive and finite so long as the $k_1$ integral is finite.  This requires $R_{(2)}[k]$ be cutoff at large $k$.  If all these conditions are satisfied, then non-linearities add white noise on scales larger than the sound horizon.  If the linear theory power spectrum is sub-Poissonian, i.e. if $P_{(2)}^\mathrm{lin}[0^+]=0$, then this white noise from non-linear corrections will dominate on the largest scales.\\

%%%%%%%%%%%%%%%%%%%%%%%%
\vskip-25pt
\label{sec:ConservativeRelics}
%\noindent \textbf{Conservative Relics}
%\vskip2pt
\subsubsection{Conservative Relics}

Linear and conservative systems do not produce LSWN (see \S\ref{sec:InstantaneousGeneration}), so if the leading-order non-linearity is conservative, then LOBA will yield zero LSWN. It could be that higher-order non-linearities are non-conservative and do produce LSWN, which would dominate at the largest scales.  Eq.~(\ref{eq:SelfSimilarp2n2}) is conservative only when $\zeta_3=\zeta_4=\zeta_5=0$ and $\zeta_6=\zeta_7$, so
\begin{equation}
    0=
\frac{\partial^2q}{\partial h^2}
+\frac{\zeta_1}{h}\,\frac{\partial q}{\partial h}
+\frac{\zeta_2}{h^2}\,q
-\nabla^2q
+\frac{1}{2}\,\zeta_6\,\nabla^2q^2 \,.
\end{equation}
In this case, $\mathfrak{W}_\pm=0$ and thus the LSWN estimates of the previous section are correctly zero.  For rapid convergence, the large $|\vec{r}_1|$ limit is
\begin{align}
{\cal N}[\varphi,\vec{r}_1,\hat{\vec{n}}-\vec{r}_1]
&\approx-\zeta_6\,(\hat{\vec{n}}\cdot\vec{r}_1)\,
{\cal Q}_+[|\vec{r}_1|\,\varphi]^2 \,,
\nonumber\\
\mathfrak{W}_\pm
&\equiv
2^{-1\pm\nu}\,\Gamma[\nu]^{\pm1}\,\nu^{-\frac{1\mp1}{2}}\,
\zeta_6\,\int_0^\infty d\psi\,\,\psi^{1-\mu\mp\nu}\,
{\mathcal Q}_+[\psi]^2 \,,
\nonumber\\
{\cal M}[\varphi,\vec{r}_1,\hat{\vec{n}}-\vec{r}_1]
&\approx
\sum_\pm\,\mathfrak{W}_\pm\,
(\hat{\vec{n}}\cdot\vec{r}_1)\,|\vec{r}_1|^{\mu\pm\nu-2}\,{\cal Q}_\pm[\varphi]
\approx
\mathfrak{W}_+\,
(\hat{\vec{n}}\cdot\vec{r}_1)\,|\vec{r}_1|^{\mu+\nu-2}\,
{\cal Q}_+[\varphi] \,,
\nonumber\\
{\cal U}[\varphi,\hat{\vec{n}},\vec{r}_1]&\approx
2\,{\mathfrak{W}_+}^2\,
(\hat{\vec{n}}\cdot\vec{r}_1)^2\,
|\vec{r}_1|^{2\,(\mu+\nu-2)}\,{\cal Q}_+[\varphi]^2 \,,
\label{eq:aLOBAconservative}
\end{align}
so that
\begin{eqnarray}
P_{(2)}[h,k]&\approx&
\left(
R_{(2)}[k]+R_{(2)}^\mathrm{nl}[k]
\right)\,{\cal Q}_+[k\,h]^2 \,,
\nonumber\\
R_{(2)}^\mathrm{nl}[k]
&\approx&
\frac{2\,{\mathfrak{W}_+}^2}{k^{2(\mu+\nu-1)}}\,
\frac{V_{S^{d-1}}}{d}\,
\int_0^\infty dk_1,{k_1}^{2(\mu+\nu-1)+d-1}\,
R_{(2)}[k_1]^2 \,,
\label{eq:2PointFunctionsLOBAselfSimilarAsymptotic1}
\end{eqnarray}
where we have used 
$\int d^{d-1}\hat{\vec{n}}\,(\hat{\vec{n}}\cdot\vec{k}_1)^2
=|\vec{k}_1|^2\,V_{S^{d-1}}/d$.
This is nearly identical to the power spectrum modifications for non-conservative systems in 
Eq.~(\ref{eq:2PointFunctionsLOBAselfSimilarAsymptotic})
except that $R_{(2)}^\mathrm{nl}[k]$ has a different power law. At the longest wavelengths,
\begin{equation}
\lim_{k\,h\ll1}
P_{(2)}^\mathrm{nl}[k]\approx
2\,
\left(
\frac{\mathfrak{W}_+}{2^\nu\,\nu\,\Gamma[\nu]}
\right)^2\,k^2\,
V_{S^{d-1}} \int_0^\infty d k_1\,{k_1}^{d-1}\,
(k_1\,h)^{2\,(\mu+\nu-1)}\,R_{(2)}[k_1]^2 \,,
\end{equation}
so the spectral index is 2 rather than 0. Thus, while there is no LSWN, if the spectral index of $R_{(2)}[k]$ is greater than 2, this large scale (blue) noise will still dominate at the largest scales.\\

%%%%%%%%%%%%%%%%%%%%%%%%
\vskip-25pt
\label{sec:CosmicConfusion}
%\noindent \textbf{Cosmic Confusion}
%\vskip2pt
\subsubsection{Cosmic Confusion}

The LOBA results of the above subsections give an additive correction, $R_{(2)}^\mathrm{nl}[k]$, to the initial power, $R_{(2)}[k]$ on larger scales. 
If $R_{(2)}[k]$ falls off faster at small $k$ than $R_{(2)}^\mathrm{nl}[k]$, then it is the non-linear correction which will dominate. If one only measures $P_{(2)}[k]$ on scales where the non-linear corrections dominate, one might mistakenly conclude that the power observed on these scales represents linear growth of initial conditions with a very different spectrum from the actual initial conditions.  Thus one can attribute the power observed on these scales to {\it either} initial conditions or non-linearities on scales much smaller than those observed.  We refer to this ambiguity as {\it cosmic confusion}.

The large scale power spectra we have obtained from LOBA are power laws with spectral index
$d\ln[P_{(2)}^\mathrm{nl}]/d\ln[k]\approx0$ (LSWN) unless the non-linearities are conservative, in which case $d\ln[P_{(2)}^\mathrm{nl}]/d\ln[k]\approx2$. We do not expect deviations from these power laws with non-self-similar non-linearities, but proving this will require further investigation.  Thus, not observing a power law might be a hint that what one is observing is not a product of non-linearities.

The No-No-Scale Theorem requires that for non-linear systems, the initial power spectrum cannot be a power law on all scales.  However, a system can have a power law on large scales so long as there is a cutoff on small scales. The requirement of a cutoff on unobserved small scales does not resolve any cosmic confusion. 

Cosmic confusion does not imply that the initial conditions cannot be inferred from $q[\vec{x},t]$ at any subsequent time. In fact, LOBA with Gaussian initial conditions is an easily invertible, quadratic, mapping between $P_{(2)}[h,k]$ and $R_{(2)}[k]$.  However, the inversion can only be done with large enough $k$-space coverage, covering at least all the scales that had power initially.  If one only resolves scales much larger than this, then cosmic confusion will persist.  One might hope that the confusion could be resolved by higher order correlation functions than the 2-point function encoded in $P_{(2)}$ --- e.g.~by the bi-spectrum $P_{(3)}$. Generally one would also need to resolve the small scales where the non-linearities were generated with $P_{(3)}$, $P_{(4)}$, etc.

%%%%%%%%%%%%%%%%%%%%%%%%
\label{sec:DependentVariableChoice}
\subsection{Choice of Dependent Variable}

Our predictions for LSWN from non-linearities depend on the choice of dependent variable $q$.  A non-linear change of variable is called a deformation because it will change the LSWN properties (see \S\ref{sec:Deformations}).  For example the amplitude of relic power spectrum will be differ for different choices of $q$. Consider a general mapping between one variable and another $q=F[\underline{q}]$ where $F[\underline{q}]$ is invertible $F[0]=0$ and $F'[0]\ne0$.  The EoM will remain self-similar under such a mapping. Define $\alpha\equiv F''[0]/F'[0] $ and since a linear rescaling is trivial henceforth only consider the case $F'[0]=1$ so that $q=\underline{q}+\alpha\,\underline{q}^2+{\cal O}[\underline{q}^3]$
and
$\underline{q}=q-\alpha\,q^2+{\cal O}[q^3]$.  Taylor expanding one finds the 7 parameters characterizing the self-similar EoM become
\begin{equation}
\underline{\zeta}_1=\zeta_1 \quad\ 
\underline{\zeta}_2=\zeta_2 \quad\ 
\underline{\zeta}_3=\zeta_3-\frac{1}{2}\,\alpha \quad\ 
\underline{\zeta}_4=\zeta_4 \quad\ 
\underline{\zeta}_5=\zeta_5+\alpha \quad\ 
\underline{\zeta}_6=\zeta_6 \quad\ 
\underline{\zeta}_7=\zeta_7-\alpha
\label{eq:zetadeformation}
\end{equation}
for $\underline{q}$.  Since the two parameters characterizing the linear EoM, $\zeta_1$ and $\zeta_2$ the growing and decaying modes, ${\cal Q}_\pm$ are unchanged.  Three of the five parameters characterizing the non-linear part of the EoM are changed if $\alpha\ne0$.  The LOBA normalization of the non-conservative late time relics is also changed
\begin{equation}
\underline{\mathfrak{W}}_+=\mathfrak{W}_+
+\alpha\,\left(
2^{-1+\nu}\,\Gamma[\nu]
\int_0^\infty d\psi\,\,\psi^{1-\mu-\nu}
\left(
\left(\frac{1}{2\,\psi^2}+1\right)\,{\cal Q}_+[\psi]^2-{\cal Q}_+'[\psi]^2
\right)
\right) \ .
\end{equation}
From this we see that there is usually an $\alpha$ and therefore choices of $\underline{q}$ where $\underline{\mathfrak{W}}_+=0$.  We call such a {\it suppressed LSWN variable}.  While generically one would expect that $\mathfrak{W}_+\sim{\cal O}[1]$ we now seed that it can be very small or zero even zero.  For a given physical system the measured amplitude of the LSWN, $\propto{\underline{\mathfrak{W}}_+}^2$, may or may not be very small depending on what one is measuring. This is another manifestation of the tyranny of observables (\S\ref{sec:Tyranny}).

%%%%%%%%%%%%%%%%%%%%%%%%
\label{sec:PerturbedInitialConditions}
\subsection{Perturbative Initial Conditions}
%\noindent \textbf{Perturbative Initial Conditions}
%\vskip2pt

The dynamical systems we are modeling are reversible.  Thus any final state can be obtained by appropriate choice of initial conditions.  All claims we have made about the existence or non-existence of LSWN are dependent on our assumptions about these initial conditions.  The assumption we have made in much of this paper is that the initial conditions are pure growing mode initial conditions, which we now review.

Any perturbation expansion depends on the choice of perturbation variable(s).  These are not defined by the EoMs themselves.  Perturbation expansions for different variables may (or may not) converge to the same true solution but, order by order, the expansion for different perturbation variables will differ.  Here we claim that LOBA, which uses $n^{\rm th}$ order perturbation theory where $n$ is the leading order of non-linearities, is accurate.  Even though the expansions differ in detail for different perturbation variables the order $n$ approximation of each may be accurate.  A crucial ingredient for consistency for different perturbation variables is that  one specify equivalent initial conditions to  $n^{\rm th}$ order.  These initial conditions will be different for different perturbation variable.

If an EoM is a $p^{\rm th}$ order PDE then initial conditions for this PDE requires specification of $p$ quantities, $\tilde{q}_0$, $\dot{\tilde{q}}_0, \ldots$ for each Fourier mode . In $n^{\rm th}$ order perturbation theory one must specify these $p$ quantities at each order i.e. one must specify $p\times n$ quantities. For example, solving $2^{\rm nd}$ order PDEs to order 2 requires
$\{{}_{(1)}\tilde{q}_0,\,{}_{(1)}\dot{\tilde{q}}_0,\,
{}_{(2)}\tilde{q}_0,\,{}_{(2)}\dot{\tilde{q}}_0\}$.
Redistributing the total $q_0$ or $\dot{q}_0$ among different orders of perturbations in non-linear equations can lead to significantly different approximate solutions.

It is important to emphasize here that the initial conditions are a choice not determined by the EoMs.  When describing a physical system one must use physics to decide the appropriate initial conditions. The three types of initial conditions for perturbation theory described in the previous paragraph are not exhaustive and the appropriate initial condition might be something very different. There are many choices among the $p\times n$ parameters for each mode. Applications of this formalism of most interest here are for systems which are statistical ensembles of initial conditions. One should distinguish the properties of initial conditions of a particular instance and properties of all instances of the ensemble.  This adds another level of complexity.

%%%%%%%%%%%%%%%%%%%%%%%%
\label{sec:GuidingVariables}
\subsubsection{Guiding Variables at the Initial Singularity}
%\noindent \textbf{Initial Conditions at the Singularity}
%\vskip2pt

The above prescription becomes is obscure when the initial conditions are at an initial singularity.  This is the case of most interest to us!  One might {\it erroneously} think that when the growing mode becomes arbitrarily small at the initial singularity that the linear theory becomes more and more accurate and therefore one need only set the initial conditions at linear order.  This might lead one to believe that all choices of perturbation variable that are equal at linear order, e.g.~$\underline{q}=q-\alpha\,q^2+\cdots$, have the same initial conditions.  Furthermore one might conclude that all $\underline{q}$ can have sub-Poissonian initial conditions. All this is wrong for the simple reason that increasingly small difference in solutions as one approaches the singularity are amplified by subsequent growth and not negligible.

To see this we may use the fact that the gradient terms in Eq.~(\ref{eq:SelfSimilarp2n2}) become irrelevant as one approach the initial singularity at $h=0$. This is a property of self-similar systems. In this limit, Eq.~(\ref{eq:SelfSimilarp2n2}) reduces to the hyperlocal equations considered in \S\ref{sec:MatterLike}.  As noted there, by change of the temporal variable $t\rightarrow\mathfrak{t}$ and dependent variable $q\rightarrow\mathfrak{q}$, one can convert to an equivalent system in canonical form, where the linear order operator is ${}_{(1)}\hat{\mathcal{D}}\,\mathfrak{q}=\ddot{\mathfrak{q}}[\mathfrak{t},\vec{x}]$.  The canonical form will differ for deformations of the dependent variable, $\mathfrak{q}\rightarrow\underline{\mathfrak{q}}=\mathfrak{q}-\alpha\,\mathfrak{q}^2+\cdots$, but the linear operator remains the same. Thus, both $\mathfrak{q}$ and $\underline{\mathfrak{q}}$ have growing and decaying modes $\mathfrak{t}$ and $1$. If $\mathfrak{q}$ has pure growing mode initial conditions, then near the singularity  
$\mathfrak{q}[\mathfrak{t},\vec{x}]
=\mathfrak{q}_0[\vec{x}]\,\mathfrak{t}+\mathcal{O}[\mathfrak{t}^3]$ while $\underline{\mathfrak{q}}$ exhibits very different behavior,
$\underline{\mathfrak{q}}[\mathfrak{t},\vec{x}]
=\mathfrak{q}_0[\vec{x}]\,\mathfrak{t}
-\alpha\,\mathfrak{q}_0[\vec{x}]^2\,\mathfrak{t}^2+\mathcal{O}[\mathfrak{t}^3]$.  The $\mathfrak{q}_0^2$ terms is the 2nd order part of the initial condition for $\underline{\mathfrak{q}}$ that must not be neglected.  If $\alpha\ne0$ there are at least three properties in which $\mathfrak{q}$ and $\underline{\mathfrak{q}}$ {\it must} differ:
\begin{enumerate}
    \item $\mathfrak{q}$ and $\underline{\mathfrak{q}}$ cannot both be guiding variables, 
    \item $\mathfrak{q}$ and $\underline{\mathfrak{q}}$ cannot both be conservative,
    \item $\mathfrak{q}$ and $\underline{\mathfrak{q}}$ cannot both have sub-Poissonian initial conditions
\end{enumerate}
because
\begin{enumerate}
    \item $\underline{\mathfrak{q}}$ is not a guiding variable due to the quadratic term just mentioned,
    \item a conservative quantity must have $\zeta_3=0$, and $\alpha\ne0$ means that $\zeta_3\ne\underline{\zeta}_3$,
    \item since the square of a sub-Poissonian function is a Poissonian function, the $\mathfrak{q}_0[\vec{x}]^2\mathfrak{t}^2$ difference means that 
    $\ddot{\tilde{\mathfrak{q}}}[0,\vec{0}^+]\ne
    \ddot{\tilde{\underline{\mathfrak{q}}}}[0,\vec{0}^+]$.
\end{enumerate}
This has a variety of interesting implications, among them:
\begin{itemize}
    \item one must specify among quantities which are identical to 1$^{\rm st}$ order which (if any) have pure growing mode initial conditions to leading non-linear order ($n$), and
    \item a quantity that is conservative or has a linear EoM will develop LSWN {\it unless} it has {\it both} sub-Poissonian initial conditions {\it and} pure growing mode initial conditions.
\end{itemize}
Thus, protection from LSWN hinges on initial conditions too!

Many of our results up to this point in \S\ref{sec:LSWN}-\ref{sec:LargeScaleNoise} use the assumption that $q$ is a guiding variable, meaning it has growing mode initial conditions at order 1 and zero initial conditions at order 2 to $n$. For example if $q$ is a guiding variable in the relic systems of \S\ref{sec:LateTimeRelics} and if $q$ has sub-Poissonian initial conditions then if $q$ has a non-linear non-conservative EoM and will generate LSWN.  One can then compute the power spectrum of the deformation $\underline{q}=F[q]$ using the function $F[q]=q-\alpha q^2+\cdots$ to determine the power spectrum for $\underline{q}$.  One will find that $\underline{q}$ will also have LSWN whether or not it has a linear or conservative EoM i.e.~$\underline{q}$ "inherits" LSWN from $q$.  Only the $\mathfrak{W}_+$ of the guiding variable really matters.  A guiding variable plays a special role in determining the large scale evolution of all of its deformations which is why we use "guiding".

For physical systems, one must know the initial conditions in detail in order to predict how it manifests non-linearities. One must know what are the guiding variables and which have sub-Poissonian initial conditions. 

%%%%%%%%%%%%%%%%%%%%%%%%%%%%%%%%%%%%%%%%%%%%%%%%%%%%%%%%%%%%
%%%%%%%%%%%%%%%%%%%%%%%%%%%%%%%%%%%%%%%%%%%%%%%%%%%%%%%%%%%%
\section{Planar Newtonian Cosmology as an Illustration}
\label{sec:PlanarNewtonianCosmology}

We now illustrate the formalism developed in the previous sections by applying it to a particular class of partial differential equations (PDEs). Here we choose a system related to (but simpler than) real world cosmology: Newtonian cosmology with planar inhomogeneities. As shown in Ref.~\cite{Stebbins2025}, general relativistic geometrodynamics is closely analogous to Newtonian fluid dynamics, so this has more than a little bearing on real world cosmology.  Furthermore, we should emphasize that this is a realizable physical system (just not one which describes our universe). The phenomenology is thus real physical phenomenology.

The pseudo-linear phenomenology we find will closely mimic that of 3D cosmology with non-planar inhomogeneities, whether Newtonian or general relativistic; \textit{i.e.}~it will mimic real world cosmology. Planar symmetry is amenable to low resource numerical simulations with very large dynamic range, and it is easy to visualize and analyze the results. Given their close correspondence with real world cosmology and tractability of the system, we have found planar Newtonian cosmology to be an excellent ``sandbox'' in which to study the consequences of non-linearities in cosmology. There is much more that can be learned from playing in this sandbox than we have room for here.  A more thorough exposition will be given in a subsequent paper.

Newtonian gravity involves action at a distance, and so the dynamics is (generally) non-local, not described solely by a PDE, and therefore outside the class of dynamical systems we consider here.  Furthermore, one would expect that mass conservation could protect the system from generating Large Scale White Noise (LSWN, see \S\ref{sec:InstantaneousGeneration}) in the density distribution in Eulerian coordinates.  To circumvent these issues, we first consider planar inhomogeneities in Lagrangian coordinates which move with the matter. In this case, the non-locality is hidden and the EoM is a PDE. Admittedly, Lagrangian coordinates are less observationally relevant since one would not normally be measuring the Fourier power spectrum in Lagrangian coordinates. Nevertheless, it is not too difficult to deform results in Lagrangian coordinates to results in Eulerian coordinates. Furthermore, in Ref.~\cite{Stebbins2025} it is argued that cosmic inhomogeneities on super-horizon scales are best described in Lagrangian coordinates since in general relativity there is no global Eulerian coordinate system; rather, the matter defines the space-time.  

The Newtonian equations of motion for a perfect fluid are the continuity, Euler, and Poisson equations,
\begin{align}
&\frac{\partial}{\partial t}\rho[t,\vec{x}]
+\vec{\nabla}\cdot(\rho[t,\vec{x}]\,\vec{v}[t,\vec{x}])=0 \,,
\nonumber \\
&\frac{\partial}{\partial t}\vec{v}[t,\vec{x}]+(\vec{v}[t,\vec{x}]\cdot\vec{\nabla})\,\vec{v}[t,\vec{x}]
=-\frac{\vec{\nabla}p[t,\vec{x}]}{\rho[t,\vec{x}]}-\nabla{\Phi}[t,\vec{x}] \,,
\nonumber \\
&\nabla^2\Phi[t,\vec{x}]=4\pi\,G\,\rho[t,\vec{x}] \ ,
\label{eq:ContinuityEulerPoisson}
\end{align}
where $\rho$ is the density, $p$ the pressure, $\vec{v}$ the velocity with respect to a Galilean frame, $\Phi$ the gravitational potential, and $\vec{x}$ an Eulerian coordinate. These equations are invariant under Galilean transformations. In addition, they are invariant under adding a uniform acceleration, which can be absorbed in the addition of a uniform gradient in $\Phi$.  Thus, $\Phi$ is only defined up to an additive constant and uniform gradient.  $\Phi$ is clearly not a locally measurable quantity. These equations are self-similar in the sense of \S\ref{sec:SelfSimilar} for a barotropic fluid with a constant squared speed of sound, $c_\mathrm{s}^2=p/\rho$, which we henceforth assume.

Newtonian gravity is non-local because of the elliptical nature of the Poisson equation.  One manifestation of this is that nearly any changes in $\rho$ result in instantaneous changes in $\Phi$ throughout space.  Such a non-local system is not of the type we have thus far considered in this paper, other than some exceptional cases (see \S\ref{sec:PlanarNewtonianLagrangian}).

By Newtonian cosmology, we mean perturbations about a classical self-gravitating fluid expanding homologously, 
$\vec{v}_1-\vec{v}_2=H[t]\,|\vec{x}_1-\vec{x}_2|$.  For a self-gravitating fluid, Newtonian gravity only allows Hubble laws, $H[t]$, from a dust cosmology. The simplest of these is the ``flat'' universe Einstein-de Sitter
$H[t]=\frac{2}{3\,t}$, where the background density is 
$\bar{\rho}[t]=\frac{3\,H[t]^2}{8\pi\,G}=\frac{1}{6\pi\,G\,t^2}$, which we take as the ``background'' flow.

Newtonian planar cosmology assumes inhomogeneities in only one direction, which we take to be the $x$-direction, so
\begin{equation}
\rho[t,\vec{x}]=\frac{1}{6\pi\,G\,t^2}+\delta\rho[t,x] \,,
\qquad
\vec{v}[t,\vec{x}]=\frac{2}{3\,t}\,\vec{x}+v_\mathrm{p}[t,x]\,\hat{x} \,,
\qquad
\Phi[t,\vec{x}]=\frac{1}{9}\,\frac{|\vec{x}|^2}{t^2}+\phi[t,x] \,,
\label{eq:PlanarAssumptions}
\end{equation}
where $\hat{x}$ is the unit vector in the $x$-direction and $v_\mathrm{p}$ is the {\it peculiar velocity} in the $x$-direction.  The $x$ coordinate can be rescaled to a {\it comoving coordinate}, $x_\mathrm{co}\equiv x\,(t_\mathrm{fid}/t)^{2/3}$ which expands with the background flow, or with a Lagrangian coordinate $\chi$ which expands with the actual fluid.  We consider the latter.

%%%%%%%%%%%%%%%%%%%%%%%%%%%%%%%%%%%%%%%%%%%%%%%%%%%%%%%%%%%%
\subsection{Lagrangian Coordinates}
\label{sec:PlanarNewtonianLagrangian}

A planar fluid consists of planes of fluid elements each moving along some Eulerian trajectory $x=\mathbb{X}[t,\chi]$ where $\chi$ is a label for the different planes, \textit{i.e.}~$\chi$ is a Lagrangian coordinate.  Henceforth in this subsection we use $\dot{f}$ to denote partial derivatives with respect to $t$ for fixed $\chi$ and $f'$ to denote partial derivatives with respect to $\chi$ for fixed $t$.\footnote{In Eulerian coordinates 
$\dot{f}=(\frac{\partial}{\partial t}+\vec{v}\cdot\vec{\nabla})\,f$ and 
$f'=\mathbb{X}'\,\frac{\partial}{\partial x}f$}.
The fluid velocity in the $x$ direction is $\dot{\mathbb{X}}[t,\chi]$ but in the transverse directions follows the unperturbed Hubble flow such that 
$\dot{\rho}/\rho=-\vec{\nabla}\cdot\vec{v}
=-\frac{4}{3\,t}-\frac{\dot{\mathbb{X}}'[t,\chi]}{\mathbb{X}'[t,\chi]}$.  
Since the mass in each plane is conserved, one can choose $\chi$ such that $d\chi$ is proportional to the surface density.  The continuity equation then tells us that
$\mathbb{X}'=(t/t_\mathrm{fid})^{2/3}\,\bar{\rho}/\rho$ for some choice of a fiducial time, $t_\mathrm{fid}$.  It is these relations between $\vec{v}$, $\mathbb{X}$, and $\rho$ which allow one to eliminate the velocity from the equations of motion in Lagrangian coordinates, which can be written solely in terms of the density.

One can parameterize the inhomogeneity in the density in a variety of ways, three of which are
\begin{equation}
\rho[t,\mathbb{X}[t,\chi]]
=\bar{\rho}[t]\,(1+\delta[t,\chi])
=\bar{\rho}[t]\,e^{\delta\lambda[t,\chi]}
=\frac{\bar{\rho}[t]}{1-\mathfrak{d}[t,\chi]} \ .
\label{eq:OverdensityParameterization}
\end{equation}
Parameterizing the inhomogeneity using $\delta$ is the conventional choice, using $\delta\lambda$ has the nice feature that the EoM is non-singular for all finite values, and using $\mathfrak{d}$ yields the simplest EoM. To linear order in amplitude, these are all the same: $\delta+{\cal O}[\delta^2]=\delta\lambda+{\cal O}[\delta\lambda^2]=\mathfrak{d}+{\cal O}[\mathfrak{d}^2]$. To higher order, however, they differ.  One can in principle use any of these parameters as dependent variables because they are perturbation variables, meaning zero ($\rho=\bar{\rho}$) is a solution to the equations of motion. They thus obey homogeneous PDEs. Since these different dependent variables are not linearly related to each other, they are deformations of each other (see \S\ref{sec:Deformations}) and may have different levels of LSWN.  As explained in \S\ref{sec:GuidingVariables} we can choose one of these to be a guiding variables i.e.~to have pure growing mode initial conditions, and this means that others will not.  The subsequent evolution of the system can be quite different depending on the choice of guiding variable.

The EoMs are
\begin{subequations}
\begin{equation}
0=\ddot{\delta}+\frac{4}{3\,t}\,\dot{\delta}
-2\frac{\dot{\delta}^2}{1+\delta} 
-\frac{2}{3\,t^2}\,(1+\delta)\,\delta
-c_\mathrm{s}^2\,
\left(\frac{t_\mathrm{fid}}{t}\right)^{4/3}\,(1+\delta)^2\,\delta'' \,,
\end{equation}
\begin{equation}
0=\ddot{\delta\lambda}
+\frac{4}{3\,t}\,\dot{\delta\lambda}-\dot{\delta\lambda}^2
-\frac{2}{3\,t^2}\,\left(e^{\delta\lambda}-1\right)
-c_\mathrm{s}^2\,\left(\frac{t_\mathrm{fid}}{t}\right)^{4/3}\,
e^{2\,\delta\lambda}\,
\left(\delta\lambda''+{\delta\lambda'}^2\right) \,,
\end{equation}
\begin{equation}
0=\ddot{\mathfrak{d}}
+\frac{4}{3\,t}\,\dot{\mathfrak{d}}
-\frac{2}{3\,t^2}\,\mathfrak{d}
-c_\mathrm{s}^2\,\left(\frac{t_\mathrm{fid}}{t}\right)^{4/3}\,
\left(\frac{\mathfrak{d}''}{(1-\mathfrak{d})^2}
    +2\,\frac{{\mathfrak{d}'}^2}{(1-\mathfrak{d})^3}
\right) \,,
\end{equation}
\label{eq:PlanarLagrangianEoMs}
\end{subequations}
which are all $2^{\rm en}$ order ($p=2$) with quadratic ($n=2$) leading order non-linearities.  In spite of the fact that this describes a 3D system, the planar symmetry classifies these PDEs as $d=1$. They are also all self-similar PDEs (see \S\ref{sec:SelfSimilar}). Differing leading order non-linearities give differing $\zeta_i$; in \S\ref{sec:RadiationLike} for  $c_\mathrm{s}^2\ne0$ and in \S\ref{sec:MatterLike} for $c_\mathrm{s}^2=0$.  The $\zeta_i$ parameters determine the amount of LSWN generated as a function of the initial conditions. If $c_\mathrm{s}^2=0$, than the EoM  of $\mathfrak{d}$ is linear while if $c_\mathrm{s}^2\ne0$, its EoM is conservative.  In either case, one should not expect to generate LSWN though non-linearities.

As emphasized in \S\ref{sec:Deformations} \& \S\ref{sec:DependentVariableChoice}, these three parameterizations of inhomogeneity ($\delta$, $\delta\lambda$, $\mathfrak{d}$), though equal in linear theory, are non-linearly related to each other, and thus are not equivalent in the initialization of their non-linear evolution nor in the LSWN they produce. It is instructive to see the differences in behavior due to non-linearities, given that there are no differences in linear theory. 
$\zeta_1$ and $\zeta_2$ give the linearized EoM, which are the same for all $q$.  $\zeta_3$, $\zeta_4$ and $\zeta_5$ give the leading order non-linearity, which differs between different $q$. The linear theory growing and decaying modes are 
$Q_+[t]=(t/t_\mathrm{fid})^{2/3}$ and $Q_-[t]=t_\mathrm{fid}/t$ (see Eq.~(\ref{eq:GrowDecayNoSound})).

%%%%%%%%%%%%%%%%%%%
\subsection{Matter Era}
\label{sec:PlanarLagrangianNoSound}

The matter era is characterized by pressure-less matter which behaves like ``dust''.  This corresponds to $c_\mathrm{s}^2=0$ in Eq.~(\ref{eq:PlanarLagrangianEoMs}), and thus 
\begin{subequations}\label{eq:PlanarLagrangianNoSoundEoMs}
\begin{equation}
0=\ddot{\delta}+\frac{4}{3\,t}\,\dot{\delta}
-2\frac{\dot{\delta}^2}{1+\delta} 
-\frac{2}{3\,t^2}\,(1+\delta)\,\delta \,,
\end{equation}
\begin{equation}
0=\ddot{\delta\lambda}
+\frac{4}{3\,t}\,\dot{\delta\lambda}-\dot{\delta\lambda}^2
-\frac{2}{3\,t^2}\,\left(e^{\delta\lambda}-1\right) \,,
\end{equation}
\begin{equation}
0=\ddot{\mathfrak{d}}
+\frac{4}{3\,t}\,\dot{\mathfrak{d}}
-\frac{2}{3\,t^2}\,\mathfrak{d} \,,
\end{equation}
\end{subequations}
which are hyperlocal self-similar systems, as described in \S\ref{sec:HyperLocal2equivalentSelfSimilar}. One can Taylor expand each EoM, obtaining
\begin{equation}
0=\ddot{q}
+\frac{\zeta_1}{t  }\,\dot{q}
+\frac{\zeta_2}{t^2}\,q
+\frac{\zeta_3}{t^2}\,q^2
+\frac{\zeta_4}{t  }\,q\,\dot{q}
+      \zeta_5      \,\dot{q}^2
+{\cal O}[q^3]
\,,
\end{equation}
where the five parameters $\zeta_i$ are given in Table~\ref{tab:zetasMatterLagrangian} for the different choices of $q$.

\begin{table}[!tp]
%[!htbp]
    \centering
    \tiny
    \resizebox{0.5\textwidth}{!}{  % Adjust the width as needed (0.95 = 95% of text width)
    \begin{tabular}{|c||c|c|c|c|c|c|c|c|}
        \hline
                & $\zeta_1$     &$\zeta_2$      &$\zeta_3$ &$\zeta_4$&$\zeta_5$&
                $\mu$ & $\nu$ & $C_\mathrm{LOBA}^\mathrm{hl}$ \\
        \hline\hline
 $\delta$       & $\frac{4}{3}$ & $-\frac{2}{3}$ & $-\frac{2}{3}$ & 0 & -2 &
                $-\frac{1}{2}$ & $\frac{5}{2}$ & $-1$ \\
        \hline
 $\delta\lambda$& $\frac{4}{3}$ & $-\frac{2}{3}$ & $-\frac{1}{3}$ & 0 & -1 &
                 $-\frac{1}{2}$ & $\frac{5}{2}$ & $-\frac{1}{2}$ \\
        \hline
 $\mathfrak{d}$   & $\frac{4}{3}$ & $-\frac{2}{3}$ & 0              & 0 &  0 &
                 $-\frac{1}{2}$ & $\frac{5}{2}$ & $0$ \\
        \hline
    \end{tabular}
    }
    \caption{}  % Update this
    \label{tab:zetasMatterLagrangian}
\end{table}

The EoM of $\mathfrak{d}$ contains no non-linear terms at any order, so the linear theory solution for $\mathfrak{d}$ gives the exact solution, which can be ``deformed'' to the exact solution for all choices of $q$,
\begin{subequations}\label{eq:OverdensitySolutions}
\begin{equation}
\nonumber \\
\delta[t,\chi]=\frac{
\mathsf{g}[\chi]\,\left(\frac{t}{t_\mathrm{fid}}\right)^{2/3}+
\mathsf{d}[\chi]\,      \frac{t_\mathrm{fid}}{t}
}
{1-\mathsf{g}[\chi]\,\left(\frac{t}{t_\mathrm{fid}}\right)^{2/3}
  -\mathsf{d}[\chi]\,      \frac{t_\mathrm{fid}}{t}} \,,
\end{equation}
\begin{equation}
\delta\lambda[t,\chi]=
-\ln\left[1-\mathsf{g}[\chi]\,\left(\frac{t}{t_\mathrm{fid}}\right)^{2/3}
           -\mathsf{d}[\chi]\,\frac{t_\mathrm{fid}}{t}\right] \,,
\end{equation}
\begin{equation}
\mathfrak{d}[t,\chi]=
\mathsf{g}[\chi]\,\left(\frac{t}{t_\mathrm{fid}}\right)^{2/3}+
\mathsf{d}[\chi]\,\frac{t_\mathrm{fid}}{t} \,,
\end{equation}
\end{subequations}
where $\mathsf{g}[\chi]$ is the local amplitude of the growing mode and $\mathsf{d}[\chi]$ is the local amplitude of the decaying mode. This solution is only valid so long as $\mathfrak{d}<1$, otherwise caustics form. Thus, the solution extends to $t=0$ only if $\mathsf{d}[\chi]\le0\ \forall\chi$. To avoid caustics at early times requires $\mathsf{d}[\chi]=0$ in which case $\mathsf{d}$ is a guiding variable and therefore $\delta$ and $\delta\lambda$ cannot be (see \S\ref{sec:GuidingVariables}. Furthermore if $\mathfrak{d}$ has sub-Poissonian initial conditions then neither $\delta$ or $\delta\lambda$ can.  Even with $\mathsf{d}=0$ one may not evade singular behavior since caustics will form when  $t=t_\mathrm{fid}/\max_\chi[\mathsf{d}[\chi]]^{3/2}$.

Expanding the ($\mathsf{d}=0$) exact solutions in $\mathsf{g}$ yields
\begin{subequations}\label{eq:OverdensitySolutionsLOBA}
\begin{equation}
\delta[t,\chi]=
\mathsf{g}[\chi]\,\left(\frac{t}{t_\mathrm{fid}}\right)^{2/3}+
\mathsf{g}[\chi]^2\,\left(\frac{t}{t_\mathrm{fid}}\right)^{4/3}
+{\cal O}[\mathsf{g}^3] \,,
\end{equation}
\begin{equation}
\delta\lambda[t,\chi]=
\mathsf{g}[\chi]\,\left(\frac{t}{t_\mathrm{fid}}\right)^{2/3}
+\frac{1}{2}\,
\mathsf{g}[\chi]^2\,\left(\frac{t}{t_\mathrm{fid}}\right)^{4/3}
+{\cal O}[\mathfrak{\mathsf{g}}^3] \,,
\end{equation}
\begin{equation}
\mathfrak{d}[t,\chi]=
\mathsf{g}[\chi]\,\left(\frac{t}{t_\mathrm{fid}}\right)^{2/3} \,,
\end{equation}
\end{subequations}
for small $\mathsf{g}$ but any $t$.  This $2^{\rm nd}$ order expansion is exactly the LOBA of Eq.~(\ref{eq:HyperlocalSelfSimilarSingularLOBA}) given by the parameters from Table~\ref{tab:zetasMatterLagrangian}.  The non-linearities remain active (see \S\ref{sec:LinearRelics}) since the non-linear correction grows $\propto t^{4/3}$ and not $\propto t^{2/3}$, as predicted by linear theory. The $2^{\rm nd}$ order corrections remain small and the Taylor series remains accurate when and where 
$t\ll t_\mathrm{fid}\,|\mathsf{g}[\chi]|^{-3/2}$, \textit{i.e.}~while in the linear regime, $|\delta|,\,|\delta\lambda|,\,|\mathfrak{d}| \ll1$.

Assuming $\mathsf{g}$ is Gaussian with 1- and 2-point functions (see Eqs.~(\ref{eq:qPointFunctions})~\&~(\ref{eq:gPointFunctions})),
\begin{equation}
\langle\tilde{\mathsf{g}}[k]\rangle=0  \qquad
\langle\tilde{\mathsf{g}}[k]\,\tilde{\mathsf{g}}[k']\rangle
=2\pi\,\delta^{(1)}[k-k']\,R_{(2)}[k] \,,
\end{equation}
then under LOBA
\begin{equation}
\langle q[t,\chi]\rangle\approx
-C_\mathrm{LOBA}^\mathrm{hl}\,
\left(\frac{t}{t_\mathrm{fid}}\right)^{4/3}\,
\langle\mathsf{g}[\chi]^2\rangle \,,
\end{equation}
while from Eq.~(\ref{eq:HyperLocal2PowerSpectra})
\begin{equation}
P_{(2)}^\mathrm{LOBA}[t,k]=R_{(2)}[k]\,
\left(\frac{t}{t_\mathrm{fid}}\right)^{4/3}
+4 {C_\mathrm{LOBA}^\mathrm{hl}}^2 
\left(\frac{t}{t_\mathrm{fid}}\right)^{8/3}
\int_0^\infty dk_1\,R_{(2)}[k_1]\,R_{(2)}[|k-k_1|]\ .
\label{eq:PlanarPowerNoSound}
\end{equation}
These expressions for $\mathfrak{d}$ are exact, \textit{e.g.} $\langle\mathfrak{d}\rangle=0$, but only approximate for $\delta$ and $\delta\lambda$. In the latter case, $C_\mathrm{LOBA}^\mathrm{hl}\ne0$ and the LOBA solution has a {\it secular term} leading to a systematic increase in $\delta$ and  $\delta\lambda$, since $C_\mathrm{LOBA}^\mathrm{hl}<0$.  In Lagrangian coordinates, this is a change in the mass-weighted, not volume-weighted, quantities.  Thus, $\langle\dot{\delta}\rangle\ne0$ does not violate mass conservation.

$P_{(2)}^\mathrm{LOBA}$ is divergent for any pure power law $R_{(2)}[k]$, consistent with the No-No-Scale Theorem of \S\ref{sec:CosmologyLike}.  However, a power law with a small scale cutoff, as in \S\ref{sec:RelicLSWNIII}, is permissible.  Here we use a Gaussian cutoff
\begin{equation}\label{eq:PowerLawGaussianCutoff}
R_{(2)}[k]
=\frac{A\,\sigma}{\Gamma[\frac{1}{2}(1+n_\mathrm{s})]}\,e^{-(k\,\sigma)^2}(k\,\sigma)^{n_\mathrm{s}} \,,
\end{equation}
normalized such that $\langle\mathfrak{d}^2\rangle=A$. Here, $n_\mathrm{s}$ is the spectral index and $\sigma$ is the cutoff scale.
%and $2p$ gives long wavelength spectral index. 
The exact $P_{(2)}^\mathrm{LOBA}$ for $\mathfrak{d}$ may be written in terms of a polynomial in $k\,\sigma$ for even integer $n_\mathrm{s}\ge0$. These are given in 
Table~\ref{tab:PlanarNoSoundLOBASpectraAnalytic}.

\begin{figure}
    \centering
   \includegraphics[width=0.95\textwidth]{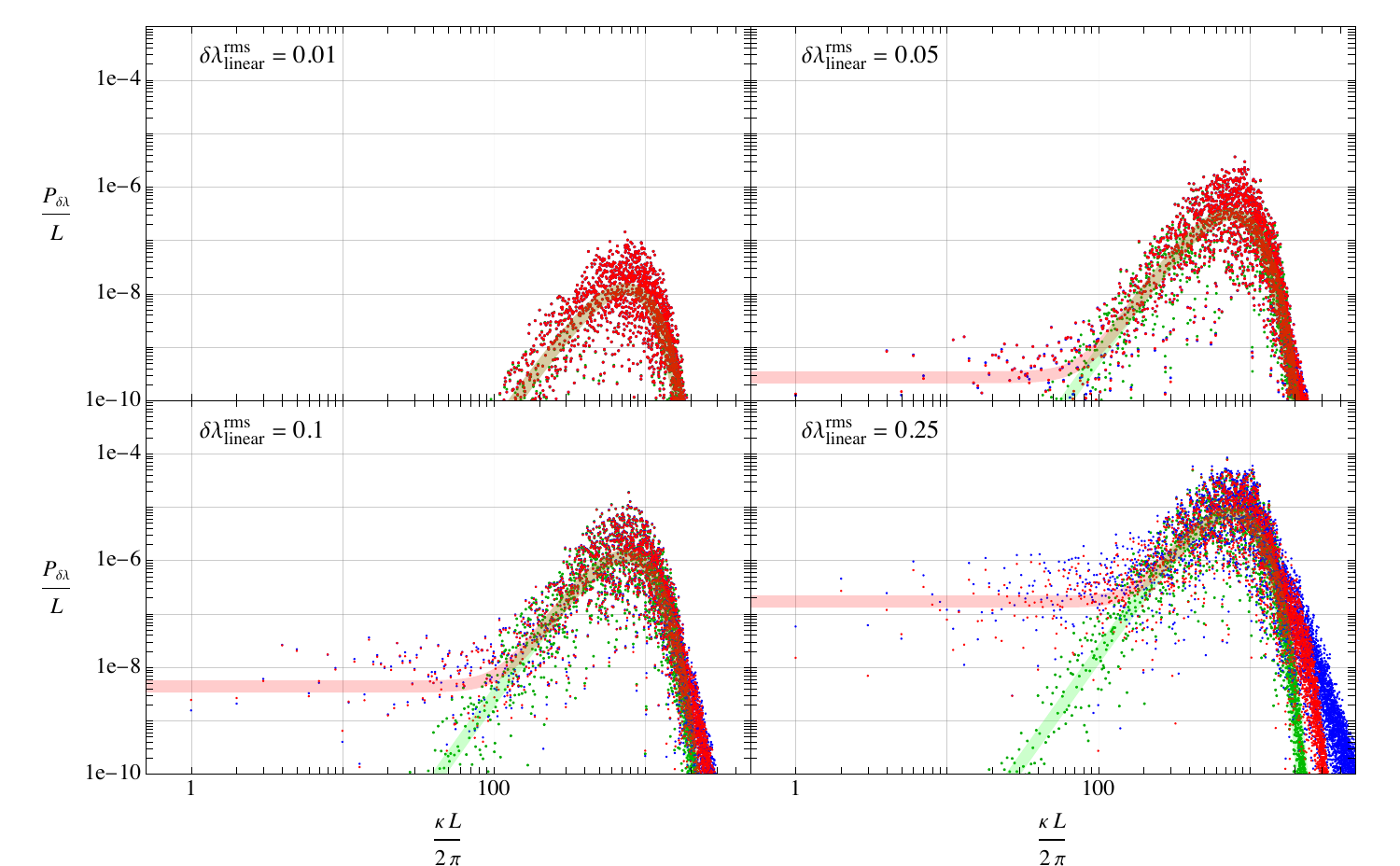}
   \caption{Evolution of dust inhomogeneities in Lagrangian coordinates using the analytic solutions of Eq.~(\ref{eq:OverdensitySolutions}) with periodic boundary conditions. Initial growing mode amplitudes, $\mathsf{g}$, drawn from a Gaussian distribution with power spectrum in Eq.~(\ref{eq:PowerLawGaussianCutoff}) with $n_\mathrm{s}=4$ and $\sigma=0.0003\,L$, where $L$ is the size of the box. Green, blue, and red points denote the power spectra of the linearized solution $\mathfrak{d}=\delta\lambda_\mathrm{linear}$, full analytic solution $\delta\lambda$, and leading order Born approximation solution $\delta\lambda_{\rm LOBA}$, respectively. 
   %We plot the power spectra of $\mathfrak{d}=\delta\lambda_\mathrm{linear}$ (green) and $\delta\lambda$ from the analytic solution (blue). The red points give the evolution of $\delta\lambda$ according to the leading order Born approximation (LOBA).  
   The green band gives the expected power spectrum for linear evolution, which should track the green points.  The pink band gives the expected power spectrum for LOBA evolution, which should track both red and blue points (provided LOBA is accurate).}
   \label{fig:PlanarDustSpectraRealization}
\end{figure}

In Fig.~\ref{fig:PlanarDustSpectraRealization}, we demonstrate the accuracy of LOBA in the linear regime.  Here we plot the evolution of the power spectra of exact solutions, Eq.~(\ref{eq:OverdensitySolutions}), for $\mathfrak{d}$ (green) and $\delta\lambda$ (blue) as well as the LOBA evolution (Eq.~(\ref{eq:LOBA})) of $\delta\lambda$ (red) from the same initial conditions.  $\mathfrak{d}$ obeys the linear theory evolution. The initial growing mode, $\mathsf{g}$, is drawn from a random Gaussian distribution with power spectrum given in Eq.~(\ref{eq:PowerLawGaussianCutoff}) with $n_\mathrm{s}=4$.  Also plotted is the expected power spectrum from linear theory (light green band) and from the LOBA formula (pink band) given by the polynomial in Table~\ref{tab:PlanarNoSoundLOBASpectraAnalytic}. The greatest discrepancy between the analytic solution and LOBA is at small scales, which we are not interested in.  Agreement at large scales remain good even as one approaches non-linearity with $25\%$ inhomogeneity. LSWN is evident in all these graphs except the first, where it is not visible because $P_\mathrm{LSWN}<10^{-10}\,L$. More simulation and analysis of planar cosmology, Newtonian and relativistic, will be given in Ref.~\cite{sandbox2026}.

\begin{table}[!tp]
%[!htbp]
    \centering
    \small
    \renewcommand{\arraystretch}{2}  % Increases row height
    \resizebox{0.99\textwidth}{!}{  % Adjust the width as needed (0.95 = 95% of text width)
    \begin{tabular}{|c||c|c|}
        \hline
        \,\,\,\textbf{$n_\mathrm{s}$}\,\,\, & 
        \textbf{$\dfrac{\sqrt{\pi}\,e^{(k\,\sigma)^2}}{A\,\sigma^2}\,R_{(2)}[k]$} & 
        \textbf{$\dfrac{\sqrt{2\pi}\,e^{\frac{1}{2}(k\,\sigma)^2}}{A^2\,\sigma}
        \int_{-\infty}^\infty dk_1\,R_{(2)}[|k_1|]\,R_{(2)}[|k-k_1|]$} \\
        \hline\hline
        $2$ & $2\,(k\,\sigma)^2$ & 
        $\dfrac{3 - 2\,(k\,\sigma)^2 + (k\,\sigma)^4}{4}$ \\
        \hline
        $4$ & $\dfrac{4}{3}\,(k\,\sigma)^4$ & 
        $\dfrac{105 - 60\,(k\,\sigma)^2 + 18\,(k\,\sigma)^4 - 4\,(k\,\sigma)^6 + (k\,\sigma)^8}{144}$ \\
        \hline
        $6$ & $\dfrac{8}{15}\,(k\,\sigma)^6$ & 
        $\dfrac{10395 - 5670\,(k\,\sigma)^2 + 1575\,(k\,\sigma)^4 - 300\,(k\,\sigma)^6 + 40\,(k\,\sigma)^8 - 6\,(k\,\sigma)^{10} + (k\,\sigma)^{12}}{14400}$ \\
        \hline
    \end{tabular}
    }
    \caption{}  % Update this
    \label{tab:PlanarNoSoundLOBASpectraAnalytic}
\end{table}

%%%%%%%%%%%%%%%%%%%
\subsection{Radiation Era}
\label{sec:PlanarLagrangianSound}

In real world cosmology during the radiation era, the matter is, to a good approximation, a barotropic perfect fluid with constant sound speed.  We can mimic this in Newtonian cosmology with a constant non-zero $c_\mathrm{s}$.  When $c_\mathrm{s}^2\ne0$, it is convenient to use the sound horizon 
\begin{equation}
h=3\,c_\mathrm{s}\,t_\mathrm{fid}^{2/3}\,t^{1/3} \,,
\label{eq:RadiationSoundHorizon}
\end{equation}
as the time variable, as suggested by Eq.~(\ref{eq:SoundHorizon}). As stated in \S\ref{app:EquivalentSystems}, using $t$ or $h$ leads to equivalent systems with the same LSWN properties (\textit{i.e.}~it is not a deformation). In this subsection, we change notation and use $\dot{f}$ to denote a partial derivative with respect to $h$ for fixed $\chi$ while $f'$ continues to denote partial derivatives with respect to $\chi$ for fixed $h$. In terms of $h$, Eqs.~(\ref{eq:PlanarLagrangianEoMs}) become
\begin{subequations}\label{eq:PlanarLagrangianSoundEoMs}
\begin{equation}
0=\ddot{\delta}+\frac{2}{h}\dot{\delta}
-\frac{2\,\dot{\delta}^2}{1+\delta}-\frac{6}{h^2}\,(1+\delta)\,\delta
-(1+\delta)^2\,\delta'' \,,
\end{equation}
\begin{equation}
0=\ddot{\delta\lambda}+\frac{2}{h}\dot{\delta\lambda}
-\dot{\delta\lambda}^2-\frac{6}{h^2}\,(e^{\delta\lambda}-1)
-e^{2\,\delta\lambda}\,(\delta\lambda''+{\delta\lambda'}^2) \,,
\end{equation}
\begin{equation}
0=\ddot{\mathfrak{d}}+\frac{2}{h}\dot{\mathfrak{d}}-\frac{6}{h^2}\,\mathfrak{d}
-\frac{\mathfrak{d}''}{(1-\mathfrak{d})^2}
-\frac{2\,{\mathfrak{d}'}^2}{(1-\mathfrak{d})^3} \,.
\end{equation}
\end{subequations}
When $c_\mathrm{s}\ne0$, $\mathfrak{d}$ is non-linear but conservative. We comment also that there is no analytic solution. These are all $2^{\rm nd}$ order ($p=2$) non-linear EoMs with quadratic ($n=2$) leading order non-linearities, which can be characterized by the seven parameters $\zeta_i$ (see \S\ref{sec:RadiationLike}) given in Table~\ref{tab:zetasRadiationLagrangian}. The different perturbation variables have the same $\zeta_1$ and $\zeta_2$, which give the linearized EoMs, but have differing leading order non-linearities given by $\zeta_3$ through $\zeta_7$. The linear parameter values $\zeta_1=2$ and $\zeta_2=-6$ correspond to $\mu=-\frac{1}{2}$ and $\nu=+\frac{5}{2}$ (see \S\ref{sec:RadiationLike}), which have growing and decaying modes
\begin{equation}\label{eq:RadiationGrowDecay}
{\cal Q}_+[\varphi]=\frac{J_{5/2}[\varphi]}{\sqrt{\varphi}}
=\sqrt{\frac{2}{\pi}}\,j_2[\varphi] \,,
\qquad
{\cal Q}_-[\varphi]=\frac{Y_{5/2}[\varphi]}{\sqrt{\varphi}}
=\sqrt{\frac{2}{\pi}}\,y_2[\varphi] \,,
\end{equation}
where $j_\ell$ and $y_\ell$ are spherical Bessel functions. 

For systems with these parameters, non-linearities at early times generate late time relics as discussed in \S\ref{sec:LateTimeRelics}. These are linear theory growing modes with power spectra very different from those of the initial conditions usually demonstrating LSWN.  If the initial conditions are sub-Poissonian, these relics will come to dominate the late time inhomogeneity. The late time (relic) power spectrum, while still in the linear regime, is given by the asymptotic leading order Born approximation (aLOBA) of Eq.~(\ref{eq:aLOBA}),
\begin{eqnarray}\label{eq:RelicRadiation}
P_{(2)}[h,k]&\approx&
\left(
R_{(2)}[k]+R_{(2)}^\mathrm{nl}[k]
\right)\,\frac{2}{\pi}\,j_2[k\,h]^2 \,,
\nonumber\\
R_{(2)}^\mathrm{nl}[k]
&\approx&
\frac{4\,{\mathfrak{W}_+}^2}{k^{4}}
\int_0^\infty dk_1\,{k_1}^4\,R_{(2)}[k_1]^2  \,,
\end{eqnarray}
where $\mathfrak{W}_+$ is defined by the integral of Eq.~(\ref{eq:aLOBA}).  For the $\zeta_1$ and $\zeta_2$ parameters (or equivalently, the $\mu$ and $\nu$ parameters) which describe the linear EoMs,
\begin{equation}
\mathfrak{W}_+=-\frac{\zeta_3+\zeta_4+3\,\zeta_5-6\,\zeta_6+6\,\zeta_7}{12\,\sqrt{2\pi}} \,,
\end{equation}
where $\zeta_3$ through $\zeta_6$ describe the leading order non-linearities.  These parameters for the three EoMs, $\delta$, $\delta\lambda$, and $\mathfrak{d}$, in Eqs.~(\ref{eq:PlanarLagrangianSoundEoMs}) are given in Table~\ref{tab:zetasRadiationLagrangian}.  Since $\mathfrak{d}$ has a conservative EoM (to all orders) LSWN cannot be generated by non-linearities and will not exist when it is not present in the initial conditions.  This is why $\mathfrak{W}_+=0$ for $\mathfrak{d}$. The relic power spectra of $\delta$ and $\delta\lambda$, determined by $\mathfrak{W}_+$, are nearly the same but with a slightly different gain, $\propto\mathfrak{W}_+$, relative to the initial conditions.  Apart from a factor, the relic power spectrum from non-linearities is generic. We describe the generic evolution of the power spectrum next.

\begin{table}[!tp]
%[!htbp]
    \centering
    \tiny
    \resizebox{0.6\textwidth}{!}{  % Adjust the width as needed (0.95 = 95% of text width)
    \begin{tabular}{|c||c|c|c|c|c|c|c|c|c|c|}
\hline
& $\zeta_1$&$\zeta_2$&$\zeta_3$& $\zeta_4$&$\zeta_5$&$\zeta_6$&$\zeta_7$&
$\mu$&$\nu$&
$\mathfrak{W}_+$ \\
\hline\hline
 $\delta$       & 2 & -6 & -6 & 0 & -2 & 2 &  0 & $-\frac{1}{2}$ & $\frac{5}{2}$
 & $\sqrt{\frac{2}{\pi}}$ \\
\hline
 $\delta\lambda$& 2 & -6 & -3 & 0 & -2 & 2 & 1 & $-\frac{1}{2}$ & $\frac{5}{2}$ 
 & $\frac{5}{4\sqrt{2\pi}}$  \\
\hline
 $\mathfrak{d}$ & 2 & -6 &  0 & 0 &  0 & 2 & 2 & $-\frac{1}{2}$ & $\frac{5}{2}$ 
 & 0 \\
        \hline
    \end{tabular}
    }
    \caption{}  % Update this
    \label{tab:zetasRadiationLagrangian}
\end{table}

%%%%%%%%%%%%%%%%%%%
%\subsubsection*{Timescales and Length scales During the Radiation Era}
%\label{sec:PlanarRadiationTimeline}
%This is the only subsubsection in this subsection so it makes sense to just integrate it into the main body

As in the matter era, non-linearities in Eq.~(\ref{eq:RelicRadiation}) diverge for any power law initial conditions, in agreement with the No-No-Scale Theorem (\S\ref{sec:NoNoScale}). We can regulate this divergence with a high-frequency cutoff. For definiteness, we will choose a Gaussian cutoff similar to Eq.~(\ref{eq:PowerLawGaussianCutoff}),
\begin{equation}
R_{(2)}[k]
=A\,\sigma\,e^{-(k\,\sigma)^2}\,(k\,\sigma)^{n_\mathrm{s}-4} \,,
\label{eq:PowerLawSpectrumGaussianTruncation}
\end{equation}
where $n_\mathrm{s}$ is the spectral index outside the horizon.  The power spectrum of Eq.~\eqref{eq:RelicRadiation} is sub-Poissonian where $n_\mathrm{s}>0$.  From this we find that
\begin{equation}
R_{(2)}^\mathrm{nl}[k]
\approx
2^{\frac{2\,n_\mathrm{s}-5}{2}}\,\Gamma[\frac{2\,n_\mathrm{s}-3}{2}]\,
\frac{\mathfrak{W}_+^2\,A^2}{k^4\,\sigma^3}
\end{equation}
so the LSWN is
\begin{equation}
P_\mathrm{LSWN}[h]\equiv
\lim_{k\rightarrow0}P_{(2)}[h,k]\approx
\lim_{k\rightarrow0}\frac{2}{\pi}\,R_{(2)}[k]\,j_2[k\,h]^2
\approx
\frac{2^{\frac{7-2\,n_\mathrm{s}}{2}}\,\Gamma[\frac{2\,n_\mathrm{s}-3}{2}]}{255\,\pi}\,\mathfrak{W}_+^2\,A^2\,
\sigma\,\left(\frac{h}{\sigma}\right)^4
\end{equation}
which grows with time according to linear theory.  The integral in Eq.~\eqref{eq:RelicRadiation} diverges at small $k$ for $n_\mathrm{s}\le\frac{3}{2}$ indicating the late time relic approximation of \S\ref{sec:LateTimeRelics} fails.  In this case there will be a contribution of active non-linearities in addition to the relics at early times.  There are a number of physical effects which regulate this divergence which will be discussed in Ref.~\cite{sandbox2026}. Now proceed assuming $n_\mathrm{s}>\frac{3}{2}$

If one were able to "see" the transitions to LSWN one would know the sound horizon, $h$, the LSWN power, $P_\mathrm{LSWN}[h]$, the spectral index $n_\mathrm{s}$, and the transition scale $k_\mathrm{LSWN}$ so could determine the cutoff scale
\begin{equation}
\sigma=
\frac{2^{}}{(255\,\pi)}
\left(
\frac{1}{255\,\pi\,\,\mathfrak{W}_+^2\,\Gamma[\frac{2\,n_\mathrm{s}-3}{2}]}
\frac{h^4\,{k_\mathrm{LSWN}}^{2\,n_\mathrm{s}}}{P_\mathrm{LSWN}[h]}
\right)^{\frac{1}{3+2\,n_\mathrm{s}}}\ .
\label{eq:sigmaEstimator}
\end{equation}
The factors of order unity have to do with the shape of the cutoff which precisely defines $\sigma$. These could vary significantly but the general scaling with $n_\mathrm{s}$ should hold so long at the cutoff is sufficiently sharp. When the inhomogeneities are very small ($A\ll1)$ the cutoff scale is much smaller than this transition scale and it might be impossible determine $\sigma$ directly.  Eq.~\eqref{eq:sigmaEstimator} allows one to use the large scale inhomogeneities as a "microscope" to infer things about very small scale inhomogeneities.

\renewcommand{\arraystretch}{1.8}  % Increase row height (vertical padding)
\setlength{\tabcolsep}{15pt}       % Increase horizontal padding (column spacing)

\begin{table}[!htbp]
    \centering
    \begin{tabular}{|l||c|c|}
        \hline
        \textbf{Epoch} & $\langle\delta^2\rangle$ & \textbf{Interval} \\
        \hline\hline
        Super-horizon growth & $\sim \delta_{\mathrm{peak}}^2 \left(\frac{h}{h_*}\right)^4$ & $h \lesssim h_*$ \\
        \hline
        Hubble damping & $\sim \delta_{\mathrm{peak}}^2 \left(\frac{h_*}{h}\right)^2$ & $h \gtrsim h_*$ \\
        \hline
    \end{tabular}
       \caption{Linear Theory Timeline}  
    \label{tab:linearTheoryTimeline}
\end{table}

\begin{table}[htbp]
    \centering
    \begin{tabular}{|l||c|c|}
        \hline
        \textbf{Epoch} 
        & $\langle\delta^2\rangle$ 
        & \textbf{Interval} \\
        \hline\hline
        Super-horizon growth 
        & $\sim \delta_{\mathrm{peak}}^2 \left(\frac{h}{h_*}\right)^4$ 
        & $h \lesssim h_*$  \\
        \hline
        Hubble damping 
        & $\sim \delta_{\mathrm{peak}}^2 \left(\frac{h_*}{h}\right)^2$ 
        & $h_* \lesssim h \lesssim \frac{h_*}{\delta_{\mathrm{peak}}^{2/3}}$ \\
        \hline
        Large scale linear growth 
        & $\sim \delta_{\mathrm{peak}}^4 \frac{h}{h_*}$ 
        & $\frac{h_*}{\delta_{\mathrm{peak}}^{2/3}} \lesssim h \lesssim \frac{h_*}{\delta_{\mathrm{peak}}^4}$ \\
        \hline
        Non-linear evolution 
        & $\gtrsim 1$ 
        & $h \gtrsim \frac{h_*}{\delta_{\mathrm{peak}}^4}$ \\
        \hline
    \end{tabular}
        \caption{Actual Timeline}
 \label{tab:actualTimeline}
\end{table}

\begin{figure}
    \centering
   \includegraphics[width=0.95\textwidth]{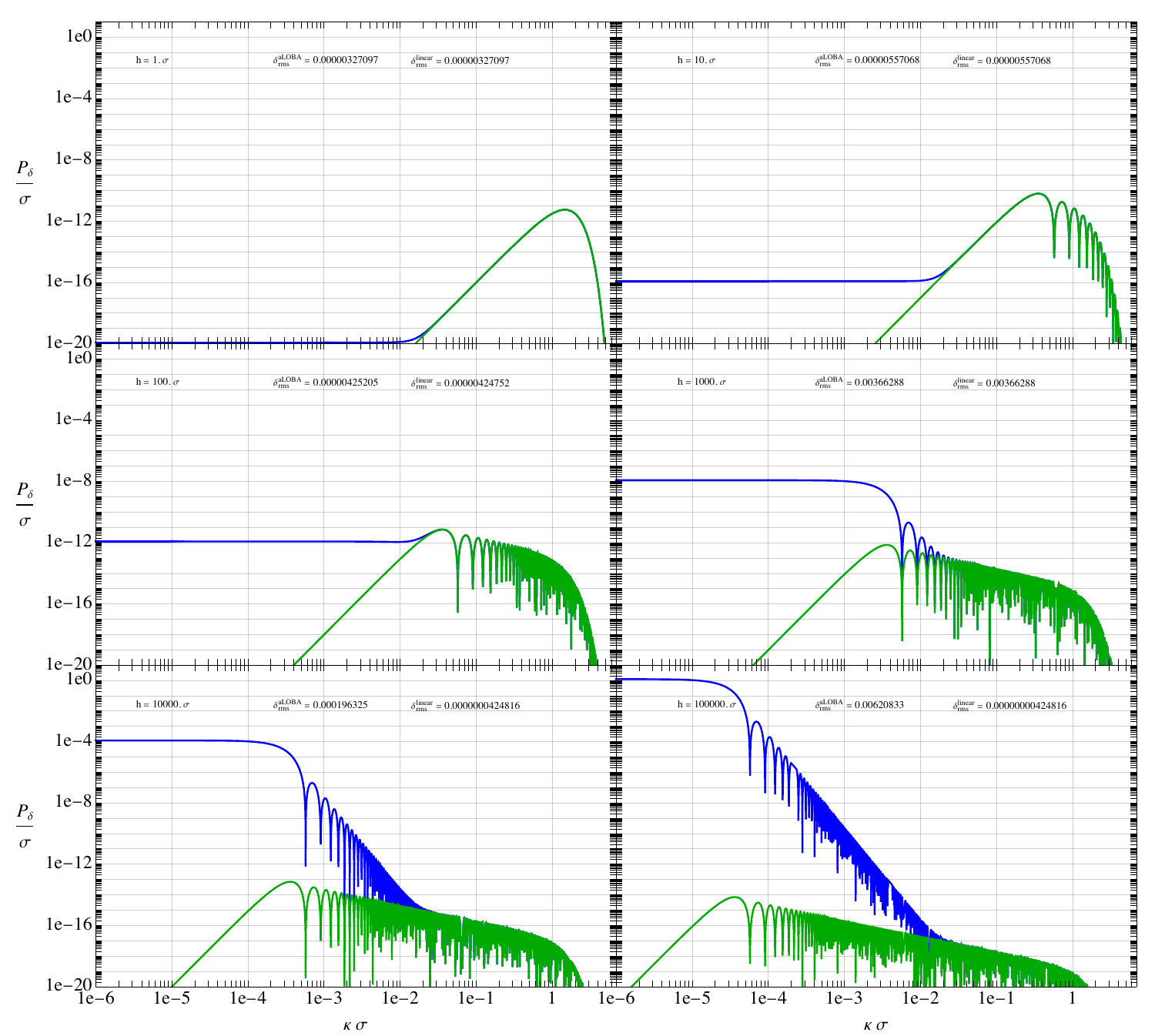}
   \caption{For a range of times parameterized by the sound horizon, $h$ as defined in Eq.~(\ref{eq:RadiationSoundHorizon}) we plot the estimated power spectrum from Eq.~(\ref{eq:RelicRadiation}) for $\delta$ in blue and the linear theory prediction in green.  Initial conditions are of the form of Eq.~(\ref{eq:PowerLawSpectrumGaussianTruncation}) with $n_\mathrm{s}=5$ and normalized such that $\delta_\mathrm{rms}^\mathrm{linear}$ peaks at $10^{-5}$. All quantities are normalized to the cutoff scale, $\sigma$.  Plotted in green is the linear theory prediction.  Evident are the acoustic oscillations on small scales and as well as large scale white noise (LSWN).  Linear theory evolution dominates initially while non-linearities generate additional large scale power which eventually also grows like linear theory, \textit{i.e.}~growth outside the sound horizon and damping inside.}
   \label{fig:NewtonianPlanarSoundDeltaTimeline}
\end{figure}

The evolution of the overdensity with non-linearities is quite different than that predicted by linear theory.  With an initial power law power spectrum, such as the Harrison-Zel'dovich spectrum as allowed in linear theory nothing noteworthy ever happens.  Modes grow out side the sound horizon and then oscillate and damp inside the horizon.  This only thing that changes is the size of the horizon.  However if one includes non-linearities the No-No Scale Theorem requires a cutoff in the spectrum at small scales. With this cutoff something does happen even in linear theory: an outline of the evolution of the mean square overdensity including a cutoff  is given in Table~\ref{tab:linearTheoryTimeline} for linear theory.  This is also illustrated by the green curve in Fig.~\ref{fig:NewtonianPlanarSoundDeltaTimeline}.  Even this varies significantly from the actual timeline which includes non-linearities which is outlined in Table~\ref{tab:actualTimeline} and is illustrated in blue in Fig.~\ref{fig:NewtonianPlanarSoundDeltaTimeline}. 

Mathematically the linear theory evolution of the overdensity is given by
\begin{equation}
\langle\delta^2\rangle_\mathrm{linear}[h]=
\frac{4}{\pi}\int_0^\infty dk\,R_{(2)}[k]\,j_2[k h]^2
\end{equation}
while under asymptotic LOBA
\begin{eqnarray}
\langle\delta^2\rangle_\mathrm{LOBA}[h]&\approx&
\langle\delta^2\rangle_\mathrm{linear}[h]+
\frac{4}{\pi}\int_0^\infty dk\,R_{(2)}^\mathrm{eff}[k]\,j_2[k h]^2
=
\langle\delta^2\rangle_\mathrm{linear}[h]+\left(\frac{h}{h_\mathrm{nl}}\right)^3
\nonumber\\
h_\mathrm{nl}&\approx&3^\mathrm{2/3}\,35^{1/3}\,2^{\frac{2\,n_\mathrm{s}-1}{6}}\,
\Gamma[\frac{2\,n_\mathrm{s}-3}{2}]^\frac{1}{3}\,
{k_\mathrm{LSWN}}^{-\frac{2\,n_\mathrm{s}}{3}}\,
\sigma^{1-\frac{2\,n_\mathrm{s}}{3}} \ .
\end{eqnarray}
According to linear theory the rms overdensity increases while the inhomogeneities are outside the horizon and after crossing the horizon when $h\sim h_*\sim{k_*}^{-1}$ the inhomogeneities are Hubble damped and $\langle\delta^2\rangle$ decreases.  At all times $\langle\delta^2\rangle\ll1$ in linear theory so one might naively think that linear theory is always accurate  This is not true! The actual evolution does follow the linear timeline, peaking at $\delta_\mathrm{peak}$ until long past horizon crossing of the initial perturbations.  However during this horizon crossing a small amplitude long wavelength tail is generated by non-linearities. These wavelengths remain outside the sound horizon, are not damped, but continue to grow.  Eventually, at a time well after horizon crossing, the contribution of the growing long wavelength inhomogeneities to $\langle\delta^2\rangle$ will exceed the linear theory prediction. This is a time when one might say that linear theory ``breaks down''.  Once the long wavelengths start to dominate $\langle\delta^2\rangle$ will continue to grow and eventually, very long after breakdown the inhomogeneities will ``go non-linear'', \textit{i.e.} $\langle\delta^2\rangle\sim1$.  At this point the flow will be supersonic and shocks will form.  As the fluid begins to go non-linear the LOBA will become increasingly inaccurate. In this paper we do not address the question of how the fluid evolves after it goes non-linear.

It is instructive to consider quantitative aspects this timeline.  For parameters used for Fig.~\ref{fig:NewtonianPlanarSoundDeltaTimeline} the rms overdensity peaks at $\delta_\mathrm{peak}\sim10^{-5}$ when the comoving horizon, $h=h_*$, equals the comoving wavelength of the initial inhomogeneities. Much later linear theory breaks down when the comoving horizon is $\sim10^{2}\,h_*$.  Much  later the fluid goes non-linear.  The comoving wavelengths which first go non-linear are $\sim10^6$ times larger than the comoving wavelengths of the initial inhomogeneities! The range of temporal scales is even more striking.  Using 
$h=3\,c_\mathrm{s}\,{t_\mathrm{fid}}^{2/3}\,t^{1/3}$ if $t_*$ is the time of horizon crossing then the time when linear theory breaks down is $10^{6}\,t_*$ and the time when the fluid goes non-linear is $10^{18}\,t_*$. The range of spatial and temporal scales would be even larger for shallower initial slope, \textit{i.e.}~$n_\mathrm{s}<5$.  A steep slope was chosen to avoid a much larger range of scales.  No matter how small the initial inhomogeneity non-linearity will eventually dominate the dynamics however one must observe a very large range of timescales and length scales see these effects.  Non-linear effects are not easily discerned when one considers a small range of scales over a limited time.  Even though cosmic inhomogeneities are initially very small the universe is a very big place which has existed for a very long time so cosmology is a system where non-linear effects have a chance to manifest.

%%%%%%%%%%%%%%%%%%%%%%%%%%%%%%%%%%%%%%%%%%%%%%%%%%%%%%%%%%%%
%%%%%%%%%%%%%%%%%%%%%%%%%%%%%%%%%%%%%%%%%%%%%%%%%%%%%%%%%%%%
\section{LSWN in Cosmology}
\label{sec:LSWNinCosmology}

The main point of this section, and indeed of the entire paper, is that large scale white noise (LSWN) is (or should be) understood as an obvious and inevitable consequence of cosmological inhomogeneities. 
%There is nothing subtle or nuanced about it.  
We have shown that the development of LSWN phenomena is inevitable in most time-dependent spatial system governed by spatially homogeneous non-linear partial differential equations (PDEs) on large or infinite volumes. If the spectrum is initially sub-Poissonian, LSWN will immediately develop and dominate on sufficiently large scales, regardless of how small the non-linearities are. This is the situation which describes cosmological inhomogeneities, so it would be surprising if LSWN did not exist in our universe on large enough scales.

In the previous sections, we have developed a general mathematical theory of LSWN, specializing to mathematical systems which are relevant to the theory of cosmological inhomogeneities.  In this section we adapt this theory  to  a general relativistic description of curvature perturbations in cosmological models.  One could also adapt it to compute LSWN in other cosmological perturbations such as gravity waves or vorticity, but this is left to future work.
 
The generic nature of LSWN allows us to draw conclusions about LSWN in the context of a standard model of cosmology from just a broad outline of this standard model. Detailed modeling of cosmological inhomogeneities involves many subtle issues, such as  precise definitions of perturbative quantities in a non-linear context and the relation of these variables to observational quantities.  None of these can invalidate the generic prediction of cosmological LSWN. Most of these details are beyond the scope of this paper but will be addressed in a follow-up paper. Nevertheless, we can estimate the amplitude of curvature LSWN from fairly simple considerations.

%%%%%%%%%%%%%%%%%%%%%%%%%%%%%%%%
%%%%%%%%%%%%%%%%%%%%%%%%%%%%%%%%
\subsection{General Relativistic Cosmological Inhomogeneities}
\label{CosmologicalInhomogeneities}

Cosmological inhomogeneities can be described as inhomogeneous metric and matter perturbations relative to a homogeneous and isotropic Friedman-Lemaitre-Robertson-Walker (FLRW) background.  These perturbations are described as a realization of a stochastic process which is itself homogeneous and isotropic, consistent with the symmetries of the underlying space-time.  FLRW space-times can be finite or infinite in spatial extent, but are empirically constrained to be sufficiently large so as to be modeled as infinite.  The spatial sections may be positively curved, negatively curved, or flat (zero curvature), but the curvature is empirically constrained to be sufficiently small that one may only consider perturbations about a flat cosmology.  Here, we assume the inhomogeneities are spatially homogeneous and isotropic in flat, infinite Euclidean spatial sections.  The LSWN formalism can be generalized to non-flat or non-infinite FLRW space-times, just not here.

Unlike classical fluid dynamics, which operates under action-at-a-distance Newtonian gravity, cosmological inhomogeneities are governed by local equations (PDEs).  Like classical fluid dynamics, the equations of motion for density, pressure, etc.~are non-linear with quadratic leading order non-linearities, just as we have found in the Newtonian cosmology illustrated in \S\ref{sec:PlanarNewtonianCosmology}.  Our universe is inferred to be nearly homogeneous at early times and on large scales, so the small inhomogeneities are usually modeled as a perturbative expansion about homogeneity.  Usually only the first order (linear theory) dynamics is considered. Pseudo-linear systems like this have been the focus of this paper so far.  As with the illustrations in \S\ref{sec:PlanarNewtonianCosmology}, one should expect that the leading order Born approximation (LOBA) developed previously should provide an accurate description of the effect of non-linearities.

%%%%%%%%%%%%%%%%%%%%%%%%%%%%%%%%
%%%%%%%%%%%%%%%%%%%%%%%%%%%%%%%%
\subsection{Standard Cosmology}
\label{sec:StandardCosmology}

Before estimating the effect of non-linearities in our universe, it is worthwhile to review the standard cosmological model. This standard model of cosmology posits a sequence of nearly distinct epochs, which in order are
\begin{itemize}
    \item[{\bf A}] an {\it inflationary era} of accelerated expansion, wherein the energy budget was dominated by the ``inflaton'' field,
    \item[{\bf B}] the {\it reheating era}, during which the inflaton decays into normal matter, 
    \item[{\bf C}] the {\it radiation era}, during which matter takes the form of an ultra-relativistic thermal gas,
    \item[{\bf D}] the {\it matter era}, during which the matter is mostly non-relativistic ``dark matter'' and baryons, and finally
    \item[{\bf E}] the {\it dark energy era}, which is another period of accelerated expansion. 
\end{itemize}
There are transitions between these eras, but for our purposes it is sufficient to consider each era separately.  We live in the beginnings of the dark energy era (or arguably the transition between the matter and dark energy eras).  There is much empirical data about the dark energy, matter, and the last stages of the radiation era. The parameters of the earlier stages of the radiation era, such as its duration, are quite uncertain. One could view the inflationary and reheating eras as optional, possibly replaced by something quite different.  There could also be a pre-inflationary era but what happened during this era is unknown and probably unimportant for cosmological observations.  The epochs most relevant for LSWN are C and D, partly because they are long and reasonably well understood (or at least modeled).

%%%%%%%%%%%%%%%%%%%%%%%%%%%%%%%%
%%%%%%%%%%%%%%%%%%%%%%%%%%%%%%%%
\subsection{Curvature Perturbations}
\label{sec:CurvaturePerturbations}

Various formalisms are used to describe cosmological inhomogeneities.  A commonly used formalism is ``gauge invariant perturbation theory'' (GIPT \cite{Bardeen:1980kt,Kodama:1984ziu}) which is generally a perturbative approach rather than a fully non-linear description. A LOBA analysis only requires perturbations to 2nd order, which is attainable with GIPT. In GIPT, one typically manipulates gravitational potentials, such as the Bardeen potentials $\Phi$ and $\Psi$. These quantities do not evolve according to PDEs; rather, they are described by integro-differential equations, which is why these potentials exhibit action-at-a-distance phenomenology in spite of describing purely local physics. Since they do not obey PDEs, we would not expect them to develop LSWN in the way described in this paper.  There are combinations of spatial derivatives of potentials which do obey PDEs and which will develop LSWN.  Generally we find using a GIPT description mathematically cumbersome and so do not take this approach here.

While GIPT quantities are ``gauge invariant'', they are not generally locally measurable (covariant) quantities, e.g. there is no way to determine the exact value of $\Phi$ or $\Psi$ from local measurements. Perturbative descriptions of space-time usually associate each space-time event in the ``perturbed'' space-time to an event in a ``background'' space-time.  This association is itself not a covariant concept.  A covariant representation of cosmological inhomogeneities (CR \cite{Hawking:1973uf,1989PhRvD..40.1804E,ellis2012relativistic,Tsagas:2007yx}) is more amenable to the LSWN analysis given in this paper.  Physics is local and locally measurable (covariant) quantities will obey causal equations of motion described by PDEs. Any quantity defined by the local Riemann curvature tensor is covariant.  For example, the center-of-momentum velocity (CM or Landau-Lifshitz frame) is the normalized future-directed time-like eigenvector of the Ricci tensor which is uniquely defined wherever normal matter ($\rho>|p|>0$) is present. Here $\rho$ and $p$ are the energy density and isotropic component of the pressure in the CM frame.  One can define the local rate of expansion of all matter by $\theta\equiv u^\alpha{}_{;\alpha}$ and with this define a quantity we call \emph{kurvature} or the \emph{kurvature density} (ref.~\cite{2026arXiv260116996S})\footnote{The two are technically the same quantity in different units, with conversion factor $8\pi\,G$.}
\begin{equation}
K\equiv\frac{8\pi\,G\,\rho}{3}-\frac{1}{9}\,\theta^2
\qquad
\Delta\rho\equiv\frac{K}{8\pi\,G}=\rho-\frac{1}{24\pi\,G}\,\theta^2 \ .
\label{eq:KurvatureDensity}
\end{equation}
We use the word ``kurvature'' to differentiate from the many other uses of curvature in cosmological perturbation theory.  $K$ is motivated by FLRW space-time, where $\theta=3\,\frac{\dot{a}}{a}$.  In this case, $K=\frac{k}{a^2}$ where $a$ is the scale factor and $k$ is the ``curvature constant'', which is zero for a flat, critical density cosmology.  Thus $K$ or $\Delta\rho$ gives a local measure of the deviation from ``flatness'' without reference to a particular background space-time. A non-zero value quantifies the ``curvature perturbation'' in a completely non-perturbative way.  There are other covariant quantities which similarly quantify curvature perturbations and reduce to $k/a^2$, but the kurvature EoM is particularly simple.

This evolution equation for the kurvature can be obtained by combining the Raychaudhuri equation (see \cite{Tsagas:2007yx}) and local energy conservation equation
\begin{eqnarray}    
\dot{\theta}&=&-\frac{1}{3}\,\theta^2
               -(2\,\sigma^2-2\,\omega^2-{\dot{u}^\alpha}_{;\alpha})
               -4\pi\,G\,(\rho+3\,p)
\nonumber\\
0&=&u^\alpha\,T_\alpha{}^\beta{}_{;\beta}
=\dot{\rho}+(\rho+p)\,\theta-\pi^{\alpha\beta}\,\sigma_{\alpha\beta}
\end{eqnarray}
yielding the two equivalent equations (ref.~\cite{2026arXiv260116996S})
\begin{subequations}
\label{eq:DeltaRhoEvolution}
\begin{equation}
K+\frac{2}{3}\,\theta\,K=
\frac{2}{9}\,\theta\,
\left(2\,\sigma^2-2\,\omega^2-\dot{u}^\alpha{}_{;\alpha}\right)
-\frac{8\pi\,G}{3}\,\pi^{\alpha\beta}\,\sigma_{\alpha\beta}
\label{eq:DeltaRhoEvolution1}
\end{equation}
\begin{equation}
\dot{\Delta\rho}+\frac{2}{3}\,\theta\,\Delta\rho=
\frac{1}{12\pi\,G}\,\theta\,
\left(2\,\sigma^2-2\,\omega^2-\dot{u}^\alpha{}_{;\alpha}\right)
-\pi^{\alpha\beta}\,\sigma_{\alpha\beta}
\label{eq:DeltaRhoEvolution2}
\end{equation}
\end{subequations}
where $\dot\square\equiv u^\alpha\square_{;\alpha}$ is the proper time rate of change along the streamlines of $u^\alpha$; $\sigma_{\alpha\beta}$ and $\omega_{\alpha\beta}$ are the rate of shear and rotation (vorticity) tensors of the CM flow; $\sigma^2\equiv\tfrac{1}{2}\,\sigma^{\alpha\beta}\,\sigma_{\alpha\beta}$ and
$\omega^2\equiv\tfrac{1}{2}\,\omega^{\alpha\beta}\,\omega_{\alpha\beta}$;
$\dot{u}_\alpha$ is proper acceleration of the $u^\alpha$ flow; and $\pi_{\alpha\beta}$ is the anisotropic stress in the CM frame.

A special property of Eq.~\eqref{eq:DeltaRhoEvolution1} is that neither $G \rho$, $G p$, or the Weyl curvature appears explicitly.  In this sense, $K$ in not directly coupled to the ``gravity'' of the matter when $\pi_{\alpha\beta}=0$, and neither is it directly coupled to the tidal gravity encoded in the Weyl curvature.  This simplifies LOBA for this system.

%%%%%%%%%%%%%%%%%%%%%%%%%%%%%%%%
%%%%%%%%%%%%%%%%%%%%%%%%%%%%%%%%
\subsection{Kurvature Perturbations Perturbatively}
\label{sec:KurvaturePerturbationsPerturbatively}

The dimensionless \emph{comoving curvature perturbation} $\mathcal{R}$ is often used in GIPT\footnote{
Do not confuse $\mathcal{R}$ used here with
ref.~\cite{Tsagas:2007yx}'s quantity
$\mathcal{R}=2\,(8\pi\,G\,\rho-\frac{1}{3}\,\theta^2+\sigma^2-\omega^2)$.} to quantify cosmological inhomogeneities. Measurements from the Planck collaboration~\cite{Planck:2018vyg} constrain the primordial inhomogeneities within the standard $\Lambda$CDM cosmological model.
%Cosmological inhomogeneities measured by Planck 2018 in ref.~\cite{Planck:2018vyg} give the primordial inhomogeneities in the current standard model of cosmology, $\Lambda$CDM.  
In fitting $\Lambda$CDM to data, these perturbations are generally treated to linear order. 
%This model fitting to data $\Lambda$CDM generally treats primordial inhomogeneities an observations to linear order. 
Since $\Lambda$CDM is therefore effectively a linearized model, it will not exhibit LSWN.

%is, in linear theory, related to $\Delta\rho$ by the for a particular time-slicing of space-time, defined by $\nabla^2_{(3)}\mathcal{R}=-\tfrac{1}{4}\,R_{(3)}$ where $\nabla^2_{(3)}$ is the Laplacian on the spatial time-slice and $R_{(3)}$ is the intrinsic curvature of the time-slice. 
%If the CM velocity has zero vorticity, $\omega=0$, there exists a time-slicing perpendicular to $u^\alpha$ and the intrinsic curvature of these time-slices is given by
%\begin{equation}
%-\nabla^2_{(3)}\mathcal{R}=\frac{1}{4}\,R_{(3)}
%=4\,\pi\,G\,\Delta\rho+\frac{1}{2}\,\sigma^2\ .
%\label{eq:DeltaRhoFromR}
%\end{equation}

%\AI{Should we comment that the above equations for $\mathcal{R}$ are also the linear order expressions? More generally ${}^{(3)}\!R = - \frac{4}{a^2} e^{- 2 \mathcal{R}} \big( \nabla^2 \mathcal{R} + \frac{1}{2} (\nabla \mathcal{R})^2 \big)$. --AI} 
%\AS{actually I don't know the "foundational" definition of $\mathcal{R}$ but if this is it we should used it.  Could you provide a reference? - AS} 
%\GB{I would  clarify that our expressions for $\mathcal{R}$ are valid at linear order. Including the full nonlinear relation seems unnecessary unless we use it later, but I can add a brief comment with a reference to the geometric definition (e.g. Maldacena 2003 or Baumann’s TASI notes) if we think it’s helpful. Would this be OK with you both}

In linear theory one can relate the kurvature to $\mathcal{R}$ by
\begin{equation}
\nabla^2_\mathbf{x}\mathcal{R}=-4\,\pi\,G\,a^2\,\Delta\rho+\mathcal{O}[2^+]
\label{eq:calRfromDeltaRho}
\end{equation}
where $\nabla^2_\mathbf{x}$ is the Laplacian with respect the comoving spatial coordinates of the unperturbed space-time. We see here how $\mathcal{R}$ is not locally measurable since one may add, $\mathcal{R}\rightarrow\mathcal{R}+\delta\mathcal{R}$, any solution to $\nabla^2_{(3)}\delta\mathcal{R}=0$.  Cosmological inhomogeneities will evolve approximately linearly in the late-radiation era since one has relic LSWN (see \S\ref{sec:LinearRelics}).  Since measurements are made in this late epoch we may use the linear relation, Eq.~\eqref{eq:calRfromDeltaRho}, to ``translate'' between $\Delta\rho$ and the $\mathcal{R}$ inferred
from observations within the linear $\Lambda$CDM model.

In an unperturbed FLRW space-time, the rate of expansion is three times the Hubble parameter, $\theta=3\,\frac{\dot{a}}{a}$, and the shear, vorticity and proper acceleration are all zero, $\sigma_{\alpha\beta}=\omega_{\alpha\beta}=\dot{u}_\alpha=0$.  In a flat ($k=0$) FLRW space-time, then, $\Delta\rho=0$. All of these quantities may be non-zero at 1st order, though. Since the EoM for $\Delta\rho$ contains terms quadratic in these linear quantities, the leading order non-linearity is $n=2$.

%{\color{purple}
%Planck \cite{Planck:2018vyg} has measured the late radiation era spectrum of $\mathcal{R}$ in the context of the linear $\Lambda$CDM model. Since any non-zero $\sigma^2$ is $\mathcal{O}[2^+]$ and is $\sim10^{-5}$ times smaller than $\Delta\rho$ in Eq.~\eqref{eq:DeltaRhoFromR} which is well below the sensitivity of these measurements.  Thus when making connections to observations we make the approximation
%}

%{\color{purple}
%This expression is valid in linear theory and since the LSWN generated in the early universe is a linear relics of early non-linearities (see \S\ref{sec:LinearRelics}) we may use this approximation at late times when comparing with observations.
%}

Unlike the systems we have considered previously, the quantity $\Delta\rho$ couples to many other perturbation variables.  We will show that averaged over large scales, the 2nd order ${}_{(2)}\Delta\rho$ does not couple to other 2nd order quantities.  It does couple to quadratic non-linearities of many other perturbation variables, but if we know the linear solution for all of these variables, we may apply LOBA to obtain $\Delta\rho$ LSWN in much the same way as was done previously.

Of all the terms on the right-hand side of Eq.~\eqref{eq:DeltaRhoEvolution} only the $\dot{u}^\alpha{}_{;\alpha}$ may contribute at $\mathcal{O}[1]$. For LOBA we will use
\begin{equation}
 {}_{(1)}\dot{u}^\alpha{}_{;\alpha} \qquad \text{and} \qquad   
{}_{(2)}\dot{u}^\alpha{}_{;\alpha}
={}_{(2,1)}\dot{u}^\alpha{}_{;\alpha}
+{}_{(2,2)}\dot{u}^\alpha{}_{;\alpha}
\end{equation}
where ${}_{(2,1)}\dot{u}^\alpha{}_{;\alpha}$ is linear in 2nd order perturbation variables and ${}_{(2,s)}\dot{u}^\alpha{}_{;\alpha}$ is quadratic in 1st order perturbation variables.  The proper acceleration, $\dot{u}^\alpha$, is given by local momentum conservation
\begin{equation}
0=u^\beta\,T_{\beta\gamma}\,\mathcal{P}^{\gamma\alpha}=
(\rho+p)\,\dot{u}^\alpha+
\mathcal{P}^{\alpha\beta}\,(p_{;\beta}+{\pi_\beta{}^\gamma}_{;\gamma})
\label{eq:MomentumConservation}
\end{equation}
where 
$\mathcal{P}_{\alpha\beta}\equiv u_\alpha\,u_\beta+g_{\alpha\beta}$ is the spatial projection tensor.  Thus perturbatively
\begin{equation}
{}_{(1)}\dot{u}^\alpha{}_{;\alpha}=
-4\pi\,G\,\frac{\nabla_\mathbf{x}\cdot
( \nabla_\mathbf{x}{}_{(1)}p
 +\nabla_\mathbf{x}\cdot{}_{(1)}\overleftrightarrow{\pi})}
      {\dot{a}^2-a\,\ddot{a}}
\qquad
{}_{(2,2)}\dot{u}^\alpha{}_{;\alpha}=
-4\pi\,G\,\frac{\nabla_\mathbf{x}\cdot
( \nabla_\mathbf{x}{}_{(2)}p
 +\nabla_\mathbf{x}\cdot{}_{(2)}\overleftrightarrow{\pi})}
      {\dot{a}^2-a\,\ddot{a}}
\label{eq:divudots}
\end{equation}
where $\overleftrightarrow{\pi}$ is the spatial matrix describing anisotropic stress. Eq.~\eqref{eq:DeltaRhoEvolution} at $\mathcal{O}[1]$ and $\mathcal{O}[2]$ is
\begin{align}
&\frac{\partial}{\partial t}{}_{(1)}\Delta\rho
+2\,\frac{\dot{a}}{a}\,{}_{(1)}\Delta\rho
+\frac{1}{4\pi\,G}\,\frac{\dot{a}}{a}\,
{}_{(1)}\dot{u}^\alpha{}_{;\alpha}=0
\nonumber\\
&\frac{\partial}{\partial t}{}_{(2)}\Delta\rho
+2\,\frac{\dot{a}}{a}\,{}_{(2)}\Delta\rho
+\frac{1}{4\pi\,G}\,\frac{\dot{a}}{a}\,{}_{(2,2)}\dot{u}^\alpha{}_{;\alpha}=S_{(2,1)}
\label{eq:DeltaLambdaPerturbative}
\end{align}
where the leading order non-linearities are
\begin{align}
 &S_{(2,1)}=
-{}_{(1)}\vec{v}\cdot\vec{\nabla}_\mathbf{x}{}_{(1)}\Delta\rho
-\frac{2}{3}\,{}_{(1)}\theta\,{}_{(1)}\Delta\rho
\nonumber\\
&\hskip38pt+\frac{1}{4\pi\,G}\,\left(\frac{\dot{a}}{a}\,
\left({}_{(2)}\sigma^2-{}_{(2)}\omega^2
-{}_{(2,1)}\dot{u}^\alpha{}_{;\alpha}\right)
-\frac{1}{3}\,{}_{(1)}\theta\,{}_{(1)}\dot{u}^\alpha{}_{;\alpha}
\right)
-{}_{(1)}\pi^{\alpha\beta}\,{}_{(1)}\sigma_{\alpha\beta}
\ .   
\end{align}
Here $\vec{v}$ is the coordinate velocity of the CM frame with respect to the coordinate frame, e.g.~it is zero for comoving/Lagrangian coordinates.  The $\vec{v}$ term is 2nd  order and a necessary computational choice for modeling the non-linearities at early times. For relic LSWN, the asymptotic late time linear inhomogeneities are independent of this coordinate choice.

%%%%%%%%%%%%%%%%%%%%%%%%%%%%%%%%
%%%%%%%%%%%%%%%%%%%%%%%%%%%%%%%%
\subsection{Scalar Vector and Tensor Kurvature Perturbations}
\label{sec:KurvaturePerturbationsSVT}

In linear theory one can decompose inhomogeneities into scalar, vector, and tensor components which evolve independently.  By contrast, non-linearities allow interactions between these components.  Nevertheless it is interesting to consider non-linearities of these different components separately assuming the others are non-zero.  For simplicity, in each case we consider a perfect fluid matter model where $\pi_{\alpha\beta}=0$ and use Lagrangian coordinates where $\vec{v}=0$. The simplest are tensor modes or gravity-waves, for which ${}_{(1)}\theta={}_{(1)}\omega_{\alpha\beta}={}_{(1)}\dot{u}_\alpha=0$ so
\begin{equation}
 S_{(2,1)}^\mathrm{tensor}
 =\frac{1}{4\pi\,G}\,\frac{\dot{a}}{a}\,{}_{(2)}\sigma^2 \ .
\end{equation}
Next simplest are vector modes, for which ${}_{(1)}\theta=0$ so
\begin{equation}
 S_{(2,1)}^\mathrm{vector}=
\frac{1}{4\pi\,G}\,\frac{\dot{a}}{a}\,
\left({}_{(2)}\sigma^2-{}_{(2)}\omega^2
-{}_{(2,1)}\dot{u}^\alpha{}_{;\alpha}\right) \ .
\end{equation}
Finally, the least simple are scalar modes, for which
${}_{(1)}\omega_{\alpha\beta}=0$ so
\begin{equation}
S_{(2,1)}^\mathrm{scalar}=
-\frac{2}{3}\,{}_{(1)}\theta\,{}_{(1)}\Delta\rho
+\frac{1}{4\pi\,G}\,\left(\frac{\dot{a}}{a}\,
\left({}_{(2)}\sigma^2
-{}_{(2,1)}\dot{u}^\alpha{}_{;\alpha}\right)
-\frac{1}{3}\,{}_{(1)}\theta\,{}_{(1)}\dot{u}^\alpha{}_{;\alpha}
\right) \ .
\label{eq:Sscalar}
\end{equation}
In standard cosmology models, it is the scalar modes which dominate, though it is observationally allowed that vector and/or tensor modes dominate $S_{(2,1)}$ on very small scales. Henceforth we concentrate on curvature generated by the standard model scalar inhomogeneities in a nearly perfect fluid during the radiation era.

%%%%%%%%%%%%%%%%%%%%%%%%%%%%%%%%
%%%%%%%%%%%%%%%%%%%%%%%%%%%%%%%%
\subsection{Super-horizon Kurvature Perturbations}
\label{sec:SuperHorizonKurvaturePerturbations}

We see from Eq.~\eqref{eq:divudots} that ${}_{(1)}\dot{u}^\alpha{}_{;\alpha}$ and ${}_{(2,2)}\dot{u}^\alpha{}_{;\alpha}$ are perfect derivatives and should average to zero over large volumes.  Super-horizon (super-sound-horizon) scales refers to wavenumbers small enough that pressure gradients can be neglected, i.e. where $\overline{{}_{(1)}\dot{u}^\alpha{}_{;\alpha}}$ and 
$\overline{{}_{(2,2)}\dot{u}^\alpha{}_{;\alpha}}$ can be neglected.  Thus averaging Eq.~\eqref{eq:DeltaLambdaPerturbative} over super-horizon scales gives 
\begin{align}
&\frac{\partial}{\partial t}\overline{{}_{(1)}\Delta\rho}
+2\,\frac{\dot{a}}{a}\,\overline{{}_{(1)}\Delta\rho}
\approx0
\nonumber\\
&\frac{\partial}{\partial t}\overline{{}_{(2)}\Delta\rho}
+2\,\frac{\dot{a}}{a}\,\overline{{}_{(2)}\Delta\rho}
\approx\overline{S_{(2,1)}} \ .
\label{eq:DeltaLambdaPerturbativeSuperHorzon}
\end{align}
Thus on super horizon scales $a^2\,\overline{{}_{(1)}\Delta\rho}\approx\text{constant}$ and Eq.~\eqref{eq:calRfromDeltaRho} tells us that $\overline{{}_{(1)}\mathcal{R}}\approx\text{constant}$, which is a well-known result of linear theory. $\overline{{}_{(2)}\mathcal{R}}$ is not constant, however, and is generated by non-linearities.  If $\Delta\rho$ is non-zero only at 1st order, then at 2nd order the generated kurvature is
\begin{equation}
\overline{{}_{(2)}\Delta\rho[t]}
=\int^t dt'\,\left(\frac{a[t']}{a[t]}\right)^2\,
\overline{S_{(2,1)}[t']}
=(1+z)^2\int_z\frac{dz'}{(1+z')^3}\,\frac{a}{\dot{a}}\,
\overline{S_{(2,1)}[z']}
\label{eq:LOBAkurvatureGeneration}
\end{equation}
where $a=1/(1+z)$. None of the terms in $S_{(2,1)}$ are spatial derivatives and thus $\overline{S_{(2,1)}}$ will not average to zero on superhorizon scales. Terms like $\sigma^2$ and $\omega^2$ are positive definite and cannot average to zero. Eq.~\eqref{eq:LOBAkurvatureGeneration} is the LOBA formula for the generation of kurvature.  If this temporal integral converges rapidly at early times then at later times linear theory becomes accurate, but with $\Delta\rho$ modified by the early time non-linearities. This we call relic kurvature (see \S\ref{sec:LinearRelics}).

%%%%%%%%%%%%%%%%%%%%%%%%%%%%%%%%
%%%%%%%%%%%%%%%%%%%%%%%%%%%%%%%%
\subsection{Sub-Poissonian Initial Conditions}
\label{sec:SubPoissonianDeltaRho}

The standard model of cosmology has quantum fluctuations during the inflationary era generating a near Harrison-Zel'dovich (HZ) scale invariant spectrum of super-horizon curvature perturbations, which at linear order where there is no temporal dependence is
\begin{equation}
\Delta^2_\mathcal{R}[k]\equiv
4\pi\,k^3\,P_\mathcal{R}[k]\,
\genfrac{}{}{0pt}{3}{\propto}{\sim}\, k^0
\label{eq:PDeltaR}
\end{equation}
where observationally $\Delta^2_\mathcal{R}[k]\sim10^{-10}$.  The relationship of Eq.~\eqref{eq:calRfromDeltaRho} means that for linear perturbations 
\begin{equation}
P_{\Delta\rho}[a,k]
=\frac{c^4 k}{(4\pi)^3 G^2 a^4}\,\Delta^2_\mathcal{R}[k]\,
\genfrac{}{}{0pt}{3}{\propto}{\sim}\,k^1
\ .
\label{eq:PDeltaRho}
\end{equation}
Thus the prediction is that $\Delta\rho$ is initially \emph{sub-Poissonian} which allows for non-linearities to generate LSWN which dominates the linear power spectrum at the largest scales.

%%%%%%%%%%%%%%%%%%%%%%%%%%%%%%%%
%%%%%%%%%%%%%%%%%%%%%%%%%%%%%%%%
\subsection{LSWN from the Radiation Era}
\label{sec:ExpectationLSWN}

In this section we \emph{estimate} LSWN in $\Delta\rho$ generated by acoustic (scalar) oscillations in the radiation using scaling arguments.  This ``informal derivation'' of the amplitude of LSWN gives a more intuitive explanation than the more formal derivations of \S\ref{sec:LargeScaleNoise}\,\&\,\ref{sec:PlanarNewtonianCosmology} and postpones a more lengthy discussion of various subtleties to later work. 

In the early radiation era, an accurate matter model is an ultra-relativistic perfect barotropic fluid with sound speed $c/\sqrt{3}$.  Such a perfect fluid obeys self-similar EoMs as described in \S\ref{sec:SelfSimilar}, \ref{sec:LargeScaleNoise},
\ref{sec:PlanarNewtonianCosmology}. LSWN is computed from non-linearities from (scalar) acoustic waves in the radiation era, which we describe by homogeneous random noise (see \S\ref{sec:HomogeneousStochasticDistribution}). One can quantify the ratio of the magnitude of non-linear to linear terms in Eq.~\eqref{eq:DeltaRhoEvolution} as the ratio of averages over realizations:
\begin{equation}
\epsilon\equiv\frac{1}{2}\,\frac{a}{\dot{a}}\,
\frac{\sqrt{\langle|S^\mathrm{scalar}_{(2,1)}|^2\rangle}}
{\sqrt{\langle|\Delta\rho|^2\rangle}}
\sim
\epsilon_\sigma\equiv
\frac{\langle\sigma^2\rangle}{8\pi\,G\,\sqrt{\langle\Delta\rho^2\rangle}}
\end{equation}
where the shear contribution to nonlinearity, $\epsilon_\sigma$, is only one component of the total, $\epsilon$. Since the system is self-similar we expect all the non-linearities to scale in the same way, which is why we approximate $\epsilon\sim\epsilon_\sigma$.  The latter is easier to estimate than the former. For small amplitude waves, the CM density is roughly the cosmological density:
$\langle\rho\rangle
\sim\tfrac{3}{8\pi\,G}\,\left(\frac{\dot{a}}{a}\right)^2$.  On sub-horizon scales, well below the Jean's length, gravity is unimportant so 
${}_{(1)}\theta\ll\frac{\dot{a}}{a}$ and 
$\Delta\rho\approx-\frac{1}{4\pi\,G}\,\frac{\dot{a}}{a}\,{}_{(1)}\theta$.  For these sub-horizon wave with period much smaller than the expansion rate
$\langle{}_{(1)}\theta^2\rangle\approx3\,\langle\sigma^2\rangle$.  Thus we find that
\begin{equation}
\epsilon\sim\epsilon_\sigma\approx
\frac{\langle{}_{(1)}\theta^2\rangle}{24\,\pi\,G\,\sqrt{\langle\Delta\rho^2\rangle}}
\sim
\frac{\sqrt{\langle{}_{(1)}\Delta\rho^2\rangle}}{\langle\rho\rangle}
\label{eq:NonlinearitySize}
\end{equation} 
i.e. the non-linear terms are roughly $\Delta\rho/\rho$ times smaller than the linear terms, as one would expect.

Like their Newtonian planar analog modeled in \S\ref{sec:PlanarNewtonianCosmology}, modes with wavelengths larger than the sound horizon (Jean's length) grow and are damped when the sound horizon grow larger than the wavelength. Fourier modes primarily generate LSWN when these modes cross the horizon, i.e. when $k\sim\dot{a}$.  Using Eq.~\eqref{eq:NonlinearitySize} we can estimate how much LSWN is generated by the cohort of modes\footnote{A cohort means modes with wavenumbers with spread $\Delta\ln[k]\sim1$ centered about $k=k_\mathrm{coh}$.} with similar $k$ which cross the horizon roughly simultaneously. The kurvature from this cohort is
$\langle\mathcal{R}^2\rangle_\mathrm{coh}
\sim{}_{(1)}\Delta^2_\mathcal{R}[k_\mathrm{coh}]
=(4\pi\,G)^2\,\,(k_\mathrm{coh}/a)^4\,\langle\Delta\rho^2\rangle_\mathrm{coh}$, and since $\langle\rho\rangle\sim3\,(\dot{a}/a)^2/(8\pi\,G)$, the average fractional non-linearity from this cohort is $\epsilon_\mathrm{coh}\sim
\sqrt{\langle\mathcal{R}^2\rangle_\mathrm{coh}}$. The actual non-linearity will vary from region to region, $\epsilon_\mathrm{coh}^2$, giving it variance within a comoving coherence volume $\sim1/k_\mathrm{coh}{}^3$.  We can use the integral of Eq.~\eqref{eq:LOBAkurvatureGeneration} to estimate the magnitude of kurvature generated in a coherence volume by restricting the redshift integral it takes the cohort to cross the horizon, $\Delta z_\mathrm{coh}\sim(1+z_\mathrm{coh})$, at the redshift of horizon crossing, $z\sim z_\mathrm{coh}$:
\begin{equation}
\left\langle\overline{{}_{(2)}\Delta\rho_\mathrm{coh}}^2\right\rangle
\sim
\left(\frac{1+z}{1+z_\mathrm{coh}}\right)^4\,\,
\left\langle\mathcal{R}^2\right\rangle_\mathrm{coh}\,
\left\langle\overline{{}_{(1)}\Delta\rho_\mathrm{coh}}^2\right\rangle\ .
\end{equation}
Since $\langle\mathcal{R}_\mathrm{coh}{}^2\rangle\ll1$, the kurvature generated at $z\sim z_\mathrm{coh}$ is a tiny fraction of the original on the scale of the coherence length.  However since $P_\mathrm{coh}$ is constant for $k\ll k_\mathrm{coh}$ it will dominate the original power for sub-Poissonian initial conditions.

Using Eq.~\eqref{eq:calRfromDeltaRho} one finds 
$\left\langle\overline{{}_{(1)}\Delta\rho_\mathrm{coh}}^2\right\rangle\sim
((1+z_\mathrm{coh})\,k_\mathrm{coh})^4\,{}_{(1)}\Delta_\mathcal{R}^2[k_\mathrm{coh}]
/(4\pi\,G)^2$ so the $\Delta\rho$ power spectrum generated by the cohort is
\begin{equation}
\Delta P_\mathrm{coh}[z,k_\mathrm{coh}]
\sim
\frac{\left\langle\overline{{}_{(2)}\Delta\rho_\mathrm{coh}}^2\right\rangle}
{4\pi\,k_\mathrm{coh}^3}
\sim
\frac{(1+z)^4\,k_\mathrm{coh}}{(4\pi)^3\,G^2}\,
{}_{(1)}\Delta_\mathcal{R}^2[k_\mathrm{coh}]^2
\ .
\label{eq:KurvatureGenerationFromCohort}
\end{equation}
We have established that the this non-linearly generated kurvature will have a white noise tail so the effect of the cohort  is to increase $P_{\Delta\rho}[z,k]$ by
$\sim\Delta P_\mathrm{coh}[z,k_\mathrm{coh}]$ for $k<k_\mathrm{coh}$ and $z<z_\mathrm{coh}$.

Late time LSWN is the cumulative increase in $P_{\Delta\rho}[z,k]$ from all cohorts to large scale ($k\gg k_\mathrm{coh}$) power. To estimate this we sum over cohorts, which, since the a cohort has $\Delta\ln[k]\sim1$, is the logarithmic integral of Eq.~\eqref{eq:KurvatureGenerationFromCohort}:
\begin{equation}
P_{\Delta\rho}^\mathrm{LSWN}[z]\sim
\frac{(1+z)^4}{(4\pi)^3\,G^2}
\int_\mathrm{rad} dk\,{}_{(1)}\Delta_\mathcal{R}^2[k]^2\ .
\label{eq:RadiationKurvatureLSWN}
\end{equation}
where the integral is over the range of all wavenumbers which enter the horizon during the radiation era.  Additional LSWN may be generated before or after the radiation era but we do not include that here.

This scaling tells us that under LOBA at late times
\begin{equation}
P_{\Delta\rho}[z,k]\sim{}_{(0)}P_{\Delta\rho}[z,k]+P_{\Delta\rho}^\mathrm{LSWN}[z]
\end{equation}
where ${}_{(0)}P_{\Delta\rho}$ is the initial (primordial) spectrum.  Or in terms of the \emph{late-time} $\mathcal{R}$ power spectrum
\begin{equation}
\Delta_\mathcal{R}^2[k]\approx{}_{(1)}\Delta_\mathcal{R}^2[k]+\frac{k_\mathrm{BH}}{k}
\qquad
k_\mathrm{BH}\sim\int_\mathrm{rad} dk'\,{}_{(1)}\Delta_\mathcal{R}^2[k']^2 \ .
\label{eq:SpectrumWithLSWN}
\end{equation}
The term $k_\mathrm{BH}/k$ is another way of representing white noise in $\Delta\rho$. Both terms are constant in time obeying linear theory evolution\footnote{This statement only applies before non-linear gravitational collapse, i.e. while $\Delta\rho\ll\rho$.}. As always with LSWN, all of the dynamics which generates LSWN yields only one observable parameter, $k_\mathrm{BH}$.  The subscript ``BH'' is used because $\Delta_\mathcal{R}^2[k_\mathrm{BH}]=1$ so if ever wavenumbers with $k\sim k_\mathrm{BH}$ enter the horizon at late times black holes will form, and in fact black holes will become ubiquitous well before then.  If dark energy is persistent these long wavelengths will never enter the horizon.

%%%%%%%%%%%%%%%%%%%%%%%%%%%%%%%%
%%%%%%%%%%%%%%%%%%%%%%%%%%%%%%%%
\subsubsection{Limits on Cosmological Scale Invariance Estimates}
\label{sec:LimitsOnScaleInvariance}
 
Measurements of the CMBR anisotropies, in particular those of the Planck mission, have determined that the inhomogeneities toward the end of the radiation era are predominantly curvature fluctuations consistent with the super-horizon power law power spectrum
\begin{equation}
\Delta^2_\mathcal{R}[k]
=A_\mathrm{s}\,\left(\frac{k}{k_0}\right)^{n_\mathrm{s}-1}
\label{eq:PlanckPowerSpectrum}
\end{equation}
with parameter values 
$\ln(10^{10} A_\mathrm{s}) =3.043\pm0.014$ and
$n_\mathrm{s}=0.9652\pm0.0042$
using the conventional pivot scale $k_0=0.05\,\mathrm{Mpc}^{-1}$~\cite{Planck:2018vyg}. The spectral index $n_\mathrm{s}$ is slightly smaller (redder) than the ``scale invariant'' Harrison-Zel'dovich $n_\mathrm{s}=1$. This power law power spectrum is part of $\Lambda$CDM model. There is no evidence of LSWN (see \cite{Barenboim2025}).

One can interpret the lack of LSWN to indicate that the observed spectrum is the primordial spectrum, ${}_{(1)}\Delta_\mathcal{R}^2[k]^2$.  In concordance with the No-No Scale theorem (\S\ref{sec:NoNoScale}) one sees that for a pure power law the LSWN would diverge.  One requires that the primordial spectrum be suppressed relative to the observed power law at small scales (large $k$). As we shall show, this cutoff must occur at astronomical scales. This is at odds with the general idea of approximating the initial distribution as fully scale invariant. There are numerous ways to achieve this suppression some of which we consider next.

%%%%%%%%%%%%%%%%%%%%%%%%%%%%%%%%
%%%%%%%%%%%%%%%%%%%%%%%%%%%%%%%%
\label{sec:CutoffModel}
\subsubsection{Cutoff Model}
%\noindent \textbf{Cutoff Model}
%\vskip2pt

A simple model for the primordial power spectrum is a power law with the measured $A_\mathrm{s}$ and $n_\mathrm{s}$ of Eq.~\eqref{eq:PlanckPowerSpectrum} and a cutoff $k\sim k_\mathrm{cut}$.  If the cutoff is sharp then Eq.~(\ref{eq:SpectrumWithLSWN}) becomes
\begin{equation}
\Delta^2_\mathcal{R}[k]\approx
A_\mathrm{s}\,
\left(\left(\frac{k}{k_0}\right)^{n_\mathrm{s}-1}+
\left(\frac{k_\mathrm{LSWN}}{k_0}\right)^{n_\mathrm{s}}\,\frac{k_0}{k}
\right)
\qquad
\begin{array}{cc}
k_\mathrm{BH}\sim{A_\mathrm{s}}^2\,
\left(\frac{k_\mathrm{cut}}{k_0}\right)^{2\,(n_\mathrm{s}-1)}\,k_\mathrm{cut}  \\
k_\mathrm{LSWN}\sim
{A_\mathrm{s}}^{\frac{1}{n_\mathrm{s}}}\,
\left(\frac{k_\mathrm{cut}}{k_0}\right)^{\frac{n_\mathrm{s}-1}{n_\mathrm{s}}}\,k_\mathrm{cut}
\end{array} \,.
\end{equation}
where $k_\mathrm{LSWN}$ is the wavenumber at which the LSWN power spectrum is equal to the amplitude of primordial power spectrum. If one approximates $n_\mathrm{s}=0.965$ by $n_\mathrm{s}=1$ then
\begin{equation}
\Delta^2_\mathcal{R}[k]\sim
A_\mathrm{s}\left(1+\frac{k_\mathrm{LSWN}}{k}\,\right) \,,
\qquad k_\mathrm{LSWN}\sim\sqrt{k_\mathrm{cut}\,k_\mathrm{BH}} \,.
\end{equation}
The three wavenumbers
$k_\mathrm{BH}\ll k_\mathrm{LSWN}\ll k_\mathrm{cut}$ are very different in magnitude, each separated by factors $A_\mathrm{s}^{-1} \sim 10^{10}$.

Observationally we are able to probe only the length range $10\,\mathrm{Mpc}\lesssim k^{-1}\lesssim10\,\mathrm{Gpc}$, a dynamic range of only $\sim10^3$.  Since in this range we see only an $n_\mathrm{s}\simeq 1$ spectrum and no transition from this scaling to $n_\mathrm{s}\simeq 0$, we have observationally determined that $k_\mathrm{LSWN} \lesssim 0.1/\mathrm{Gpc}$. This is only a rough estimate; a more precise analysis will be given in a companion work\cite{Barenboim2025}.  In this minimal model, this implies that
\begin{equation}
    k_\mathrm{cut} \lesssim \mathcal{O}(1) \, \mathrm{pc}^{-1} \,.
\end{equation}
This is well outside the range of cosmological observations, yet it is also not microscopic.  Requiring a cutoff in the power spectrum at the $\mathrm{pc}$ scale or larger has profound implications for cosmological modeling, especially restricting models of inflation. Again, a more precise determination of acceptable values of $k_\mathrm{cut}$ from the LSWN constraint will be given in a follow-up work.

The dark matter mass in this cutoff volume is $\sim10^{-7}\,M_\odot$ which is about the mass of the Moon.  In standard cosmology, the temperature of the cosmic plasma when the universe had a horizon size of $\sim\mathrm{pc}$ is 
$k_\mathrm{B}\,T\sim100\,\mathrm{MeV}$. If $k_\mathrm{cut}$ is smaller than the maximal value, the temperature when $k_\mathrm{cut}$ entered the horizon is even smaller. This suggests either some interesting low temperature phenomena in the universe or perhaps a low reheating temperature.  It is suggestive that a minimal cutoff scale very roughly coincides the horizon scale during the quark-hadron phase transition.\\

%%%%%%%%%%%%%%%%%%%%%%%%%%%%%%%%
%%%%%%%%%%%%%%%%%%%%%%%%%%%%%%%%
\label{sec:RunningModel}
\subsubsection{Running Model}
%\noindent \textbf{Running Model}
%\vskip2pt

A power law power spectrum is assumed by the $\Lambda$CDM model but, as noted above, is not expected to extend to all scales.  Furthermore, in the context of inhomogeneities produced during inflation, one expects a near but not exactly constant spectral index. This is usually expressed as an expanded model for the power spectrum
\begin{equation}
\Delta^2_\mathcal{R}[k]
=A_\mathrm{s}\,
\left(\frac{k}{k_0}
\right)^{n_\mathrm{s}-1+\frac{1}{2}\,\alpha_\mathrm{s}\,\ln[k/k_0]} \,,
\label{eq:PlanckRunningPowerSpectrum}
\end{equation}
where $\alpha_\mathrm{s}$ is the running of the spectral index. Planck observationally constrains $\alpha_\mathrm{s}=-0.006\pm0.013$, so the data does not require $\alpha_\mathrm{s}\ne0$, but a small amount of running is still possible. Negative values of $\alpha_\mathrm{s}$ will cut off the power spectrum in a very gradual way and even a small value would avoid the production of an unacceptably large amount of LSWN through non-linearities.  The running required to produce a similar or larger suppression  of LSWN as a power law cutoff at $k_\mathrm{cut}\sim 3 \, \mathrm{pc}^{-1}$ is
\begin{equation}
\alpha_\mathrm{s}\lesssim-0.015 \,.
\label{eq:alphaConstraint}
\end{equation}
This requires a {\it non-zero} value! The LSWN constraint is qualitatively different than the CMBR constraints which are consistent with $\alpha_\mathrm{s}=0$.   It seems possible that this {\it prediction} of $\alpha_\mathrm{s}$ from LSWN is large enough to be measurable from future CMBR plus large scale structure studies.   A more precise determination of acceptable values of $\alpha_\mathrm{s}$ from the LSWN constraint will be given in the follow-up work.

Requiring a small negative spectral index is a less interesting extension of $\Lambda$CDM and is within the range of expectations from inflationary cosmology. If the shape of the inflaton potential is responsible for this small negative value of $\alpha_\mathrm{s}$, this would constrain models of inflation. Note that $\alpha_\mathrm{s}$ is only the first term in a Taylor series in slow-roll inflation.  The LSWN constraint spans a much larger range of scales than the CMBR constraint. Higher order terms in the Taylor series may also be important.\\

%%%%%%%%%%%%%%%%%%%%%%%%%%%%%%%%
%%%%%%%%%%%%%%%%%%%%%%%%%%%%%%%%
\label{sec:NonMinimalModels}
\subsubsection{Non-Minimal Models}
%\noindent \textbf{Non-Minimal Models}
%\vskip2pt

It could be that the LSWN receives contributions from processes other than acoustic oscillations. The large amplitude of cosmological LSWN is largely to do with the fact that the cosmological fluid is unstable to gravitational collapse on scales above the sound horizon.  Any process that gives rise to inhomogeneities with large wavelength can set off this instability.  A generic result from~\cite{2026arXiv260116996S}\footnote{See this paper for more details and caveats.} is that the kurvature density (gravitational binding/collapse) is seeded by shear in the matter content of the universe.  On average if one starts with small $\Delta\rho$,
\begin{equation}
\overline{\Delta\rho}[z]
=\frac{(1+z)^2}{2\pi\,G}\int^z dz'\,\frac{\overline{\sigma^2}[z']}{(1+z')^3} \,,
\end{equation}
where $z$ is the redshift and $\sigma$ is the shear.  Note that the integrand is positive definite so $\overline{\Delta\rho}$ accumulates from different sources of shear.  The acoustic waves we have considered here constitute just one source of shear and there may be others, e.g. fluid motion during a first order phase transitions, gravitational waves which also cause matter shear, etc.  The lack of measured LSWN puts a constraint on the sum of all of these.  If there are other sources of shear beyond the acoustic waves, then the constraint on the acoustic waves becomes even tighter.  Thus the LSWN phenomena constrains many different process in the early universe.  These constraints will be studied further in future papers.\\

%%%%%%%%%%%%%%%%%%%%%%%%%%%%%%%%
%%%%%%%%%%%%%%%%%%%%%%%%%%%%%%%%
\label{sec:HereticalModels}
\subsubsection{Heretical Models}
%\noindent \textbf{Heretical Models}
%\vskip2pt

Implicit in our study of acoustic waves was that the spectrum of inhomogeneities on observed scales, $\sim10\,\mathrm{Mpc}$ to $10\,\mathrm{Gpc}$, is primordial rather than purely the result of non-linearities.  One usually presumes it to be the remnant of inflationary quantum fluctuations.  The argument that it is not caused by non-linearities is that the measured spectral slope is not the generic white noise, $n_\mathrm{s}=0$.  The argument {\it is not} that the measured inhomogeneities have more power on large scales than allowed by causality from small scale processes.  In fact, LSWN has more power on the largest scales than is observed, and it will be caused by processes on arbitrarily small scales. If anything, it is strange that the universe has so little power on large scales given that these scales are so susceptible to gravitational collapse.

Causality at most predicts that spatial correlations, $\xi_{\Delta\rho}[r]$, are zero above some length scale (compact support) where the matter has interacted. In our universe, the interaction scale is at least the sound horizon scale at recombination, $\sim100\,\mathrm{Mpc}$.  Requiring that, say $\Delta\rho$, has zero correlations above $100\,\mathrm{Mpc}$ does not require any particular $n_\mathrm{s}$ so long as $n_\mathrm{s}\ge0$. 
Generically correlation functions with compact support and
\begin{itemize}
\item[{\bf 0}] $\int d^3\vec{r}\,\xi_{\Delta\rho}[|\vec{r}|]>0$ implies $n_\mathrm{s}=0$ on large scales,
\item[{\bf 1}] $\int d^3\vec{r}\,\xi_{\Delta\rho}[|\vec{r}|]=0$ and
$\int d^3\vec{r}\,|\vec{r}|^2\,\xi_{\Delta\rho}[|\vec{r}|]>0$
implies $n_\mathrm{s}=2$ on large scales,
\item[{\bf 2}] $\int d^3\vec{r}\,\xi_{\Delta\rho}[|\vec{r}|]=\int d^3\vec{r}\,|\vec{r}|^2\,\xi_{\Delta\rho}[|\vec{r}|]=0$ and
$\int d^3\vec{r}\,|\vec{r}|^4\,\xi_{\Delta\rho}[|\vec{r}|]>0$
implies $n_\mathrm{s}=4$ on large scales,
\end{itemize}
etc.  One can realize the case 0 with physical processes with no conservation laws (as one would expect), case 1 with one conservation law (say $\overline{\Delta\rho}=0$, which one would \textit{not} expect), case 2 with two conservation laws, etc.  It is less obvious what physical process might produce a Harrison Zel'dovich spectrum $n_\mathrm{s}=1$, and it might be even harder to explain the observed $n_\mathrm{s}=0.96$, but these values are not forbidden by causality.  Admittedly, popular inflation models {\it predicted} $n_\mathrm{s}$ slightly less than unity, and this is an argument in favor of an inflationary origin of the observed inhomogeneities.

A {\it heretical model} would posit that at some time during the radiation era, the cosmic fluid was essentially uniform and then some physical process caused the cosmic fluid to move in a way that imprinted correlation in $\Delta\rho$ with just the right properties to leave relic large scale noise (\textit{not} white noise) with $n_\mathrm{s}\approx1$. We have not constructed such a process, but believe that one could. Getting the observed spectral index is not generic, however.  In such a scenario, an early inflationary epoch and reheating would not be necessary!  Since we do not know about such a process, there is {\it cosmic confusion} (\S\ref{sec:CosmicConfusion}) about whether such a process produced the observed inhomogeneities or whether the inhomogeneities are primordial.

It is just this sort of scenario which in linear theory is impossible, because linear theory has conservation laws which lead to an $n_\mathrm{s}\ge4$ spectrum at large scales.  However, as we have shown, when one includes non-linearities this argument is no longer valid.  Rather, large scale power as a manifestation of non-linearities on small scales is unavoidable, even in $\Lambda$CDM.  One must work to suppress it.

%%%%%%%%%%%%%%%%%%%%%%%%%%%%%%%%
%%%%%%%%%%%%%%%%%%%%%%%%%%%%%%%%
\label{sec:ModelingExtensions}
\subsection{Modeling Extensions}
%\noindent \textbf{Modeling Extensions}
%\vskip2pt

So far, we have only considered LSWN generated in the radiation era.  We feel this is conservative and justified for predictions of the CMBR since most of the curvature fluctuations during this epoch are produced well before matter-radiation equality.  During the matter era, as suggested by the planar dust models of \S\ref{sec:PlanarLagrangianNoSound}, the non-linearities in the matter era will become active and begin to grow rapidly, leading even larger LSWN.  We do not believe that the additional LSWN generated during the matter era will, on the largest observed scales, compete with that already produced during the radiation era --- though this needs further study. Furthermore, non-linearities in gravitational clustering during the matter era are well-studied, and it seems unlikely that our approach to non-linearities would lead to new phenomenology.  

In the standard cosmological paradigm, the radiation era is preceded by reheating which is preceded by inflation. Reheating is often treated as just an earlier matter era.  If that description is accurate, than the rapidly growing LSWN during this early matter era would provide stronger constraints on the LSWN produced during the radiation era. This is one reason we believe a radiation era limit is conservative.  

In a simple single field inflation model, it seems clear that the curvature inhomogeneities are strongly damped but supplemented by quantum fluctuations in the inflaton field.  It is our expectation that although LSWN will be generated during an inflationary era, this will be overwhelmed by LSWN generated in subsequent epochs.

Further study is required to give a complete picture of LSWN production during throughout the history of the universe.

%%%%%%%%%%%%%%%%%%%%%%%%%%%%%%%%
%%%%%%%%%%%%%%%%%%%%%%%%%%%%%%%%
\subsection{Simplicity}
\label{sec:Simplicity}

Explicit representation of Einstein equations are long and complex expressions, even in the case of planar symmetry.  They are non-linear to arbitrary order.  The procedure used here, which we claim is accurate in the pseudo-linear regime, greatly reduces the relevant complexity.

Scalar curvature perturbations are described by a 2nd order, $p=2$, PDE at linear order. It is clear, even from classical hydrodynamics, that the leading order non-linearities are quadratic, $n=2$.  We claim that the leading order Born approximation (LOBA) is accurate in the pseudo-linear regime.  LOBA corrects the  linear solutions to 2nd order using the linear solutions.  These require initial conditions.

Determining appropriate initial conditions is physics and may be the hardest part of the problem.  Generally one would idealize by specifying initial conditions at the initial singularity.  One would probably specify a guiding variable (see \ref{sec:GuidingVariables}). This must be a locally measurable covariant quantity e.g. $\Delta\rho$ or perhaps $\Delta\equiv\Delta\rho/\rho$ or perhaps $\delta\Omega\equiv\Delta\rho/(\rho-\Delta\rho)$.  They all differ. 

Since the 2nd order PDE must be spatially homogeneous and isotropic this reduces the complicated Einstein equations to an EoM requiring only a few temporal functions (see \S\ref{sec:p2n2}).  Furthermore in an idealized radiation era which is self-similar because of the constant sound speed  these functions reduce to only seven numbers $\zeta_1$-$\zeta_7$ to describe the dynamics. It is likely in the radiation era that non-linearities cease to be active early on and mostly manifest as a relic change of the spectrum which then evolves linearly (see \S\ref{sec:LateTimeRelics}).  This relic depends on the initial conditions, two numbers, $\mu$ and $\nu$, which describe the linear evolution plus one parameter, $\mathfrak{W}_+$, which describes the non-linearly induced LSWN.  We know how to model linear evolution so we know $\mu$ and $\nu$.  Thus GR non-linear evolution of curvature inhomogeneities in the radiation era reduce to one number $\mathfrak{W}_+$. The radiation era relic spectrum can be used as early universe initial conditions for later evolution.  In some sense we have reduced the important GR non-linear dynamics to one unknown number without specifying any of the details of GR. This number only serves to normalize the correction to the early radiation era power spectrum.

\subsection{Why not $\delta\rho/\rho$?}
\label{sec:deltarhobyrho}

Here we have discussed LSWN of the kurvature inhomogeneity $\Delta\rho$ rather than the more commonly used $\delta\equiv\delta\rho/\rho$, which is a more intuitive measure of cosmic inhomogeneity: 
$\delta[t,\vec{x}]\equiv
(\rho[t,\vec{x}]-\bar{\rho}[t])/\bar{\rho}[t]$ where $\rho[t,\vec{x}]$ is the local density and $\bar{\rho}[t]$ is the density in the background FLRW space-time at coordinate time $t$. As has long been realized, $\delta$ is a gauge (coordinate) dependent quantity since for a given physical point in space-time the time coordinate, $t$, assigned to it is a gauge choice.  This is possible because $t$ and $\bar\rho[t]$ are not locally measurable.  Nevertheless $\delta$ is well-defined when one specifies a gauge and is often used to describe cosmological inhomogeneities.  Using quantities which are not locally measurable is problematic for LSWN because their EoMs are generally not local, they do not evolve according to a PDE, and the LSWN analysis described here does not apply directly to such quantities.  Expression of the LSWN phenomena in the language of the non-local quantities of GIPT will be given in subsequent work, not here.

By definition, observations are of locally measurable quantities, and given a gauge these can be related to non-local quantities such as $\delta$.  Furthermore, amongst reasonable gauge choices the value of $\delta$ on sub-horizon scales is only weakly dependent on the gauge choice.  This is because in flat space, not in GR, $\delta$ is well defined, which is why it was used in the analysis of LSWN in Newtonian cosmology in \S\ref{sec:PlanarNewtonianCosmology}. The relic LSWN considered here is, in the radiation era, asymptotically a linear theory growing mode for which $\Delta\rho$ determines $\delta$ in the gauge of your choice. Late time matter era non-linearities, which we do not discuss, can be treated accurately in Newtonian gravity since it occurs on scales well below the horizon.

%%%%%%%%%%%%%%%%%%%%%%%%%%%%%%%%%%%%%%%%%%%%%%%%
%%%%%%%%%%%%%%%%%%%%%%%%%%%%%%%%%%%%%%%%%%%%%%%%
\section{Related Phenomena}
\label{sec:RelatedPhenomena}

The non-linear combining of signals is ubiquitous in both natural and man-made phenomena. Whenever this happens one will usually have the ``mixing'' described in \S\ref{sec:InstantaneousGeneration} where short frequencies/wavenumbers combine to create long frequencies/wavenumbers.  When this is done on purpose\footnote{There is also inadvertent down conversion of AM signals which allows one can to ``hear'' AM audio signal in electronics not designed for that purpose.} in electronics this is known as ``down conversion'', which is the basis of heterodyning and which is used in AM radio to decode/encode kHz audio signals from/to GHz electro-magnetic waves. The cosmological down conversion described here is the conversion of short wavelength acoustic wave into longer wavelength acoustic waves. Unlike in AM radio where the down-converted signal has the structure of music or speech, the cosmological down-converted acoustic waves is just noise.

Spatial down conversion of cosmological acoustic waves can also be described by scattering theory, \textit{e.g.}~using Feynman diagrams. Unlike in the most common applications of scattering theory the cosmological acoustic wave scattering takes place in a temporally evolving medium where there is no energy conservation and thus individual waves do not have a well defined frequency/energy. The temporal waveform of spatial sinusoids are temporal Bessel functions not temporal sinusoids.\footnote{Sinusoids are special in that products of sinusoids are equal to finite sums of sinusoids.}  There are no Lorentz or Galilean boost symmetries either but there is spatial homogeneity which gives conservation of momentum.  In the absence of energy conservation cosmological scattering allows for processes which are normally kinematically forbidden.  The leading order quadratic non-linearities of most interest here correspond to the scattering of two waves into one, which if there were energy conservation, would necessarily have a frequency equal to the sum of the frequencies of the two incoming waves and therefore a not a low frequency wave.  However in cosmology this is not the case but rather two high frequency waves will scatter into a low frequency wave if the momentum of the two are nearly opposite. The down scattering, which produces large scale white noise (LSWN), predominantly occurs by acoustic waves at horizon crossing when the Bessel function waveforms differ significantly from sinusoids.

In \S\ref{sec:LSWN} it was  shown how LSWN generically arises from non-conservative non-linear PDEs, equations which govern the dynamics of many physical systems.  Therefore we expect and do find that cosmological LSWN has parallels in the phenomenology of many other physical systems.  The non-linearities which are central to these phenomena go by different names, \textit{e.g.} ``interactions'' in particle physics, ``mode couplings'' in wave mechanics, etc. Similarly the words used to describe the growth of power on large scales/long wavelengths also vary: ``down conversion'', ``upscale transfer'', ``inverse cascade'', etc. While the  details vary considerably, there are common elements to all of these phenomena. 

A foundational example is found in hydrodynamic turbulence. Here, non-linear interactions cause the redistribution of power over scales but in this case the presence of conserved quantities, such as energy and enstrophy, constrain how power can be transferred across scales and in particular prevent the production of white noise on large scales \cite{EyinkSreenivasan2006} just as shown in \S\ref{sec:LSWN}.  However transfer of power to large scales does still happen \cite{batchelor1967introduction}.  In fact for 2D turbulence, in contrast to 3D turbulence, the power is primarily transferred to large scales \cite{Fjortoft1953} which is one reason one finds large eddies on the (2D) surface of Jupiter \cite{Lindborg_Nordmark_2022}. 

Wave turbulence presents a closely related picture. Non-linearities in wave systems naturally lead to spectral broadening, where energy is redistributed across scales. This mirrors the mode-coupling mechanism we analyze, in which short-wavelength interactions generate long-wavelength power. In particular, magnetohydrodynamic (MHD) turbulence provides a concrete illustration: weak turbulence in MHD leads to spectra that develop low-$k$ tails from initially compact support, closely resembling the emergence of LSWN in cosmological inhomogeneities \cite{Nazarenko2011}.

The emergence of power-law or flat (white noise) spectra is a hallmark of non-linear wave systems and reinforces the point that LSWN is a generic outcome of spectral broadening via non-linear interactions. This broadening reflects universal physics shared across diverse systems.

Further analogy can be drawn with the inverse cascade in turbulence, where energy flows from small to large scales—a rare feature in physical systems. While cosmology lacks a direct analog to the traditional energy cascade seen in fluids, the underlying mechanism of power transfer toward large scales via non-linear interactions provides a useful conceptual parallel. This highlights the physical plausibility of LSWN formation through non-linear mode mixing \cite{Frisch1995}.

Similar spectral broadening effects also appear in non-linear optics and signal processing. For instance, in non-linear fiber optics, Kerr non-linearity can transform a narrow input spectrum into one with broad, noise-like tails—a behavior directly analogous to LSWN generation \cite{Agrawal2001}. In time-frequency analysis, non-linear systems often exhibit broadband, noise-like output components, which likewise reflect a redistribution of power in spectral space \cite{Cohen1989}.

In conclusion, LSWN in cosmology exemplifies a broader principle seen across non-linear systems: mode coupling and spectral transfer routinely lead to large-scale, uncorrelated fluctuations. The parallels to turbulence, non-linear wave theory, optics, and signal processing provide a robust physical context for understanding LSWN as a universal feature of non-linear dynamics.

\section{Synopsis}
\label{sec:Synopsis}

White-noise fluctuations are a familiar concept in physics and engineering, representing uncorrelated stochastic variations that can obscure underlying signals. Cosmology is not immune to this effect. In fact, early measurements of the cosmic microwave background radiation (CMBR) initially attributed certain features to instrumental or environmental white noise---only to later reveal them as genuine cosmological signals. In this work, we explore the inverse phenomenon: how dynamically generated stochastic contributions can mimic or mask real physical effects. This highlights the necessity of distinguishing them from primordial or deterministic signals.

We have shown that non-linear evolution generically redistributes power from small to large scales, producing an effective white-noise contribution at long wavelengths---a phenomenon we term Large Scale White Noise (LSWN). This effect arises even when the primordial spectrum is strongly suppressed at low $k$, demonstrating that large-scale fluctuations need not be present initially in order to appear later. Instead, they are generated dynamically through mode coupling between short- and long-wavelength modes via quadratic and higher-order non-linearities in the equations of motion.

In regimes where the primordial spectrum contains ample small-scale power but little large-scale power---as is the case for sub-Poissonian initial conditions with spectral index $n_s > 0$---the induced white-noise component can compete with, or even dominate over, the primordial long-wavelength contribution. This result highlights that large-scale modes are not guaranteed to be a simple linear evolution of the initial spectrum: they also carry the imprint of non-linear dynamics operating at much smaller scales.

The framework developed here---particularly the leading-order Born approximation (LOBA)---provides an accurate and tractable method for computing LSWN in the pseudo-linear regime. We have demonstrated its validity through comparison with exact solutions and numerical simulations of planar Newtonian cosmology in both radiation and dust-dominated eras. The key findings include:

\begin{itemize}
\item \textbf{Inevitability of LSWN:} Any non-linear, non-conservative system with sub-Poissonian initial conditions will generate LSWN, regardless of how small the amplitude of inhomogeneities remains.

\item \textbf{The No-No-Scale Theorem:} Systems exhibiting relic LSWN cannot be truly scale-free.  A power law power spectrum that does not diverge on large scales must be cutoff on small scales for finite non-linearities.  This leads to a characteristic scale above which the power is generated by non-linearities.

\item \textbf{Cosmic Confusion:} On sufficiently large scales, LSWN can be observationally indistinguishable from primordial fluctuations, creating an ambiguity in interpretation that can only be resolved by measurements spanning a wide dynamic range.

\item \textbf{Observational Constraints:} The non-observation of LSWN on the largest observable scales ($k \lesssim 0.1~\text{Gpc}^{-1}$) places stringent constraints on the small-scale cutoff of the primordial power spectrum. For the standard cosmological model with $n_s \approx 0.965$ and $A_s \approx 3 \times 10^{-10}$, we estimate that the primordial spectrum must be cut off at comoving scales $k_{\text{cut}} \lesssim 3~\text{pc}^{-1}$, or equivalently, must exhibit sufficient running of the spectral index ($\alpha_s \lesssim -0.015$).
\end{itemize}

These findings suggest that care is warranted when interpreting signals on the largest scales. Although in standard scenarios the induced contribution is expected to remain subdominant, in models with suppressed primordial power at low $k$---or with deviations from the conventional inflationary paradigm---the non-linear white-noise component may represent the leading source of long-wavelength fluctuations. The largest scales accessible to observation may therefore contain both primordial and dynamically generated contributions, and distinguishing between them is essential for a robust understanding of the origin of cosmological structure.

The constraint on small-scale power has profound implications for inflationary model building. A cutoff at $\sim 1$ pc corresponds to modes entering the horizon at temperatures $k_BT \sim 100$ MeV, suggestively near the quark-hadron phase transition, and to enclosed dark matter masses comparable to the Moon. This scale is far below current observational reach yet far above microphysical scales, pointing to unexplored low-temperature phenomena in the early universe or possibly a low reheating temperature.

Beyond the standard model, LSWN provides a window into early-universe physics that is complementary to CMBR and large-scale structure observations. Any process that generates shear in the cosmic fluid---acoustic oscillations, first-order phase transitions, gravitational waves, or vorticity---contributes to the curvature density and thus to LSWN. The non-observation of LSWN therefore constrains the integrated effect of all such processes, offering a novel probe of physics at scales and epochs that are otherwise inaccessible.

The universality of LSWN extends well beyond cosmology. We have identified analogous phenomena in hydrodynamic turbulence, wave turbulence, non-linear optics, and signal processing. In each case, non-linear mode coupling leads to spectral broadening and the emergence of power on scales where it was initially absent. This universality underscores that LSWN is not an artifact of a particular cosmological model, but rather a generic feature of non-linear dynamical systems.

\bigskip

{\it The discovery of the cosmic microwave background began with a mistake: a hiss in the receiver, dismissed at first as noise—irrelevant, instrumental, to be removed Penzias and Wilson sought clarity, yet in doing so uncovered the afterglow of the universe’s birth. What they thought was meaningless static turned out to be a message from the beginning of time. In a sense, the story of large scale white noise in cosmology follows a similar arc, but in reverse. Where the CMBR emerged from the small-scale thermal chaos of the early universe, LSWN arises not from primordial randomness, but from the inexorable unfolding of structure under gravity’s hand It is not imposed at the start, but generated as the universe evolves. Yet like that early static, it may masquerade as mere noise—subtle, overlooked, woven into the fabric of data we already have. The question now is not whether noise is present, but whether we’ve learned to listen for it in the right register.}

%%%%%%%%%%%%%%%%%%%%%%%%%%%%%%%%%%%%%%%%%%%%%%%%%%%%%%%%%%%%
%%%%%%%%%%%%%%%%%%%%%%%%%%%%%%%%%%%%%%%%%%%%%%%%%%%%%%%%%%%%
\section*{Acknowledgments}

 GB is supported by the Spanish grants  PID2020-113775GB-I00 
(AEI/10.13039/501100011033), CIPROM/2021/054 (Generalitat Valenciana), and by the European ITN project HIDDeN (H2020-MSCA-ITN-2019/860881-HIDDeN). AI is supported by NSF Grant PHY-2310429, Simons Investigator Award No.~824870, DOE HEP QuantISED award \#100495, the Gordon and Betty Moore Foundation Grant GBMF7946, and the U.S.~Department of Energy (DOE), Office of Science, National Quantum Information Science Research Centers, Superconducting Quantum Materials and Systems Center (SQMS) under contract No.~DEAC02-07CH11359. AS is supported by FermiForward Discovery Group, LLC.

This manuscript has been authored by FermiForward Discovery Group, LLC under Contract No. 89243024CSC000002 with the U.S. Department of Energy, Office of Science, Office of High Energy Physics.

%\bibliographystyle{JHEP}
%\bibliography{biblio}
%\printbibliography

\newpage
\begin{appendix}
%%%%%%%%%%%%%%%%%%%%%%%%%%%%%%%%%%%%%%%%%%%%%%%%%%%%%%%%%%%%
%%%%%%%%%%%%%%%%%%%%%%%%%%%%%%%%%%%%%%%%%%%%%%%%%%%%%%%%%%%%
\section{Conventions}\label{app:conventions}

\subsection{Definitions}

In this paper, we consider partial differential equations (PDEs) for a quantity $q(t,\vec{x})$, i.e. one real dependent variable $q$ which is a function of a temporal coordinate $t$ and a $d$ dimensional Euclidean spatial coordinates $\vec{x}=\{x_1,\cdots,x_d\}\in\mathbb{R}^d$.  Allowing for arbitrary $d$ adds little complexity. Spatial integral are
\begin{equation}\label{xintegral}
    \int d^d\vec{x}\,f(\vec{x}) \equiv \int_{-\infty}^\infty dx_1\, \cdots\int_{-\infty}^\infty dx_d\,f(\vec{x})
\end{equation}
and similarly for the wavenumber $\vec{k}\equiv\{k_1, \cdots, k_d\}$.  If the integrand depends only on the modulus of the wavenumber, $|\vec{k}|\equiv\sqrt{\sum_{i=1}^d {k_i}^2}$, then
\begin{equation}
    \int d^d\vec{k}\,f(|\vec{k}|) = V_{S^{d-1}}\int_0^\infty dk\,k^{d-1}\,f(k) \,,
\end{equation}
where $V_{S^{d-1}}=2\pi^{d/2}/\Gamma[d/2]$ is the volume of the unit $(d-1)$-sphere, e.g. $2,\,2\pi,\,4\pi$ for $d=1,2,3$.  We make extensive use of the identity
\begin{equation}\label{eq:deltaidentity}
    \int d^d\vec{x}\,e^{i\,\vec{k}\cdot\vec{x}} =(2\pi)^d\,\delta^{(d)}(\vec{k}) \quad \mathrm{where} \quad \delta^{(d)}(\vec{k}) \equiv \delta(k_1) \cdots \delta(k_d) \,,
\end{equation}
i.e. $\delta^{(d)}(\vec{k})$ is the $d$-dimensional $\delta$-function. Derivatives with respect to $t$ are denoted by overdots, e.g. $\dot{q}(t,\vec{x}) = \frac{\partial}{\partial t}\,q(t,\vec{x})$, $\ddot{q}(t,\vec{x})=\frac{\partial^2}{\partial t^2}\,q(t,\vec{x})$, etc. We will also used the notation that the $0^{\rm th}$ time derivative of a function is just that function, i.e. $\frac{\partial^0}{\partial t^0}q(t,\vec{x}) \equiv q(t,\vec{x})$.

We only consider {\it spatially homogeneous} PDEs, i.e. equations where if $q(t,\vec{x}) = f(t,\vec{x})$ is a solution then so is $q(t,\vec{x}) = f(t,\vec{x} + \vec{\Delta})$. We also only consider {\it homogeneous} PDEs, where $q(t,\vec{x}) = 0$ is a solution. We will use the acronym shhPDE for any homogeneous spatially homogeneous and homogeneous PDE. We will choose temporal variables where the PDE is non-singular in the interval $t_\mathrm{i}<t<\infty$.

We define the Fourier transform of $q$ and its inverse by
\begin{subequations}
\begin{equation}
    q(t,\vec{x}) = \bar{q}(t) + \int \frac{d^d k}{(2\pi)^{d/2}}\, e^{+i\,\vec{k}\cdot\vec{x}}\,\tilde{q}(t,\vec{k}) \,,
\end{equation}
\begin{equation}
    \tilde{q}(t,\vec{k}) = \int \frac{d^d x}{(2\pi)^{d/2}}   \, e^{-i\,\vec{k}\cdot\vec{x}} \, \left(q(t,\vec{x}) -\bar{q}(t) \right) \,,
\end{equation}
\end{subequations}
where $\bar{q}(t)$ is the spatial average value of $q(t,\vec{x})$. This transform plays a special role for spatially homogeneous PDEs because $e^{i\,\vec{k}\cdot\vec{x}}$ are eigenfunctions of spatial translations. Note that by definition $\int d^d k\,\tilde{q}(t,\vec{k})=0$. 

We define the statistical power spectrum $P_q(t,\vec{k})$ by the average over realizations
\begin{equation}
    \langle \tilde{q}(t,\vec{k})\,\tilde{q}(t,\vec{k}')\rangle 
    = (2\pi)^d \delta^{(d)}(\vec{k}+\vec{k}')\, P_q(t,\vec{k}) \,.
\end{equation}
The real-space statistical correlation function is
\begin{equation}
    \xi_q(t,\vec{x}-\vec{x}') \equiv \langle
    (q(t,\vec{x}) -\langle q(t,\vec{x}) \rangle)\,
    (q(t,\vec{x}')-\langle q(t,\vec{x}')\rangle)\rangle \\
    =\int d^d k \, P_q(t,\vec{k})\,
    e^{i\,\vec{k}\cdot(\vec{x}-\vec{x}')} \,,
\end{equation}
where by homogeneity $\langle q(t,\vec{x}) \rangle = \langle \bar{q}(t) \rangle$. 
A white noise power spectrum is $\vec{k}$-independent,
\begin{equation}
    P_q^{\rm WN}(t,\vec{k}) = N(t) \,,
\end{equation}
which corresponds to the real-space correlation function
\begin{equation}
    \xi_q^{\rm WN}(t,\vec{x}-\vec{x}'|) 
    =(2\pi)^d\,N(t)\,\delta^{(d)}(\vec{x}-\vec{x}') \,.
\end{equation}
A white noise spectrum then implies that fluctuations at separated spatial points are uncorrelated. Typically, physical systems which exhibit white noise spectra only do so up to a certain frequency, or equivalently down to a certain scale. White noise can exist on arbitrarily large scales but does not usually extend to arbitrarily small scales. 

It will also be useful to classify the distribution of the scalar field based on its large scale $k \rightarrow 0$ properties. Consider the long-wavelength limit of the power spectrum,
\begin{equation}
    P_q(t,\vec{0}^+) \equiv \lim_{\vec{k} \rightarrow \vec{0}} P_q(t,\vec{k}) \,.
\end{equation}
If $0 <P_q(t,\vec{0}^+)< \infty$, the distribution is said to be {\it Poissonian}, or ``white noise on large scales''. Meanwhile a distribution with $P_q(t,\vec{0}^+)=0$ is {\it sub-Poissonian} while a distribution with $P_q(t,\vec{0}^+) = \infty$ is {\it super-Poissonian}. As we will see in the main text, an initially sub-Poissonian distribution can dynamically evolve into a Poissonian distribution (and hence produce large scale white noise) in the presence of non-linearities in the equations of motion. 

\subsection{Notational Simplification}

For 2-point statistics of leading order $m$ non-linearities ne will need to manipulate two sets of $m$ wavenumbers.  We compactly denote these manipulations by making the substitutions
\begin{eqnarray}\label{eq:CompactIntegral}
    \vec{k}_1,\cdots,\vec{k}_m \rightarrow{\bf K} \,, \qquad \vec{k}_1',\cdots,\vec{k}_m'\rightarrow {\bf K}' \,, &\qquad& \sum_{i=1}^m \vec{k}_i \rightarrow\vec{K} \,, \qquad \sum_{i=1}^m \vec{k}_i'\rightarrow\vec{K}' \,, \nonumber 
    \nonumber \\
    \int\frac{d^d k_1}{(2\pi)^{d/2}} \cdots \int \frac{d^d k_m}{(2\pi)^{d/2}} f(\vec{k}_1,\cdots,\vec{k}_m) & \rightarrow & \int d^m{\bf K}\,f({\bf K})
    \nonumber \\
    \int\frac{d^d k_1'}{(2\pi)^{d/2}} \cdots \int\frac{d^d k_m'}{(2\pi)^{d/2}} f(\vec{k}_1',\cdots,\vec{k}_m') &\rightarrow& \int d^m{\bf K}'\,f({\bf K}') \,.
\end{eqnarray}
The order $m$ non-linearity will depend on Fourier amplitudes and their time derivatives, $\tilde{q}^{(l)}(t,\vec{k}) \equiv
\frac{\partial^l}{\partial t^l}q(t,\vec{k})$ for $0 \le l < p$. This set of quantities we denote compactly as ${\bf Q}$.

The Fourier transform of the order $m$ term in the Taylor series is
\begin{equation}
{}_{(m)}\hat{\tilde{\cal D}}\,\tilde{q}[t,\vec{k}]
\equiv\int\frac{{d^d\vec{x}}}{(2\pi)^{d/2}}\,
e^{-i\,\vec{k}\cdot\vec{x}}\,{}_{(m)}\hat{\cal D}\,q[t,\vec{x}] \ .
\label{eq:FourierOperators}
\end{equation}
For a {\it regular} differential operators this each element of the operator Taylor series can be decomposed into a sum of terms each of which can be factorized as follows 
\begin{eqnarray}
{}_{(m)}\hat{\tilde{\cal D}}\,\tilde{q}[t,\vec{k}]
&=&(2\pi)^{d/2}
\int d^m{\bf K}\,\delta^{(d)}[\vec{k}-\vec{K}]\,
N_{(m)}[t,{\bf Q},{\bf K}] \nonumber \\
N_{(m)}[t,{\bf Q},{\bf K}]&\equiv&
\sum_{j=1}^{n_m}c_{(m,j)}[t]\,d_{(m,j)}[{\bf K}]\,
\prod_{i=1}^m\tilde{q}^{(l_{(m,j,i)})}[t,\vec{k_i}]
\label{eq:NonlinearityExpansion}
\end{eqnarray}
where $\hat{d}_{(m,j)}[\bf{K}]$ is a multinomial in the spatial  of components the $\vec{k}_i$ and 
$\tilde{q}^{(l)}[t,\vec{k}]\equiv
\frac{\partial^l}{\partial t^l}q[t,\vec{k}]$. The number of terms in the sum, $n_m$, depends on the form of the non-linearity.

%%%%%%%%%%%%%%%%%%%%%%%%%%%%%%%%%%%%%%%%%%%%%%%%%%%%%%%%%%%%
%%%%%%%%%%%%%%%%%%%%%%%%%%%%%%%%%%%%%%%%%%%%%%%%%%%%%%%%%%%%
\section{Equivalent Systems}\label{app:EquivalentSystems}

A system is a PDE for a function $q(t,\vec{x})$ plus a statistical distribution of initial conditions at $t=t_\mathrm{i}$.  Here we define the equivalence class of systems by a restricted set of changes of variable.  Allowed transformations for equivalent systems are:
\begin{itemize}
\item[{\bf T1}] any {\it monotonically increasing smooth} change of the temporal variable:
$t\rightarrow\underline{t}[t]$.  For uniformity of presentation we will also require that $\underline{t}[\infty]=\infty$.
\item[{\bf T2}] any {\it linear time independent} change of the of spatial variable: $\vec{x}\rightarrow\underline{\vec{x}}
=\overleftrightarrow{T}\cdot\vec{x}+\vec{x}_0$ where $\overleftrightarrow{T}$ is a non-singular $d\times d$ matrix and $\vec{x}_0$ is a $d$-vector.
\item[{\bf T3}] any {\it linear time dependent smooth} change of the dependent variable:
$q[t,\vec{k}]\rightarrow\underline{q}[t,\vec{k}]=T[t]\,q[t,\vec{k}]$ where $0<T[t]<\infty$.
\item[{\bf T4}] any change of basis solutions of the linear equation: $Q_i[t,\vec{k}]\rightarrow\underline{Q}_i[t,\vec{k}]
=\sum_j T_{ij}[\vec{k}]\,Q_j[t,\vec{k}]$ where $T_{ij}[\vec{k}]$ is a non-singular $p\times p$ $\vec{k}$-dependent matrix.
\item[{\bf T5}] any {\it linear $k$-dependent rescaling} of the definition of the initial growing mode:
$G[t,\vec{k}]\rightarrow\underline{G}[t,\vec{k}]
=\mathcal{T}[\vec{k}]\,G[t,\vec{k}]$ where ${\cal T}[\vec{k}]\ne0$.
\end{itemize}
Smooth means that $\underline{t}$ and $T[t]$ are $C^p$.  Essentially all functions used in this section must be adjusted for these transformations but the three integer parameters $d$, $p$ and $n$ are not changed.  For {\bf T1} one can transform the fastest growing mode as
$F[t]\rightarrow\underline{F}[\underline{t}]
=T[\infty]^2\,\mathfrak{T}^2\,F[{\underline{t}}]$.
All of these transformations can be made for convenience, e.g. simplification, and in some cases can be thought of as simply a change of units.

Maintaining isotropy imposes the additional requirement that
\begin{itemize}
\item
$\overleftrightarrow{T}=\mathfrak{T}\,\overleftrightarrow{O}$where $\overleftrightarrow{O}$ is an orthogonal $d\times d$ matrix and $\mathfrak{T}>0$.
\end{itemize}
For isotropic transformations wavenumbers transform as
$k\equiv|\vec{k}|\rightarrow\underline{k}=\mathfrak{T}\,k$.
If a non-isotropic system is equivalent to an isotropic system it is probably simpler to use the isotropic form.

Non-linear transformations of $q$ or $\vec{x}$ are not allowed as they are the deformations discussed previously in \S\ref{sec:Deformations}.  Deformations fundamentally alter the meaning of the power spectrum as well as the nature of the non-linearities.  For example while they do not change $d$ or $p$ they can change $n$.

The $\tilde{g}$ and $\tilde{q}$ power spectra LSWN transform as
\begin{eqnarray}
R_{(2)}[k]\rightarrow R_{(2)}[\underline{k}]
&=&\mathcal{T}[\underline{k}/\mathfrak{T}]^2\,\mathfrak{T}^d
\,R_{(2)}[k]
\nonumber \\
P_{(2)}[t,\vec{k}]\rightarrow
\underline{P}_{(2)}[\underline{t},\vec{\underline{k}}]
&=&T[t[\underline{t}]]^2\,\mathfrak{T}^d\,
P_{(2)}[t[\underline{t}],\mathfrak{T}\,\underline{k}]
\label{PowerSpectraTransformationLaw}
\end{eqnarray}
The spectral shape of $R_{(2)}$ can be transformed arbitrarily as it depends on the arbitrary normalization of the initial growing mode functions, $G[t,k]$.  In contrast the shape of $P_{(2)}$ (on a log[$P_{(2)}$]-log[$k$] plot) is not transformed, but is shifted horizontally and vertically by a time dependent amount.  The asymptotic relic power transforms like
\begin{eqnarray}
P_\mathrm{rLSWN}[t]\rightarrow
\underline{P}_\mathrm{rLSWN}[\underline{t}]
&=&T[\infty]^2\,\mathfrak{T}^d\,
P_\mathrm{rLSWN}[t[\underline{t}]]
\label{PowerSpectraTransformationLaw}
\end{eqnarray}
i.e. is only changed by a constant factor.   The dimensionless $\mathcal{P}_\mathrm{aLSWN}$ is unchanged with the suggested transformation of the fastest growing mode.

%%%%%%%%%%%%%%%%%%%%%%%%%%%%%%%%%%%%%%%%%%%%%%%%%%%%%%%%%%%%
%%%%%%%%%%%%%%%%%%%%%%%%%%%%%%%%%%%%%%%%%%%%%%%%%%%%%%%%%%%%
\section{LSWN Kernel For Self-similar Local Systems}
\label{sec:calManalytic}

In \S\ref{sec:SelfSimilar}, we define the functions ${\cal M}_\pm[\varphi]$  and
${\cal M}_\pm[\varphi]={\cal M}_+[\varphi]+{\cal M}_-[\varphi]$ which give the LSWN for $n=p=2$ self similar systems.  These are defined the integrals of Eq.~\eqref{eq:M0GKGpm} using Eq.s~\ref{eq:Psidef}\,\&\,\ref{eq:SelfSimilarModes}.  These integrals can be performed analytically:
\begin{equation}
{\cal M}_\pm[\varphi]
=\sum_{a=1}^4\varpi_\mathrm{a}\,C_\pm^\mathrm{a}[\varphi]
\end{equation}
where
\begin{equation}
\varpi_1\equiv\zeta_3
\qquad
\varpi_2\equiv\zeta_4+2\,\mu\,\zeta_5
\qquad
\varpi_3\equiv\zeta_5
\qquad
 \varpi_3\equiv\zeta_7-\zeta_6 \ ,
\end{equation}
\begin{equation}
\begin{split}
&C_+^1[\varphi]\equiv
\frac{\Gamma[\frac{\nu+1}{2}]\Gamma[\frac{\mu+\nu}{2}]}{4\sqrt{\pi}\,\nu}\,
F_\mathrm{A}[\varphi]\,
\varphi^{2(\mu+\nu)}
\\
&C_+^2[\varphi]\equiv\frac{
\Gamma[\frac{\nu+1}{2}]\Gamma[\frac{\mu+\nu}{2}]}
{32\sqrt{\pi}\nu}
\left(4F_\mathrm{B}[\varphi]
+8\,\mu\, F_\mathrm{A}[\varphi]
-(\mu-\nu)(1+2\nu)\,F_\mathrm{C}[\varphi]\,\varphi^2
\right)\,\varphi^{2(\mu+\nu)}
\\
&C_+^3[\varphi]\equiv
\frac{\Gamma[\frac{\nu+1}{2}]\Gamma[\frac{\mu+\nu}{2}]}
{128\sqrt{\pi}\,\nu}\left(
\frac{16}{2\nu-1}F_\mathrm{D}[\varphi]
+\frac{2+\mu+\nu}{\mu-\nu}F_\mathrm{E}[\varphi]\varphi^2
\right)\varphi^{2(\mu+\nu)}+\frac{1}{4}\frac{\varphi^{2(\mu +1)}}{\nu(\nu-\mu)}
J_{\nu -1}[\varphi]J_{\nu +1}[\varphi]
\\
&C_+^4[\varphi]\equiv
\frac{
\Gamma[\frac{\nu+1}{2}]\Gamma[\frac{\mu+\nu}{2}]}
{8\sqrt{\pi}\,\nu}
F_\mathrm{F}[\varphi]\,\varphi^{2(\mu +\nu)}
\end{split} \ ,
\end{equation}
\begin{equation}
\begin{split}
&F_\mathrm{A}[\varphi]\equiv
\,_2\bar{F}_3
[{\scriptstyle\frac{\mu+\nu}{2},\frac{2\nu+1}{2}};
{\scriptstyle\frac{2+\mu+\nu}{2},1+\nu,1+2\nu};-\varphi^2]
\\
&F_\mathrm{B}[\varphi]\equiv
\,_2\bar{F}_3
[{\scriptstyle\frac{\mu+\nu}{2},\frac{2\nu +1}{2}};
{\scriptstyle\frac{2+\mu+\nu}{2},2\nu,1+\nu};-\varphi^2]
\\
&F_\mathrm{C}[\varphi]\equiv
\,_2\bar{F}_3
[{\scriptstyle\frac{2+\mu+\nu}{2},\frac{2\nu+3}{2}};
{\scriptstyle\frac{4+\mu +\nu}{2},2+\nu,2(1+\nu)};-\varphi^2]
\\
&F_\mathrm{D}[\varphi]\equiv
\,_2\bar{F}_3
[{\scriptstyle\frac{\mu+\nu }{2},\frac{2\nu-1}{2}};
{\scriptstyle\frac{2+\mu+\nu}{2},\nu,2\nu-1};-\varphi^2]
\\
&F_\mathrm{E}[\varphi]\equiv
\,_2\bar{F}_2
[{\scriptstyle\frac{4+\mu+\nu }{2},\frac{2\nu +3}{2}};
{\scriptstyle\frac{6+\mu +\nu }{2},2+\nu ,3+2\nu};-\varphi^2]
\\
&F_\mathrm{F}[\varphi]\equiv
\,_2\bar{F}_3
[{\scriptstyle\frac{2+\mu+\nu}{2},\frac{2\nu+1}{2}};
{\scriptstyle\frac{4+\mu+\nu}{2},1+\nu,1+2\nu};-\varphi^2]
\end{split} \ ,
\end{equation}
and
\begin{equation}
\begin{split}
_2\bar{F}_3[a_1,a_2;b_1,b_2,b_3;x]&\equiv
\frac{_2F_3[a_1,a_2;b_1,b_2,b_3;x]}
{\Gamma[b_1]\,\Gamma[b_2]\,\Gamma[b_3]} \\
_2F_3[a_1,a_2;b_1,b_2,b_3;x]&\equiv
\sum_{k=1}^\infty
\frac{
\frac{\Gamma[a_1+k]}{\Gamma[a_1]}\,
\frac{\Gamma[a_2+k]}{\Gamma[a_2]}}
{
\frac{\Gamma[b_1+k]}{\Gamma[b_1]}\,
\frac{\Gamma[b_2+k]}{\Gamma[b_2]}\,
\frac{\Gamma[b_3+k]}{\Gamma[b_3]}}
\end{split}\ .
\end{equation}
${}_pF_q$ is the generalized hypergeometric function and ${}_p\bar{F}_q$ is
the regularized generalized hypergeometric function preferred by Mathematica.

For small and large  $\varphi$
\begin{equation}
\begin{split}
\lim_{\varphi\rightarrow0}
&\,_2\bar{F}_3[a_1,a_2;b_1,b_2,b_3;-\varphi^2]
=\frac{1}{\Gamma[b_1]\,\Gamma[b_2]\,\Gamma[b_3]}\,\\
\lim_{\varphi\rightarrow\infty}
&\,_2\bar{F}_3[a_1,a_2;b_1,b_2,b_3;-\varphi^2]
=\frac{\cos[2\varphi+\frac{\pi}{4}\,c]}
{\sqrt{\pi}\,\Gamma[a_1]\,\Gamma[a_2]}\,\varphi^c +\\
&\frac{\Gamma[a_2-a_1]\,\varphi^{-2a_1}}
{\Gamma[a_2]\,\Gamma[b_1-a_1]\,\Gamma[b_2-a_1]\,\Gamma[b_3-a_1]}
+\frac{\Gamma[a_1-a_2]\,\varphi^{-2a_2}}
{\Gamma[a_1]\,\Gamma[b_1-a_2]\,\Gamma[b_2-a_2]\,\Gamma[b_3-a_2]}
\end{split}
\end{equation}
where $c\equiv\frac{1}{2}+\sum_{i=1}^2a_i-\sum_{j=1}^3b_j>0$. Depending on the values of $a_1$, $a_2$ and $c$ the asymptotic form of $\bar{F}$ is either oscillatory or a power law. 

The ${\cal M}_\pm[\varphi]$ are weighted sum of these $\bar{F}$'s with varying values of the $a_i$ and $b_i$.  Finite 
${\cal M}_\pm$ requires that $\nu>0$ and $\mu+\nu>0$ which makes the oscillatory asymptotes subdominant.  The small and large $\varphi$ limits are
\begin{equation}
\begin{split}
\lim_{\varphi\rightarrow0}     {\cal M}_\pm[\varphi]=&
\varsigma_0^\pm\,\varphi^{2(\mu+\nu)}
\quad\mathrm{if}\quad\varsigma_0^\pm\ne0
\\
\lim_{\varphi\rightarrow\infty}{\cal M}_\pm[\varphi]=&
\begin{cases}
\varsigma_1^\pm\,\varphi^{2\mu+1}    &
\quad\mathrm{ if}\quad1+\mu>\pm\,\nu
\quad\mathrm{and}\quad\varsigma_1^\pm\ne0\\
\varsigma_2^\pm\,\varphi^{\mu\pm\nu} &
\quad\mathrm{ if}\quad1+\mu<\pm\,\nu
\quad\mathrm{and}\quad\varsigma_2^\pm\ne0
\end{cases}
\end{split}
\end{equation}
where
\begin{equation}
\begin{split}
\varsigma_0^\pm\equiv&\pm
\frac{\zeta_3+(\mu+\nu)(\zeta_4+(\mu+\nu))\zeta_5}
{2^{1+2\nu}\,\nu\,(\mu+2\,\nu\mp\nu)\,\Gamma[1+\nu]^2}
\\
\varsigma_1^\pm\equiv&
\pm\frac{1}{2\pi\,\nu}
\frac{\zeta_5-\zeta_6+\zeta_7}{1+\mu\mp\nu} \\
\varsigma_2^\pm\equiv&
-\frac{
\Gamma[\frac{\mu\mp3\nu}{2}]
\Gamma[\frac{-1-\mu\pm\nu}{2}]
\Gamma[\frac{\mu\pm\nu}{2}]
}{
32\,\pi^{3/2}\Gamma[\frac{2-\mu\pm\nu}{2}]
}
\sin[\frac{\pi}{2}(\mu-(2\pm1)\nu)]
\left(
\begin{matrix}
4(1+\mu\mp\nu) \\
2(\mu\pm\nu)(1+\mu\mp\nu) \\
(\mu^2-\nu^2)(2+\mu\pm\nu) \\
+(\mu^2-\nu^2)(\mu\mp3\nu) \\
-(\mu^2-\nu^2)(\mu\mp3\nu)
\end{matrix}
\right)\cdot
\left(
\begin{matrix}
\zeta_3 \\
\zeta_4 \\
\zeta_5 \\
\zeta_6 \\
\zeta_7
\end{matrix}
\right)
\end{split} \ .
\label{eq:varsigmas}
\end{equation}
These limiting forms are for "generic" parameter values which doesn't include special cases, e.g. values of $\nu$, $\mu$ and $\zeta_i$ where $\varsigma_i^\pm=0$.

For the integral definition of ${\cal M}_\pm[\varphi]$ (Eq.~\eqref{eq:M0GKG0}) to converge we require that $\nu>0$ and $\mu+\nu>0$.  The condition that 
$1+\mu<-\nu$ is never satisfied and hence $\varsigma_2^-$ is irrelevant.  It is however possible that $1+\mu<\nu$ if 
$\nu>\frac{1}{2}$ so $\varsigma_2^+$ can be relevant.  One finds that
\begin{equation}
|\varsigma_0^+|>|\varsigma_0^-|\quad
|\varsigma_1^+|>|\varsigma_1^-| \quad
\varsigma_0^+\,\varsigma_0^-<0 \quad
\varsigma_1^+\,\varsigma_1^-<0 \quad
\varsigma_2^+\,\varsigma_2^-<0 \quad
|\varsigma_0^+|\sim|\varsigma_0^-|\quad
|\varsigma_1^+|\sim|\varsigma_1^-| \quad
\end{equation}
so that if $1+\mu>\nu$ for both large and small $\varphi$
\begin{itemize}
\item ${\cal M}_+[\varphi]{\cal M}_-[\varphi]<0$ 
\item $|{\cal M}_+[\varphi]|>|{\cal M}_-[\varphi]|$
\item $|{\cal M}_+[\varphi]|\sim|{\cal M}_-[\varphi]|$.
\end{itemize}
where $\sim$ indicates they are the same order of magnitude.  If $1+\mu<\nu$ then 
${\cal M}_+[\varphi]\propto\varphi^{\mu+\nu}$ and 
${\cal M}_-[\varphi]\propto\varphi^{2\mu+1}$,  The exponent of the 1st is greater than the exponent of the 2nd so
\begin{itemize}
\item $|{\cal M}_+[\varphi]|\gg|{\cal M}_-[\varphi]|$
\end{itemize}
in this case.

Using the previous results on finds that for both large and small $\varphi$ that ${\cal M}_+[\varphi]$, ${\cal M}_-[\varphi]$ and ${\cal M}[\varphi]$ are power laws in $\varphi$. For the sum
\begin{equation}
\begin{split}
&\lim_{\varphi\rightarrow0}{\cal M}[\varphi]
={\mathfrak A}\,\varphi^{2(\mu+\nu)} \qquad
\lim_{\varphi\rightarrow\infty}{\cal M}[\varphi]
={\mathfrak B}\,\varphi^{\alpha}
\\&
{\mathfrak A}=\varsigma_0^++\varsigma_0^-
=\frac{\zeta_3+(\mu+\nu)(\zeta_4+(\mu+\nu)\zeta_5)}
{2^{2\nu}\,(\mu+\nu)\,(\mu+3\,\nu)\,\Gamma[1+\nu]^2}
\\&
{\mathfrak B}=
\begin{cases}
\varsigma_1^++\varsigma_1^-
=\frac{(1+\mu)\,\,(\zeta_5-\zeta_6-\zeta_7)}
{\pi\,\nu\,((1+\mu)^2-\nu^2)} &  1+\mu>\nu\\
\varsigma_2^+& 1+\mu<\nu 
\end{cases}
\\&
\alpha=
\begin{cases}
2\,\mu+1 & 1+\mu>\nu \\
\mu+\nu& 1+\mu<\nu 
\end{cases}
\end{split} \ .
\label{eq:calMasymptoticGeneral}
\end{equation}
The power law exponent is larger at small $\varphi$ and smaller at large $\varphi$  (producing a "knee" in ${\cal M}[\varphi]$). In no case is the exponent smaller for $\varphi\ll1$ than for $\varphi\ll1$ (producing an "ankle" in ${\cal M}[\varphi]$) nor can the exponents at $\varphi\ll1$ and at $\varphi\gg1$ be the same.

\begin{equation}
\langle\prod_{i=1}^n\tilde{g}[\vec{k}_i ]\,
       \prod_{j=1}^n\tilde{g}[\vec{k}_j']\rangle
=(2\pi)^d\,\delta^{(d)}[\vec{K}+\vec{K}']\,
R_{(n,n)}[{\bf K},{\bf K}'] \ .
\end{equation}

\end{appendix}

\bibliography{LSWN}

@article{EyinkSreenivasan2006,
  author = {Eyink, Gregory L. and Sreenivasan, K. R.},
  title = {Onsager's conjecture and the energy dissipation in turbulence},
  journal = {Reviews of Modern Physics},
  year = {2006},
  volume = {78},
  pages = {87--135},
  doi = {10.1103/RevModPhys.78.87}
}

@book{Nazarenko2011,
  author = {Nazarenko, Sergey},
  title = {Wave Turbulence},
  publisher = {Springer},
  year = {2011},
  address = {Berlin},
  doi = {10.1007/978-3-642-15942-8}
}

@book{Frisch1995,
  author = {Frisch, Uriel},
  title = {Turbulence: The Legacy of A.N. Kolmogorov},
  publisher = {Cambridge University Press},
  year = {1995},
  address = {Cambridge},
  isbn = {978-0-521-55749-1}
}

@book{Agrawal2001,
  author = {Agrawal, Govind P.},
  title = {Nonlinear Fiber Optics},
  publisher = {Academic Press},
  year = {2001},
  address = {San Diego},
  isbn = {978-0-12-121732-0}
}

@book{Cohen1989,
  author = {Cohen, L.},
  title = {Time-Frequency Analysis},
  publisher = {Prentice Hall},
  year = {1989},
  address = {Upper Saddle River, NJ},
  isbn = {978-0137448284}
}

@article{Lindborg_Nordmark_2022,
    title={Two-dimensional turbulence on a sphere},
    volume={933},
    DOI={10.1017/jfm.2021.1130},
    journal={Journal of Fluid Mechanics},
    author={Lindborg, Erik and Nordmark, Arne},
    year={2022},
    pages={A60}}

@book{batchelor1967introduction,
  author = {Batchelor, G. K.},
  doi = {10.1017/CBO9780511800955},
  publisher = {Cambridge University Press},
  title = {An Introduction to Fluid Dynamics},
  url = {https://www.cambridge.org/core/books/an-introduction-to-fluid-dynamics/18AA1576B9C579CE25621E80F9266993},
  year = 1967
}

@article{Fjortoft1953,
    author = "Fj{\o}rtoft, Ragner",
    title = "{On the changes in the spectral distribution of kinetic energy
    for two-dimensional, nondivergent flow}",
    journal={Tellus},
    volume={5},
    number={3},
    pages={225},
    month = "4",
    year = "1953",
    publisher={Svenska Geofysiska F{\o}reningen}
}

@book{Peebles1971,
  author    = {Peebles, P. J. E.},
  title     = {Physical Cosmology},
  publisher = {Princeton University Press},
  address   = {Princeton, NJ},
  year      = {1971},
  isbn      = {9780691080896}
}

@article{Zeldovich1972,
  author  = {Zel'dovich, Ya. B.},
  title   = {A Hypothesis, Unifying the Structure and the Entropy of the Universe},
  journal = {Monthly Notices of the Royal Astronomical Society},
  volume  = {160},
  pages   = {1P--3P},
  year    = {1972},
  doi     = {10.1093/mnras/160.1.1P}
}

@book{Peebles1980,
  author    = {Peebles, P. J. E.},
  title     = {The Large-Scale Structure of the Universe},
  publisher = {Princeton University Press},
  address   = {Princeton, NJ},
  year      = {1980},
  isbn      = {9780691082401}
}

@book{Weinberg2008,
  author    = {Weinberg, Steven},
  title     = {Cosmology},
  publisher = {Oxford University Press},
  address   = {Oxford, UK},
  year      = {2008},
  isbn      = {9780198526827}
}

@article{Planck:2018vyg,
    author = "Aghanim, N. and others",
    collaboration = "Planck",
    title = "{Planck 2018 results. VI. Cosmological parameters}",
    eprint = "1807.06209",
    archivePrefix = "arXiv",
    primaryClass = "astro-ph.CO",
    doi = "10.1051/0004-6361/201833910",
    journal = "Astron. Astrophys.",
    volume = "641",
    pages = "A6",
    year = "2020",
    note = "[Erratum: Astron.Astrophys. 652, C4 (2021)]"
}

@misc{2026arXiv260116996S,
    author = "Stebbins, Albert",
    title = "{A Space-Time Fluid (Unabridged)}",
    eprint = "2601.16996",
    archivePrefix = "arXiv",
    primaryClass = "physics.gen-ph",
    reportNumber = "FERMILAB-PUB-25-0256-T",
    month = "1",
    year = "2026"
}

@unpublished{sandbox2026,
    author = {Gabriela Barenboim and Aurora Ireland and Albert Stebbins},
    title = {Planar Cosmology as a Sandbox for Understanding Non-linearities},
    year = 2026,
    note = {in preperation}
}

@misc{Barenboim2025,
    author = "Barenboim, Gabriela and Ireland, Aurora and Stebbins, Albert",
    title = "{The Noisy Universe}",
    eprint = "2511.15803",
    archivePrefix = "arXiv",
    primaryClass = "astro-ph.CO",
    reportNumber = "FERMILAB-PUB-25-0855-PPD",
    month = "11",
    year = "2025"
}

@article{Bardeen:1980kt,
    author = "Bardeen, James M.",
    title = "{Gauge Invariant Cosmological Perturbations}",
    doi = "10.1103/PhysRevD.22.1882",
    journal = "Phys. Rev. D",
    volume = "22",
    pages = "1882--1905",
    year = "1980"
}

@article{Kodama:1984ziu,
    author = "Kodama, Hideo and Sasaki, Misao",
    title = "{Cosmological Perturbation Theory}",
    doi = "10.1143/PTPS.78.1",
    journal = "Prog. Theor. Phys. Suppl.",
    volume = "78",
    pages = "1--166",
    year = "1984"
}

@book{Hawking:1973uf,
    author = {S.~W.~Hawking and G.~F.~R.~Ellis},
    title = {The Large Scale Structure of Space-Time},
    publisher = {Cambridge University Press},
    series = {Cambridge Monographs on Mathematical Physics},
    month = {2},
    year = {2023}
}

@ARTICLE{1989PhRvD..40.1804E,
       author = {{Ellis}, G.~F.~R. and {Bruni}, M.},
        title = "{Covariant and gauge-invariant approach to cosmological density fluctuations}",
      journal = {\prd},
         year = 1989,
        month = sep,
       volume = {40},
       number = {6},
        pages = {1804-1818},
          doi = {10.1103/PhysRevD.40.1804},
       adsurl = {https://ui.adsabs.harvard.edu/abs/1989PhRvD..40.1804E}
}

@book{ellis2012relativistic,
title={Relativistic Cosmology},
author={Ellis, George F. R. and Maartens, Roy and MacCallum, Malcolm A. H.},
year={2012},
publisher={Cambridge University Press},
address={Cambridge},
isbn={9780521381154}
}

@article{Tsagas:2007yx,
    author = "Tsagas, Christos G. and Challinor, Anthony and Maartens, Roy",
    title = "{Relativistic cosmology and large-scale structure}",
    eprint = "0705.4397",
    archivePrefix = "arXiv",
    primaryClass = "astro-ph",
    doi = "10.1016/j.physrep.2008.03.003",
    journal = "Phys. Rept.",
    volume = "465",
    pages = "61--147",
    year = "2008"
}

\end{document}